%% file: prom20.tex
\documentclass[12pt]{report}

\usepackage{a4}
\usepackage{epsfig}
\usepackage{epsf}
\usepackage{rotating}

\def\parag            {\vspace{2ex}}
          
\def\equal            {\leftrightarrow }

\def\ul               {\underline }     

\renewcommand{\Re}    {{\rm Re}}

\def\1,2              {{1\over 2}}

\def\r[#1]            {\quad {\rm #1} \quad}
\def\nn               {\nonumber}
\def\id               {\mbox{\large\bf 1}}

\def\ma[#1,#2,#3,#4]  {{\left( \matrix{ #1  & #2 \cr
                                        #3  & #4 \cr } \right)}}
\def\ve[#1,#2]        {{\left( \matrix{ #1  \cr
                                        #2  \cr } \right)}}

\def\gsim             {\stackrel{\textstyle>}{\sim}}

\begin{document}

\begin{titlepage}

\thispagestyle{empty}
\hbox{}
\begin{center}
\vspace*{-0.4cm}

\renewcommand{\thefootnote}{\fnsymbol{footnote}}
\setcounter{footnote}{1}

{\LARGE 
The Local Bosonic Algorithm applied to the massive Schwinger model}

\vspace*{0.9cm}
{\large D i s s e r t a t i o n} \\ 

\vspace{0.4cm}
{\large
zur Erlangung des akademischen Grades \\
doctor rerum naturalium \\
{\normalsize (Dr.~rer.~nat.)}
}

\vspace{0.4cm}
{\large
{\normalsize eingereicht am} \\
Institut f\"ur Physik\\
der Mathematisch--Naturwissenschaftlichen Fakult\"at~I\\
der Humboldt~-- Universit\"at zu Berlin\\
\vspace{0.5cm}
{\normalsize von}\\
Dipl.-Phys. Stephan Elser \\
geb. am 5.~Februar~1968 in Gaildorf\\
\vspace{0.9cm}
Pr\"asident der Humboldt-Universit\"at zu Berlin\\
Prof. Dr. H. Meyer \\
\vspace{0.5cm}
Dekanin der Math.-Nat. Fakult\"at I \\
Prof. Dr. V. Bonacic-Kouteck\'y \\
\vspace{0.7cm}
\begin{tabular}{cl}
Gutachter~:& 1.  {\sl  ..................................... }\cr
           & 2.  {\sl  ..................................... }\cr
           & 3.  {\sl  ..................................... }\cr
\end{tabular}
\vfill
Tag der m\"undlichen Pr\"ufung~:~ {\sl ....................... }
}

\end{center}
\pagebreak

\vfill

\end{titlepage}

\begin{abstract}

We investigate 
various variants of the Hermitean version
of the Local Bosonic Algorithm
proposed by M. L\"uscher.
The model used is  two-dimensional Quantum Electrodynamics
(QED) 
with two flavours of massive
Wilson fermions.
The simplicity of the model allows for high statistics simulations 
close to
the chiral and continuum limit.

To find optimal
CPU cost behaviour,
we 
carefully scan a 3-dimensional para\-meter subspace
varying the approximation polynomial parameters
$n$ and $\epsilon$ as well as
the number of over-relaxation steps within each update
trajectory.
We find flat behaviour around the optimum
and a modest gain with respect to 
the Hybrid Monte Carlo algorithm
for all variants.
Generally, the gain is slightly smaller
for the reweighting method
than for the acceptance step variants
and quite
different for 
plaquette-like and correlator-like observables.

On the technical side,
we demonstrate that
a noisy Metropolis acceptance step
is possible also for the Hermitean variant.
The 
numerical instabilities appearing in the
evaluation of the factorized form for the
approximating Chebyshev  polynomial
are investigated.
We propose a quantitative criterion 
for these instabilities
and a reordering scheme of the roots
reducing the problem.

A different formulation
of the Hermitean Local Bosonic Algorithm 
using an acceptance step 
which generally 
avoids these instabilities
was recently proposed.
We compare the CPU cost to that of Hybrid Monte Carlo and 
to the more standard Local Bosonic Algorithm variants
and find compatible results.

The more physical 
problem of topological 
charge sectors and metastability is addressed.
We find no plateau in the
effective pion mass
if metastabilities become too large.

\end{abstract}

\thispagestyle{empty}
\setcounter{page}{0}
\vspace*{2.75truecm}
\begin{center}
{\bf Zusammenfassung}
\end{center}
\vspace{0.15cm}
Gegenstand der Untersuchung sind
Varianten der Hermiteschen Version des
von M. L\"uscher vorgeschlagenen {\em Local Bosonic Algorithm}. 
Das dabei benutzte Modell ist die
zweidimensionale Quantenelektrodynamik
(QED) 
mit
2 massiven Flavours von Wilson-Fermionen.
Die Einfachheit des Modells
erm\"oglicht
Simulationen mit hoher Statistik
nahe am chiralen und Kontinuumslimes.

Um Bereiche mit optimalem CPU-Kostenverhalten
zu finden,
untersuchen wir
einen 3-dimensionalen
Parameter-Unterraum,
wobei wir die Parameter des 
N\"ahe\-rungs\-poly\-noms $n$ und $\epsilon$
sowie die Anzahl der \"Uberrelaxationsschritte pro 
Update-Trajektorie variieren.
Wir erhalten dabei 
bei allen Varianten
ein flaches Verhalten um das Optimum
und einen
kleinen Kostengewinn
im Vergleich zum Hybrid Monte Carlo Algorithmus.
Der Gewinn ist generell
f\"ur die Reweighting Methode
etwas kleiner 
als f\"ur die Ak\-zep\-tanz\-schritt-Varianten
und deutlich 
unterschiedlich
f\"ur Plaquette-\"ahnliche bzw.
Korrelator-\"ahnliche Observablen.

Auf der  technischen Seite
zeigen wir,
da{\ss}
ein stochastischer
Metropolis-Ak\-zep\-tanz\-schritt
auch in der Hermiteschen Variante m\"oglich ist.
Die numerischen Instabilit\"aten,
die bei der Berechnung
der Chebyshev-N\"ahe\-rungs\-poly\-nome
in der faktorisierten Form auftreten,
werden untersucht.
Wir schlagen
ein quantitatives Kriterium 
f\"ur die Instabilit\"aten 
und ein Umordnungsschema f\"ur die Nullstellen
des Polynoms,
das dieses Problem reduziert, vor.

Eine alternative Formulierung des Hermiteschen 
{\em Local Bosonic Algorithm}
mit Akzeptanzschritt,
die diese Instabilit\"aten generell vermeidet,
wurde vor kurzen vorgeschlagen.
Wir vergleichen die CPU Kosten
mit denen des Hybrid Monte Carlo Algorithmus
und der \"ublichen {\em Local Bosonic Algorithm}
Varianten und finden kompatible Resultate.

Im Weiteren wird das mehr physikalische
Problem der topologischen Ladungssektoren
und der Metastabilit\"at behandelt.
Wir finden kein Plateau der effektiven Pionenmasse,
falls Metastabilit\"aten zu gro{\ss} werden.


\tableofcontents


\chapter{Introduction}

The task of physics
is the determination of all relevant 
quantities of Nature from 
basic ``principles''.
In the case of elementary particle physics
the relevant guidelines
which have proven to be successful
are
the symmetry and unification principles.
Their application  leads to the so-called
standard model \cite{standard_model},
incorporating 3 of the 4 known basic forces,
namely the 
electro-magnetic, weak and strong interactions.
Basic degrees of freedom
of the standard model
are 
the
fundamental
fermions (quarks and leptons)
and 
their interaction with each other is
described via the 
exchange of bosonic gauge quanta.
An important part is 
formed by the Higgs mechanism,
generating mass
for fermions and massive gauge bosons
and introducing the 
scalar Higgs particle
into the model.
The theory
is formulated
as a 
quantised 
relativistic local
gauge theory
coupling the symmetry principle with
the dynamics of fields \cite{gauge_theory,cheng_book}.
Main building blocks of the
standard model are
Quantum Electrodynamics (QED) 
with symmetry group $U(1)$
(as part of the electro-weak sector $U(1) \otimes SU(2)$)
and
Quantum Chromodynamics (QCD),
relying on the symmetry group $SU(3)$.

Yet the standard 
model leaves a lot to be explained.
It has as many as 19 free parameters \cite{book_standard_model}
and no reason is
given why a symmetry of $SU(3) \otimes SU(2) \otimes U(1)$
is realized in nature.
Thus we would like a more advanced theory
with fewer parameters and
an even more symmetric structure.
Possible ways of extending the standard model
include
the search for more fundamental constituents
of quarks and leptons,
string theory including supersymmetric partners
to the known elementary particles
and grand unified theories.
In these higher dimensional symmetry groups
are applied at very large energy scales,
which for lower energies are broken
down to the familiar standard model structure \cite{unification}.

One main research focus 
is the relativistic $SU(3)$ 
gauge theory including dynamical fermions,
Quantum Chromodynamics (QCD) \cite{qcd},
which describes hadrons and nuclei
under strong interaction with the help of
quarks and gluons
as fundamental degrees of freedom.
The beta-function is known at
3-loop level,
leading in the high energy limit
(or equivalently at small distances)
to asymptotic freedom,
i.e. the appearance 
of almost unbound quarks.
In that case,
the running coupling constant
is small enough
to allow perturbative methods.
Applying these,
there exist in this regime
convincing 
predictions
of the theory,
which agree well with  experiment
\cite{qcd_book}.

On the energy scale of the hadrons themselves,
i.e. for the 
long distance part of the strong interaction,
quasi-free quarks were never observed.
The quark coupling
in this regime is too large
for perturbation theory around the free solution
to be sensible.
We therefore need non-perturbative
methods 
to shed light on e.g. the questions
of quark confinement
and
hadron masses.

Among the attempts to reach 
non-perturbative insight
into the structure of QCD 
ap\-proa\-ches using
QCD sum rules,
large $N$ expansions,
the Bethe-Salpeter equation
or Discretized Light-Cone Quantisation \cite{dlcq_overview}
can be mentioned
besides effective purely phenomenological models.
For an overview discussion we
refer to the textbook of Narison \cite{sumrules_book}.

Compared to these approaches,
lattice gauge calculations
offer a way to results
stemming directly from the
use of first principles.
Lattice gauge theory
uses a
path integral representation
with a space-time grid
regularisation 
with lattice spacing $a$
\cite{wilson}.
The theory is set up in such a way
that
the naive continuum limit $a \to 0$
yields 
the desired continuum Lagrangian to leading 
order in the lattice spacing.
This of course offers a certain freedom
in the choice of the lattice theory.
In this approach all 
quantities are measured in units
of the lattice spacing or inverses thereof.
Results have to be extrapolated to the
continuum limit,
in which
the lattice spacing $a$ is set to zero and
the dimensionless
lattice size $L$ to infinity.
In order to still have meaningful results,
a continuous phase transition point has to be chosen
for this procedure.
At this point,
all correlation functions diverge,
so that results in physical units can still be finite.
In this limit,
we will
therefore
retain 
ratios of (e.g. mass) observables
corresponding to relations
between renormalised quantities.
An important point is
that certain continuum symmetries
(like e.g. Lorentz symmetry)
can be broken on a space time grid.
The restoration of these symmetries has to
be checked in order to justify extrapolation results.
For completeness we mention that
in order to exclude a 
contamination of the results
by finite size effects,
one would like to send also the
physical lattice extent $a\cdot L$ to infinity,
a limit called the thermodynamic limit.

One non-perturbative 
approach
to lattice gauge theory
obtains results 
with the help of Monte Carlo methods. 
Introductions can be found in \cite{creutz_book, diplomarbeit, montvay_book}.
To illustrate the success
of this approach,
we mention that a recent work 
was able to determine 
ratios of hadron masses to
an error of $2\%$ \cite{hadron_masses}.
Nevertheless, we are still
far away
from a complete understanding of the
physics of
hadronic systems using lattice gauge theory.

A severe problem appearing in
simulations
is the phenomenon of
critical slowing down,
i.e.
the fact that the CPU cost
of a simulations grows 
more than proportional to the
space-time volume of the lattice \cite{monte_carlo}.
Especially
fermion simulations
are 
very time consuming \cite{algorithms_overview}.
They face the problem that
the fermion interaction is inherently
non-local,
being described by
a determinant in the path integral.
Thus, up to now
most large-scale calculations
apply the so-called quenched approximation,
replacing this determinant by a constant.
Thereby the dynamics of fermionic vacuum fluctuations 
interacting
with a bosonic gauge field are
ignored.

The standard algorithm used to include dynamical fermions
is the so-called Hybrid Monte Carlo algorithm \cite{hmc}.
Here a trajectory of small size update steps
is generated introducing a fictitious computer time
coordinate,
conjugate momenta and 
the corresponding classical equations of motion.
At the end of the trajectory
an acceptance step is used to make
the algorithm exact.
Other approaches, e.g. based on
Kramers equation \cite{horowitz,kramers}, 
have not shown significantly better behaviour
than Hybrid Monte Carlo.

Recently,
an alternative approach was proposed 
by M. L\"uscher \cite{lba}.
This so-called Local Bosonic Algorithm (LBA)
applies an $n$-th order polynomial approximation
to the fermion determinant
and thus makes it possible
to rewrite the determinant as a set of $n$
local and bosonic integrals
plus a non-local correction term.
As simulations of locally coupled 
degrees of freedom
are much easier, 
this could potentially
reduce the computational task.
This algorithm has 
generated considerable interest
\cite{lba_improvements1,lgt_overview,lba_exact,kramersboson,algorithms_developments,wolff}.
Examples for applications of the 
local bosonic algorithm are
Monte Carlo simulations of lattice QCD
\cite{lgt_overview,algorithms_developments}, 
Supersymmetry
\cite{susy},
the 
Hubbard model \cite{hubbard_model}
and the Schwinger model 
both with staggered \cite{lba_staggered}
and Wilson fermions \cite{els_lattice96,els_lattice97}. 
 
Of course,
this new algorithm has to be tested and
its efficiency
compared to the standard
Hybrid Monte Carlo algorithm (HMC)
\cite{lba_analysis,lba_improvements1,
lba_improvements2}.
In this context it may be noted that
it is well known
how to tune the parameters of
Hybrid Monte Carlo routines,
whereas this is still under investigation
for the local bosonic case.

For the proposed studies of algorithms for dynamical fermions that 
are rather time-consuming,
we 
want to advocate 
the massive 2-flavour Schwinger model (2-dimensional QED)
\cite{schwinger,2f}
as a low-cost laboratory \cite{testbed}.
With the appearance of an axial anomaly,
confinement,
light pseudo-scalar and heavy scalar mesons,
it has  
enough rich physics to be similar to QCD.
One of the aims is to set up
a program package
facilitating further studies.

This work 
introduces the Schwinger model variants
commonly used,
and gives some details on the
lattice 2-flavour Schwinger model.
We will
briefly describe the Hybrid Monte Carlo algorithm applied
to set the scale for the CPU cost comparison
and
the four variants of the Hermitean version of the local bosonic
algorithm
we are testing.
We will
discuss some problems
regarding numerical instabilities and
topological metastabilities
encountered in our simulations. 
Finally,
we give results
of our investigation of the CPU cost of
the plaquette and the pion correlator.

Throughout this work,
figures and tables are defered
to the end of each chapter.

\chapter{The model: Schwinger model -- 2D QED}

\section{Continuum}

Quantum Electrodynamics (QED) 
in one space and one time
dimension ($d=2$) with $N_f$ flavours
of fermions having mass $m_a$, $a=1 \dots N_f$,
is generally defined by the
continuum Lagrangian
in Euclidean space
\begin{eqnarray}
\label{e_pathintegral}
{\cal L}
=
{1 \over 4} F^2
+
\sum_{a=1}^{N_f} \Bigl[
\bar\psi^a ( / \hspace*{-0.195truecm} \partial 
+ i e / \hspace*{-0.25truecm} A + m_a)
\psi^a
\Bigr]
\; ,
\end{eqnarray}
where we use the standard conventions
$
\hbar =c=1 $.
The
fermion fields $\psi$ 
are Grassmann 2-spinors having dimension $[m]^{\1,2 }$
and the electromagnetic tensor 
\begin{eqnarray}
F^{\mu\nu} = \ma[0,-E,E,0] 
\end{eqnarray}
in 1+1 dimensions 
only includes the electric field $E$,
as because of the missing transverse
directions no magnetic field is existing.
We point out that the mass and the electric charge are of the same
dimension.

For an introduction to  Euclidean space 
we defer the reader to \cite{montvay_book}.
Conventions are 
generally collected in  App.~\ref{c_conventions}.

\subsection{Massless 1-flavour model}

2-dimensional 1-flavour
QED
with massless fermions, i.e. $N_f=1, m_1=0$,
was presented by 
Julian Schwinger 
as an analytically soluble model in 1962 \cite{schwinger}.
This version is generally called the Schwinger model. 

Schwinger was able to show that a theory
of a massless vector gauge field
does not prevent the existence of
a massive particle.
The analytic solution
allows only uncharged states.
The thus formulated theory
was therefore regarded as a model for
complete charge screening
by vacuum polarisation \cite{screening}.
This so-called ``quark'' {\it trapping}
became an interesting object of study
in view of quark {\it confinement} \cite{screening}.
The fact that no electron excitations 
exist was confirmed in numerous works \cite{greens_functions,anomaly2}.

An equivalent formulation of the Schwinger model in
boson fields using the     
mass parameter 
\begin{eqnarray}
\label{e_boson_mass}
m_B = e / \sqrt{\pi}
\end{eqnarray}
is given by a Lagrange density of the form \cite{coleman_sinegordon,
coleman_bosonization,
coleman_theta,anomaly2}
\begin{eqnarray}
{\cal L} =\1,2 \; ( \partial_\mu \phi
\partial^\mu \phi -m_B ^2 \phi^2)
\; ,
\end{eqnarray}
thus enabling us to interpret the spectrum as 
that of free bosons,
the one-particle state with mass $m_B $ corresponding
to a quark-antiquark pair, i.e. a meson.
Above that we find a continuum of
two meson, three meson $\dots$ states
beginning at the minimum energies of $2 m_B$, $3 m_B, \dots$,
where the corresponding mesons
are at rest relative to each other.
A pedagogical introduction can be found in \cite{anomaly_book}.

A further feature of the Schwinger model\
is the appearance of a non-vanishing
vacuum condensate
$\langle \bar\psi \psi \rangle$. Naively we would expect
conservation of electric as well as axial charge.
Yet the phenomenon of an axial anomaly is
observed \cite{anomaly1,anomaly2}.

\subsection{Massive 1-flavour model}

Much interest was concentrated on the variant
of the Schwinger model
with massive fermions, $N_f=1, m_1=m \neq 0$,
often called the 
massive Schwinger model.
Using quasi-classical approximations,
Coleman 
was able to confirm that
also in this model no charged particles
are allowed as asymptotic states \cite{coleman_bosonization}.
One finds in the massless case
decoupled, degenerate vacua
upon which for each value of the parameter $\theta $,
corresponding to a constant electric background field
$E_c={e \over 2\pi} \theta $,
the same spectra are built up.
In four dimensions this field would be eliminated
due to pair creation.
In the case of two dimensions this is not possible;
electron-positron pairs are only created until
the background field falls below a critical value of
$E_c^{\rm crit} = {e \over 2}$,
so that $\theta$ 
can be chosen to lie in the range $\left[ -\pi,\pi \right]$.
Coleman showed in a semi-classical calculation
that also in the massive case the
background electric field is decisive for
the structure of the spectrum.
In the weak coupling limit we find
for
the number of stable states
\begin{eqnarray}
\label{col-eq}
N = {4 m ^2 \over \pi e ^2 } \, \cdot \,
{1 \over {1- {\theta^2 \over \pi^2} }}
\, \Bigl( 2 \sqrt{3} - 
{\rm \ln}(2 + \sqrt{3}) \Bigr) + {\cal O}(1) 
\; ,
\end{eqnarray}
which at vanishing coupling diverges as expected.
In the strong coupling limit
the result is one, two and three stable particles
for the ranges
$|\theta|  >{\pi \over 2}$, 
$0<|\theta|  <{\pi \over 2}$ and 
$\theta=0 $ respectively \cite{coleman_bosonization}. 

Basic to the expansion around the massless case
is the proof that this limit is allowed
and leads to the soluble massless Schwinger model \cite{anomaly2}.
The connection of these phenomena
and topology
was investigated in \cite{chiral_limit,manton}.

The massive Schwinger model\ can equivalently
be described by an interacting boson field
with the Lagrange density\ \cite{coleman_bosonization}
\begin{eqnarray}
{\cal L} =\1,2 \left[ \partial_\mu \phi \partial^\mu \phi
-m_B ^2\phi^2 +
m m_B {e^\gamma \over \pi} \cos (2\sqrt{\pi}\phi)
\right]
\; ,
\end{eqnarray}      
where $\gamma=0.577\dots$ is the Euler constant
and $m_B$ the mass parameter Eq.~\ref{e_boson_mass}.
An expansion
in the fermion mass yields through comparison of
the coefficients of $\phi^2$ for the mass
of the lowest state \cite{anomaly2,chiral}
\begin{eqnarray}
\label{mass2kogut}
M_1^2=m_B ^2 \Bigl( 1 + 2{m \over m_B } e^\gamma 
\Bigr) +{\cal O}(m^2)
\; .
\end{eqnarray}
In the framework of lattice gauge theory 
the massive Schwinger model\ was investigated numerically
using Hamiltonian methods \cite{anomaly2,hamiltonian}.
Here besides mass eigenvalues of the lowest-lying states 
also the coefficients of the linear potential
in the limit of small mass were obtained.
The results are in good agreement of some percent error 
with
the continuum theory \cite{hamiltonian}.
Discretized Light-Cone Quantisation \cite{dlcq_overview}
was able to calculate whole mass spectra in some 
Fock space approximation \cite{eller,els_dlcq}.
These data were used to study 
finite temperature quantities 
and critical exponents of the massive
Schwinger model 
\cite{els_finite}.

\subsection{Massless $N$-flavour model}

The $N$-flavour model can be solved analytically
in the limit $m_a=0, \; \forall a=1 \dots N$ \cite{hetrick}.
Main predictions are the existence of a massive 
pseudoscalar isosinglet state
with mass
\begin{eqnarray}
m
=
\sqrt{N} {e \over \sqrt{\pi}}
\end{eqnarray}
and $N^2-1$ massless (Goldstone-like) states.
We would like to point out that earlier work 
identifies only $N-1$ pion states \cite{wrong_nflavour}.
Both the fermion condensate $\langle \bar\psi \psi \rangle$
and the pseudoscalar density $\langle \bar\psi \gamma^5 \psi \rangle$
are  predicted to be zero.

\subsection{Massive 2-flavour model}

Most useful for lattice investigations 
with Wilson fermions is
the
massive degenerate 2-flavour model $N_f=2, m_1=m_2=m$.
It is first of all convenient 
for technical reasons
as the effective probability 
after integration of the fermionic degrees
of freedom is manifestly positive.
On the other hand,
the 2-flavour model 
has the nice feature of being
rather QCD-like,
as was observed in several classic papers 
\cite{coleman_bosonization,2f,2f_confinement}.

The model 
describes 
light 
pseudoscalar isotriplet states ($\pi$,
analogous to the Goldstone particles of QCD),
a pseudoscalar isosinglet state ($\eta$, much like the $\eta'$ of QCD)
and scalar mesons ($a_0,f_0$).
For most of these
the mass perturbation theory is known to first order
\cite{2f,2f_strong_coupling}
\begin{eqnarray}
m_\pi
&=&
2.066 m^{2\over 3} e^{1\over 3} 
\nn \\
m_{f_0} 
&=& 
\sqrt{3} m_\pi
\nn \\
m_\eta
&=&
\sqrt{2} {e \over \sqrt{\pi}}
\; .
\end{eqnarray}
The rich meson physics
therefore
allows 
to set the scale via the pion mass and
measure physical mass ratios,
e.g. ${m_\pi \over m_\eta}$.
Fermionic densities
are generally zero with
the exception of the fermion condensate,
which is found to be real.

\section{Lattice formulation}
\label{s_lattice}

The lattice version 
of the Schwinger model
with two flavours of fermions
of identical mass
is defined using the
positive effective distribution for the compact $U(1)$ link variables
$U_{x,\mu}$
\begin{eqnarray}
P_{\rm eff}[U] 
&\propto& 
{\rm det} M^2 \, e^{-S_g[U]} 
\end{eqnarray}
one obtains after
integration over the fermionic Grassmann
variables in the path integral Eq.~\ref{e_pathintegral}
and discretization of the fields \cite{wilson}.
We use the standard plaquette action
\begin{eqnarray}
S_g[U] = \beta \Re \; \sum_P ( 1 - U_P )
\end{eqnarray}
with $\beta={1\over e^2 a^2}$ the dimensionless
lattice gauge coupling 
and
plaquette variables $U_{P\;x}$
defined starting at the lower left corner
of each plaquette 
\begin{eqnarray}
\label{e_plaquette}
U_{P\;x}
=
U^\dagger_{x,\mu} \;
U^\dagger_{x+\hat{\mu},\bar{\mu}}\;
U        _{x+\hat{\bar{\mu}},\mu}\;
U        _{x,\bar{\mu}}              
\; ,
\end{eqnarray}
so that the lowest order expansion in 
the lattice spacing matches to the naive continuum limit.
Barred Greek symbols signify the orthogonal direction.
The Wilson fermion matrix is given by
\begin{eqnarray}
M_{x,y}
=
\delta_{x,y}
-\kappa  \sum_\mu
\Bigl(
\delta_{x-\hat\mu,y} (1+\gamma^\mu) U_{x-\hat\mu,\mu}
+
\delta_{x+\hat\mu,y} (1-\gamma^\mu) U^\dagger_{x,\mu}
\Bigr)
\; ,
\end{eqnarray}
where $\kappa={1\over 2(ma+d)}$
is the dimensionless
Wilson parameter.
Gamma matrices and details of notation are given in
App.~\ref{s_fermion_matrix}.
We remark that the lattice spacing $a$ is 
only explicitly shown in exceptional cases to avoid confusion.
Generally it is
set to 1 
in all formulae.

Chiral symmetry
is explicitly broken
by the Wilson term.
We remark that, as $-1$ is in the centre of the $U(1)$ group,
periodic or anti-periodic boundary conditions for the links
should be irrelevant.

The 2-flavour case
ensures that the 
effective distribution
to be simulated is manifestly positive.
We work with the
Hermitean
fermion matrix \cite{lba}
\begin{eqnarray}
\label{e_Q}
Q 
&=&
c_0 \gamma^5 M
\\
&=&
c_o \gamma^5 \delta_{x,y}
-
c_o \kappa \sum_\mu
\Bigl(
\delta_{x-\hat\mu,y} \gamma^5 (1+\gamma^\mu) U_{x-\hat\mu,\mu}
+
\delta_{x+\hat\mu,y} \gamma^5 (1-\gamma^\mu) U^\dagger_{x,\mu}
\Bigr)
\nn
\; ,
\end{eqnarray}
scaled so that its 
eigenvalues are in $[-1,1]$
using
the scaling factor
\begin{eqnarray}
c_0 
&=& {m+d \over m+2d} {1\over c_M} 
=
{1 \over 1+2d\kappa} {1\over c_M} 
\quad;\quad c_M \ge 1 
\; .
\end{eqnarray}
Explicit formulae are given in
App.~\ref{s_fermion_matrix}.

\label{s_kappa_c}
The critical Wilson parameter
is known in the weak coupling limit
\begin{eqnarray}
\kappa_c(\beta\to\infty) = {1\over 4}
\end{eqnarray}
and expected
to be larger for lower $\beta$. 
It is still an open question
whether simulations above $\kappa_c$ 
are possible.
As this was not the focus of this work,
we did not investigate this point.
Preliminary results (e.g. Sec.~\ref{critical})
suggest that reasonable results are difficult
but possible
in this regime.
The second order phase transition point
(and thus the continuum limit)
is reached for $\beta \to \infty$
with lines of constant physics
fixed via the demand to keep certain (e.g. mass) 
ratios constant.

\vspace{0.3cm}
\noindent{\bf Observables}
\\
A measurement of the average plaquette 
$ {1\over LT} \sum_{x} U_{P\;x}$
and of the temporal and spatial
Polyakov loops
averaged over the orthogonal direction
$
P_L 
= 
{1\over T} \sum_{x_2=1}^T \prod_{x_1=1}^L U_{x, 1}
$ and 
$
P_T 
= 
{1\over L} \sum_{x_1=1}^L \prod_{x_2=1}^T U_{x, 2} 
$
is directly possible from the definitions.

Averages of local fermion bilinears
$\langle \bar\psi \Gamma \psi \rangle$ 
can be measured
applying
a noisy estimator
scheme, using random spinors $\eta(x)$ as sources for the 
solver as shown in App.~\ref{s_mesons}. 
One expects the pseudoscalar density (also called the chiral condensate)
$\langle \bar\psi \gamma^5 \psi \rangle $
and the further densities
$\langle \bar\psi \gamma^\mu \psi \rangle $
to vanish within errors.
The fermion condensate $\langle \bar\psi \psi \rangle $ remains 
non-zero even
in the massless limit because Wilson fermions break chiral 
invariance explicitly \cite{wilson}.

For the determination of meson masses 
operators with various quantum
numbers are defined
\begin{itemize}
\item
flavour triplet -
$( \bar\psi \gamma^5 \tau \psi )$ `$\pi$', 
$( \bar\psi \tau \psi )$ `$a_0$',
\item
flavour singlet -
$( \bar\psi \gamma^5 \psi )$ `$\eta$', 
$( \bar\psi \psi )$ `$f_0$' 
.
\end{itemize}
Moreover, insertion of a $\gamma^0$,
e.g. $( \bar\psi \gamma^0 \gamma^5 \tau \psi )$ for the $\pi$,
leads to alternative operators with
the same quantum numbers in the rest frame. 
This exhausts the Dirac algebra
in two dimensions. The result is summed up in Tab.~\ref{t_operators}.

The calculation of temporal correlators $C^{\Gamma T}(\Delta t)$ involves
point sources at randomly chosen positions $(x,t)$ and summation over
spatial displacements $y$ in the other time slice $(y,t+\Delta t)$ to
project out the zero momentum states. As to the flavour--singlet channels,
the disconnected piece is subtracted with the aid of the noisy inversions
performed earlier for the measurement of the condensate
\cite{disconnected}.
For more details consult App.~\ref{s_mesons}.

All measurements
are analysed
taking into account 
covariances and auto-correlations of the correlators
\cite{datenanalyse_book}.
A binning method is applied to reduce data size.

The scaling idea we are using
is based on the following strategy.
Imagine
that $m_\pi$ and the ratio ${m_\pi \over m_\eta}$ are 
fixed, e.g. by experiment.
Working at a certain value of $\beta$,
we will then have to adjust $\kappa$ such that
the ratio is the physical value,
while $m_\pi$
sets the scale, determining the 
lattice spacing in physical units.
The size of the lattice has to be chosen so 
that finite size effects are small.

Repeating this procedure at larger $\beta$
results in a
smaller pion mass in lattice units,
which 
again sets the scale for the lattice spacing and thus
determines the scaling factor achieved.

We remark that
motivated by the continuum results
we expect the masses to 
scale like
$
m_\eta \approx {1 / \sqrt{\beta}}
$,
so that we have to increase the value of $\beta$
enormously
if we want to double the 
lattice size and reduce the lattice spacing by a factor of 2,
keeping physical ratios constant by
adjusting the $\kappa$ value.
This 
unfortunately 
limits the usefulness of the
Schwinger model,
as very high $\beta$ simulations
encounter topological metastabilities 
\cite{topology_joos,topology_dilger,els_topology}.

\chapter{The standard way: Hybrid Monte Carlo}
\label{c_hmc}

We are ultimately interested in 
deciding
whether
the new local bosonic algorithm can be faster
than the ``workhorse'' Hybrid Monte Carlo (HMC) \cite{hmc}
used for more than 10 years.
We thus
implemented a 
HMC code
working with the Hermitean matrix $Q$
to set the scale.
A brief description of the formulae is given in
Sec.~\ref{s_implementation_hmc}.

In order to make a fair comparison
we 
ulilize the optimisation procedures
now common in the application  
of the HMC.
Namely,
the implementation
includes 
trajectory length set by
{ $n \cdot \Delta\tau = 1$}
and 
{ acceptance $\approx 70 \%$}.
We further use the fact
that
extending the solver 
solution in the trajectory
via the condition $\stackrel{\dots}{x}=0$ to generate
an optimal new start vector for the inverter
results in a 
gain of approximately 20\%. 

The 
inverter algorithm applied throughout
is a standard Conjugent Gradient algorithm (CG)
\cite{cg1_book,cg2_book}
used with
precision $10^{-6}$.
The decision
not to use advanced inverters like
BiCGgamma5 for the observables
was taken in order to 
have a guaranteed convergence
for all simulation parameters.
In order to avoid trouble with bad pseudo-random numbers,
we use L\"uscher's high-quality random 
number generator (RG) \cite{random_generator},
which has been shown to have
a very long recursion period and efficient decorrelation properties.
It is used in 
a vectorized form applying
257 generators in parallel.

The standard way
to get portable results for
CPU cost analyses
is to count in matrix multiplication operations.
These are the most CPU-time
intensive
parts of any
realistic QCD calculation.
We describe our explicit 
counting in Sec.~\ref{s_counting_hmc}.

Generally,
one uses this information
to define the cost of an algorithm
by quantifying the cost
needed to generate two independent (i.e. decorrelated)
configurations on which measurements are
executed.
Assuming an exponential decorrelation
of configurations \cite{datenanalyse_book}
an estimator for 
the distance of 
independent configurations
is given by
twice the integrated auto-correlation time $\tau_{\rm int} \geq 1$.
Thus this quantity
measured in units of molecular dynamics time or respectively
matrix multiplications,
\begin{eqnarray}
2 \tau_{\rm int} 
[{\rm Q \; ops}]
=
2 \tau_{\rm int} 
\cdot N_{\rm Q \; ops / update}
\; ,
\end{eqnarray}
can be regarded
as a reasonable estimator for the CPU cost of an algorithm.
We would like to point out that although the 
autocorrelation time is dependent on the observable,
it is generally of the same order of magnitude
characterizing the update algorithm.

\section{Implementation}
\label{s_implementation_hmc}

The idea of HMC is based
upon rewriting
the partition function
given by
\begin{eqnarray}
Z
=
\int D[U]\;
{\rm det} Q^2\;
e^{-S_G(U)}
\end{eqnarray}
using complex pseudo-fermion fields
(neglecting constant factors)
\begin{eqnarray}
Z
=
\int D[U] D[\eta] D[\eta^\dagger]
e^{-S_G(U) - \eta^\dagger Q^{-2} \eta}
\end{eqnarray}
and introducing real valued momenta $\pi_{x,\mu} $
conjugate to the links $U_{x,\mu} $
with a kinetic term which is not changing the path integral
\begin{eqnarray}
Z
=
\int D[U] D[\eta] D[\eta^\dagger] D[\pi]
e^{-S_G(U) - \eta^\dagger Q^{-2} \eta
- {1\over 2} \pi^2}
\; .
\end{eqnarray}
The thus defined action
is called in analogy to classical mechanics the Hamiltonian
of the system
\begin{eqnarray}
H
=
S_G(U) + \eta^\dagger Q^{-2} \eta
+ {1\over 2} \pi^2
\; .
\end{eqnarray}
Applying this distribution,
the complete dynamical update 
of the link configuration is done 
in a three step procedure
\begin{enumerate}
\item
\begin{itemize}
\item
a complex Gaussian update for the
field
$\chi = Q^{-1} \eta$ 
\item
a real Gaussian update for the
field $\pi$
\end{itemize}

\item
micro-canonical steps
via 
discretised but reversible leapfrog integration
of the equations of motion
in fictitious 
computer time $\tau$
leaving the Hamiltonian approximately constant
\begin{eqnarray}
\dot{H} = { \partial H \over \partial \tau} \approx 0
\end{eqnarray}

\item
a Metropolis acceptance step
correcting for
discretisation errors in the 
leapfrog integration
with
\begin{eqnarray}
{\rm min} \{ 1, e^{-\Delta H} \}
\end{eqnarray}

\end{enumerate}

\subsection{Micro-canonical update}

We need the update equations, also called
equations of motion (EOMs),
for the fields $U_{x,\mu}$
and $\pi_{x,\mu}$.
The Hamiltonian EOMs can be chosen to fulfil
the two requirements
\begin{itemize}
\item
$U_{x,\mu}$
should stay in $U(1)$,
\item
the Hamiltonian is constant in computer time.
\end{itemize}
yielding a simple update
process for the links
\begin{eqnarray}
\dot{U}_{x,\mu} = i \pi_{x,\mu} U_{x,\mu} 
\end{eqnarray}
while the equations for the 
momenta are fixed via
\begin{eqnarray}
\dot{H} = 0
\; .
\end{eqnarray}
Writing this out, 
we find
\begin{eqnarray}
\dot{H}
=
\dot{S}_G(U)
+
\eta^\dagger {\partial \over \partial \tau} [
Q^{-2} ] \eta
+
\pi \dot{\pi}
\; .
\end{eqnarray}
Our aim is to express 
$\dot{S}_G(U)$
and 
$\eta^\dagger {\partial \over \partial \tau} [
Q^{-2} ] \eta$
in terms of the derivatives
$\dot{U}_{x,\mu}$ and $\dot{\pi}_{x,\mu}$
to allow numerical integration of the EOMs.
We rewrite
the second term
\begin{eqnarray}
\eta^\dagger
{\partial \over \partial \tau} 
\left[
Q^{-2} \right]
\eta
=
- \eta^\dagger Q^{-2} 
\left[ \dot{Q} Q + Q \dot{Q} \right]
Q^{-2} 
\eta
\end{eqnarray}
and introduce
abbreviations for
parts independent of the derivatives
\begin{eqnarray}
\omega 
= 
Q^{-2} \eta
\quad {\rm and} \quad
\xi
=
Q \omega
\end{eqnarray}
to rewrite
\begin{eqnarray}
\dot{H}
=
\dot{S}_G(U)
-
2 \Re [
\xi^\dagger 
\dot{Q} \omega]
+
\pi \dot{\pi}
\; .
\end{eqnarray}
We now calculate the dependency
of the first term $\dot{S}_G$
of the derivatives
\begin{eqnarray}
\dot{S}_G
&=&
-\beta 
\sum_x \Re 
(
\dot{U}_{x,1} U_{x+\hat{1},2} U^\dagger_{x+\hat{2},1} U^\dagger_{x,2} 
+
U_{x,1} \dot{U}_{x+\hat{1},2} U^\dagger_{x+\hat{2},1} U^\dagger_{x,2} 
\nn \\
&&
\phantom{-\beta \sum_x }
+
U_{x,1} U_{x+\hat{1},2} \dot{U}^\dagger_{x+\hat{2},1} U^\dagger_{x,2} 
+
U_{x,1} U_{x+\hat{1},2} U^\dagger_{x+\hat{2},1} \dot{U}^\dagger_{x,2}
)
\; .
\end{eqnarray}
Using the EOMs we can rewrite
\begin{eqnarray}
\dot{S}_G
&=&
-\beta 
\sum_x \Re 
(
  i \pi_{x,1} U_{x,1} U_{x+\hat{1},2} U^\dagger_{x+\hat{2},1} U^\dagger_{x,2}
+ U_{x,1} i\pi_{x+\hat{1},2} {U}_{x+\hat{1},2} U^\dagger_{x+\hat{2},1} U^\dagger_{x,2} 
\nn \\
&&
- U_{x,1} U_{x+\hat{1},2} i\pi_{x+\hat{2},1} {U}^\dagger_{x+\hat{2},1} U^\dagger_{x,2} 
- U_{x,1} U_{x+\hat{1},2} U^\dagger_{x+\hat{2},1} i\pi_{x,2} {U}^\dagger_{x,2} )
\end{eqnarray}
and reorder
\begin{eqnarray}
\dot{S}_G
&=&
-\beta 
\sum_x 
\pi_{x,1}
\Re 
i (
  U_{x,1} U_{x+\hat{1},2} U^\dagger_{x+\hat{2},1} U^\dagger_{x,2} 
- U_{x-\hat{2},1} U_{x-\hat{2}+\hat{1},2} {U}^\dagger_{x,1} U^\dagger_{x-\hat{2},2} 
)
\nn \\
&&\phantom{-\beta \;}
+
\pi_{x,2}
\Re i
(
  U_{x-\hat{1},1}  {U}_{x,2} U^\dagger_{x-\hat{1}+\hat{2},1} U^\dagger_{x-\hat{1},2} 
- U_{x,1} U_{x+\hat{1},2} U^\dagger_{x+\hat{2},1} {U}^\dagger_{x,2} )
)
\nn \\
&=&
\sum_{x,\mu} 
\pi_{x,\mu}
\Re \Phi_{x,\mu}
\end{eqnarray}
defining thus another abbreviation $\Phi$
independent on the derivatives
\begin{eqnarray}
\Phi_{x,1}
=
-i\beta
(U_{P\;x} - U_{P\;x-\hat{2}} )
\quad {\rm and} \quad
\Phi_{x,2}
=
-i\beta
(U_{P\;x-\hat{1}} - U_{P\; x})
\; .
\end{eqnarray}
The dependence of $\dot{Q}$ on the derivatives
is given by
\begin{eqnarray}
\dot{Q}
&=&
-
c_o \kappa \sum_\mu
\Bigl(
\delta_{x-\mu,y} \gamma^5 (1+\gamma^\mu) \dot{U}_{x-\mu,\mu}
+
\delta_{x+\mu,y} \gamma^5 (1-\gamma^\mu) \dot{U}^\dagger_{x,\mu}
\Bigr)
\nn \\
&=&
-
c_o \kappa \sum_\mu
\Bigl(
\delta_{x-\mu,y} \gamma^5 (1+\gamma^\mu) i \pi_{x-\mu,\mu} {U}_{x-\mu,\mu}
-
\delta_{x+\mu,y} \gamma^5 (1-\gamma^\mu) i \pi_{x,\mu} {U}^\dagger_{x,\mu}
\Bigr)
\nn
\; .
\end{eqnarray}
In the Hamiltonian this is
used in the combination
\begin{eqnarray}
-
2 \Re [
\xi^\dagger 
\dot{Q} \omega]
&=&
2 c_0 \kappa \Re \left[ i
\sum_{x,\mu} \pi_{x,\mu} 
\Bigl(
{U}_{x,\mu}
\xi^\dagger_{x+\hat{\mu}} \gamma^5 (1+\gamma^\mu) \omega_{x}
-
{U}^\dagger_{x,\mu}
\xi^\dagger_{x} \gamma^5 (1-\gamma^\mu) \omega_{x+\hat{\mu}}
\Bigr) \right]
\nn \\
&=&
\sum_{x,\mu} \pi_{x,\mu} 
\Re \; \Omega_{x,\mu}
\end{eqnarray}
defining the abbreviation $\Omega$.
The final Hamiltonian formula is therefore
\begin{eqnarray}
\dot{H}
=
\sum_{x,\mu}
\pi_{x,\mu} 
(\Re \Phi_{x,\mu} + \Re \; \Omega_{x,\mu} + \dot{\pi}_{x,\mu} )
\end{eqnarray}
giving the required
EOMs
\begin{eqnarray}
\Re \{ \Phi_{x,\mu} + \Omega_{x,\mu} \} + \dot{\pi}_{x,\mu}
=0
\; .
\end{eqnarray}

\subsection{Leapfrog integration}

Heuristically one finds optimal behaviour of
the algorithm if
the EOMs
\begin{eqnarray}
\dot{U}_{x,\mu} 
&=&
 i \pi_{x,\mu} U_{x,\mu} 
\nn \\
\dot{\pi}_{x,\mu} 
&=&
- \Re \{ \Phi_{x,\mu} + \Omega_{x,\mu} \} 
\end{eqnarray}
are integrated
for $N$ steps of length $\Delta \tau$
under the constraint
\begin{eqnarray}
N \cdot \Delta \tau \approx 1
\; .
\end{eqnarray}
Integrating the link EOMs,
we find
\begin{eqnarray}
U_{x,\mu} (\tau_0 + \delta \tau)
=
U_{x,\mu} (\tau_0) e^{i \pi_{x,\mu} \delta \tau}
\end{eqnarray}
to explicitly stay in the 
group manifold
and
\begin{eqnarray}
\pi_{x,\mu} (\tau_0 + \delta \tau)
=
\pi_{x,\mu} (\tau_0) 
- 
 (\Re \{ \Phi_{x,\mu} + \Omega_{x,\mu} \} ) \cdot \delta \tau
\; .
\end{eqnarray}
Expressing this in steps $n=1 \dots N$
in the link case and
$n=1 \dots N+1$ in the momentum case,
we can write
the result
in a computer-friendly way
\begin{eqnarray}
U_{x,\mu} (n) 
&=&
U_{x,\mu} (n-1) e^{i \pi_{x,\mu} (n) \Delta\tau} 	
\\
\pi_{x,\mu} (n)
&=&
\pi_{x,\mu} (n-1)
-
 (\Re \{ \Phi_{x,\mu} (n-1) + \Omega_{x,\mu} (n-1) \} ) \cdot
\Delta\tau (n-1)
\nn
\end{eqnarray}
with
\begin{eqnarray}
\Delta\tau (0) 
&=&
\Delta\tau (N)
= 
{\Delta\tau \over 2}
\nn \\
\Delta\tau (n) 
&=&
{\Delta\tau }
\quad\qquad\qquad \forall n \in \{ 1 \dots N-1 \}
\; .
\end{eqnarray}

\section{Counting in $Q$ operations}
\label{s_counting_hmc}

We give the number of 
$Q$ matrix multiplications necessary for one 
update step of our HMC implementation.
In the formulae we
denote with
$
N_{CG}
$
the number of matrix multiplications
needed by the Conjugent Gradient routine
applying $Q^2$.

The Hybrid Monte-Carlo
algorithm
applies
\begin{itemize}
\item
a number $N_{\rm it}$ of leapfrog iterations per trajectory

\item
a CG routine to invert the matrix $Q^2$
using $N_{\rm CG}$ iterations to invert
\end{itemize}
and thus uses
\begin{itemize}
\item
$2 N_{\rm it} N_{\rm CG}$ Q operations for the main inverter,
\item
$2 N_{\rm CG}$ Q operations for the initial inverter.
\item 
The final inverter can be omitted as
the data is known from the trajectory. 
\item
$2 N_{\rm it}$ effective Q operations to update the links
\end{itemize}
so that we finally obtain 
a total number of matrix multiplications of
\begin{eqnarray}
N_{\rm Q \; ops / update}
=
2 N_{\rm it} N_{\rm CG}
+ 2 N_{\rm CG}
+ 2 N_{\rm it}
\; .
\end{eqnarray}

\chapter{The new way: Local bosonic algorithm}
\label{c_lba}

\section{Basic idea}

As an alternative to the Hybrid Monte Carlo algorithm,
M. L\"uscher proposed a local bosonic formulation \cite{lba}.
The one main variant
of the local bosonic algorithm (LBA) 
we consider
uses the
fact that we are dealing with a rescaled 
squared Hermitean fermion matrix,
so that the eigenvalues are real and between 0 and 1.

We can then identically rewrite the path integral
\begin{eqnarray}
{\cal Z}
=
\int D[U] \det[Q^2] 
P_n(\det[Q^2]) \, \left( P_n(\det[Q^2]) \right)^{-1} e^{-S_g[U]}     
\end{eqnarray}
using for $P_{n}(s)$ any real polynomial of 
degree $n$ which approximates $1/s$ in $(0,1]$. 
Then  
\begin{eqnarray}
\det[Q^2 P_n(Q^2)] = \det[1-R] \approx 1
\end{eqnarray}
is a small correction
which can be treated via a correction scheme,
provided we can sample the remaining distribution.

We now assume
that the approximation polynomial is of the form
\begin{eqnarray}
\label{e_fact}
P_n(Q^2)
=
N_{\rm norm} \prod_{k=1}^n (Q^2-z_k)
\; ,
\end{eqnarray}
where $N_{\rm norm}$
is a normalisation constant
and 
the $z_k$ are the roots of the polynomial
coming in complex conjugate pairs
$z_k,\bar{z}_k$ 
with non-vanishing imaginary part.
They
define 
$\mu_k,\nu_k$ and $\bar\mu_k,\bar\nu_k$
via
\begin{eqnarray}
\sqrt{z_k} = \mu_k + i \nu_k 
&\to& \bar\mu_k = -\mu_k \quad \bar\nu_k=\nu_k
\end{eqnarray}
if one defines the branch of the square root to be taken
by $\nu_k > 0$.
Using this we can rewrite
\begin{eqnarray}
P(Q^2) 
= 
N_{\rm norm} \prod_{k=1}^n (Q^2-z_k)
=
N_{\rm norm} \prod_{k=1}^n \left[ (Q-\mu_k)^2 + \nu_k^2 \right]
\; .
\end{eqnarray}
As this factorises,
sampling this distribution can be done
using 
\begin{eqnarray}
\int D[\phi]D[\phi^\dagger]
e^{-{1\over 2} \phi^\dagger A\phi} \propto  (\det[A])^{-1} 
\end{eqnarray}
to transform the distribution
to a
bosonic integral
\begin{eqnarray}
P_{\rm eff}(U)
\!&\propto&\!\!
\det[1-R] \, e^{-S_g[U]}  
 \! \int \! 
{\cal D}\phi \, e^{-\sum_k [\phi_k^\dagger (Q-\mu_k)^2 \phi_k
+\nu^2_k \phi^\dagger_k \phi_k ]}
\end{eqnarray}
with $n$ complex bosonic Dirac fields $\phi_k$.

The main criteria for the polynomials $P_n(s)$
is that they should be
fast convergent to ${1 \over s}$
in $(0,1]$ and have roots coming in complex conjugate pairs.
\label{c_chebyshev}
We chose as approximation polynomials
the Chebyshev-derived 
polynomials proposed by 
Bunk et al.\cite{chebyshev}
\begin{eqnarray}
P_n(s)
=
{1+\rho_n T_{n+1}(x) \over s}
=
{1-R_n(s) \over s}
\end{eqnarray}
with 
\begin{eqnarray}
x = 2\;{s-\epsilon \over 1- \epsilon}-1 
\; ,
\end{eqnarray}
such that
$s \in [\epsilon,1]$ is mapped to $x \in [-1,1]$.
The
polynomials $T_n(z)$ of degree $n \ge 0$
are 
the standard Chebyshev polynomials given
by
\begin{eqnarray}
T_n(z) 
&=& 
\cosh n\phi
\quad
{\rm with} \quad z=\cosh\phi \; .
\end{eqnarray}
For completeness, we give some of
the first polynomials 
\begin{eqnarray}
&&T_0(z) = 1        \; , \quad
T_1(z) = z        \; , \quad     
T_2(z) = 2z^2 - 1 \; ,    
\nn \\
&&T_3(z) = 4z^3 - 3z \; , \quad    
T_4(z) = 8z^4 - 8z^2 + 1 
\; .
\end{eqnarray}
An important feature is the 
Clenshaw recursion relation
\begin{eqnarray}
\label{e_chebyshev_recursion}
T_{n+1}(z) + T_{n-1}(z) = 2z T_n(z) \quad\mbox{for}\quad n \ge 1 
\; ,
\end{eqnarray}
known to be numerically stable.
For more information 
we defer the reader to \cite{chebyshev_book}.

The polynomials $P_n(s)$
approximate $1/s$ for real $s \in [\epsilon,1]$.
The rest term $R_n(s)$ quantifies the quality of the approximation
and the normalisation factor $\rho_n$ is defined through the condition that
$P_n(s)$ exists even at $s=0$
via
\begin{eqnarray}
0 = s P_n(s) |_{s=0}
=
1 + \rho_n T_{n+1}\left( -{1+\epsilon\over 1-\epsilon} \right)
\quad \to \quad 
\rho_n 
=
{1\over T_{n+1}(-{1+\epsilon\over 1-\epsilon})}
\; .
\end{eqnarray}
The explicit expression for the error term
\begin{eqnarray}
R(s)
=
{T_{n+1}({2s\over 1-\epsilon}-{1+\epsilon\over 1-\epsilon})
\over
T_{n+1}(-{1+\epsilon\over 1-\epsilon})}
\end{eqnarray}
can be used to deduce limits for the approximation quality
\begin{eqnarray}
\hspace{-0.5cm}
|R_n(s)|
\hspace{-0.2cm}
&\le &
\hspace{-0.2cm}
 |\rho_n| = |{1 \over \cosh [(n+1)\chi ]}| \quad {\rm with}
\quad \cosh \chi = -{1+\epsilon\over 1-\epsilon}
\end{eqnarray}
yielding
\begin{eqnarray}
\hspace*{-1truecm}
|R_n(s)|
\hspace{-0.2cm}
&\le&
\hspace{-0.2cm}
 2 {1 \over \cosh [(n+1)\chi ] + \sinh [(n+1)\chi ]}
=
 2 \left( 
{1 - \sqrt{\epsilon} \over 1+\sqrt{\epsilon} } \right)^{n+1}
\approx 
2 e^{-2\sqrt{\epsilon}n}
\end{eqnarray}
proving an exponential convergence to $1/s$ for $s \in [\epsilon,1]$.
Explicitly, 
we use as the approximation quality factor
\cite{lba,chebyshev}
\begin{eqnarray}
\delta
=
 2 \left( 
{1 - \sqrt{\epsilon} \over 1+\sqrt{\epsilon} } \right)^{n+1}
\; .
\end{eqnarray}
Not necessary for the application 
in the local bosonic algorithm, but convenient is the fact that
the roots of this approximation 
polynomial can be given explicitly
observing that
\begin{eqnarray}
P_n(s) = 0
&\equal&
T_{n+1}(x) = T_{n+1}(-{1+\epsilon \over 1-\epsilon})
\end{eqnarray}
for complex $s,x$.
This yields 
\begin{eqnarray}
\label{e_roots}
z_k
=
{1\over 2} (1+\epsilon)
-
{1\over 2} (1+\epsilon)
\cos ( {2\pi k \over n+1} )
- i \sqrt{ \epsilon} \sin ( {2\pi k \over n+1} )
\; ,
\end{eqnarray}
i.e. $n$ roots on an ellipse in the complex plane
around the foci $\epsilon$ and $1$ on the real axis.

We further can deduce the global 
normalisation constant $N_{\rm norm}$
from the symmetry of the Chebyshev polynomial in
$[\epsilon,1]$
\begin{eqnarray}
T_{n+1}\left({1+\epsilon \over 2}\right) =0 \quad \to \quad
{1+\epsilon\over 2} P_{n} \left({1+\epsilon\over 2}\right) -1 =0
\end{eqnarray}
to
\begin{eqnarray}
N_{\rm norm}^{-1}
=
{1+\epsilon\over 2} \prod_{k=1}^n \left({1+\epsilon\over 2}-z_k\right)
\; .
\end{eqnarray}

The updating process consists of
exact heatbath sweeps for the $\phi$'s and $U$'s \cite{u1_heatbath},
followed by a number of over-relaxation iterations.
The formulae for the updates are given in App.~\ref{c_force}.
We confirmed in preliminary runs
that the reflection sweeps for the $\phi$'s and $U$'s have to be combined
in pairs, as was observed before \cite{lba_study,lba_comparison}.
In total, this introduces 
the 
parameters
$n$, $\epsilon$
and the number of
reflections
per heatbath 
into the algorithm.

\section{Correction step}
\label{s_correction}

Finally, the approximation 
$ \det[1-R] \approx 1$ has to be controlled. 
This 
in principle 
can either be done via a 
Metropolis
accept/reject step with
correction probability
\begin{eqnarray}
P^C_{U\phi \to U'\phi'} = 
{\rm min}[1,{\det[1-R'] \over \det[1-R]}]
\end{eqnarray}
or
by reweighting
\begin{eqnarray}
\langle O \rangle
=
{\langle O \det[1-R] \rangle_L \over
\langle \det[1-R] \rangle_L}
\; ,
\end{eqnarray}
where $\langle \dots \rangle_L$
denotes the
expectation values in the local bosonic ensemble
and primed quantities are of the new configuration.

We would like to point out
that
algorithms including an acceptance step
need the reversibility of the update trajectory
to guarantee detailed balance. This leads to a symmetrization of
the update trajectories in these cases,
which has to be included in the CPU cost derivation.

On the other hand,
the ensembles  generated 
by the local bosonic distribution
with and without correction 
are distinctly
different.
In simulations
$\langle \det[1-R] \rangle_L$
can in fact be far from 1
as will be shown in Tab.~\ref{t_correction}.
The integrated autocorrelation time can in these cases be small,
while the error from each measurement grows to very large values.
We thus can no longer use 
the autocorrelation time measured in matrix multiplications
\begin{eqnarray}
2 \tau_{\rm int} [{\rm Q \; ops}]
\end{eqnarray}
described in Ch.~\ref{c_hmc}
as a good transportable indicator for the CPU cost of the algorithms.
We have to regress to
the basic
effective cost defined by
\begin{eqnarray}
C_{\rm eff}
=
N_{\rm total \; Q \; ops} \cdot {\sigma_{\rm tot}^2 (A) \over <A>^2}
\end{eqnarray}
applicable to all measurements
as only the total work and the real result (i.e. the error
obtained for a certain observable) are used.
The cost factor is normalised by the mean value of the observable
to cancel the trivial dependency.
We remark that this cost factor is even more
observable-dependent than the standard choice $\tau_{\rm int}$,
prohibiting even a rough comparison of
different observables.

In a first step we
approximatively included the correction
via the lowest 8 eigenvalues of $Q^2$,
using them to 
apply a global Metropolis
correction step \cite{lba_qcd}.
Recently, stochastic methods
which are exact were suggested \cite{lba_exact}.
Some variants of these will be discussed in the next sections.

\subsection{Reweighting}

The correction factor
can 
be treated exactly
using a 
stochastic method.
Following a suggestion by L\"uscher \cite{reweighting},
we rewrite
\begin{eqnarray}
\langle O \rangle
=
{\langle O \det[1-R] \rangle_L \over
\langle \det[1-R] \rangle_L}
\end{eqnarray}
into 
\begin{eqnarray}
\langle O \rangle
=
{\langle \int [d\eta]\; O e^{-\eta^\dagger[1-R]^{-1}\eta}\rangle_L \over
\langle \int [d\eta]\; e^{-\eta^\dagger[1-R]^{-1}\eta}\rangle_L }
=
{\langle \int [d\eta]\; O 
e^{-\eta^\dagger[1-R]^{-1}\eta+\eta^\dagger\eta-\eta^\dagger\eta}\rangle_L \over
\langle \int [d\eta]\; 
e^{-\eta^\dagger[1-R]^{-1}\eta+\eta^\dagger\eta-\eta^\dagger\eta}\rangle_L
}
\; .
\end{eqnarray}
We then interpret the factor $e^{-\eta^\dagger\eta}$
as a Gaussian weight,
resulting in
\begin{eqnarray}
\langle O \rangle
=
{\langle O e^{\eta^\dagger \bigl( 1-[1-R]^{-1}
\bigr) \eta}\rangle_{L,\eta} \over
\langle e^{\eta^\dagger \bigl( 1-[1-R]^{-1}\bigr)\eta}\rangle_{L,\eta} }
=
{
\langle O e^{-\eta^\dagger R[1-R]^{-1} \eta} \rangle_{L,\eta} 
\over
\langle e^{-\eta^\dagger R[1-R]^{-1} \eta} \rangle_{L,\eta} 
}
\; ,
\end{eqnarray}
where $\langle \dots \rangle_{L,\eta}$
denotes the ensemble generated by the LBA distribution and the
Gaussian noise vectors.
Note that rewriting
\begin{eqnarray}
1- (1-R)^{-1}
\quad \to \quad 
-R (1-R)^{-1}  
\end{eqnarray}
is numerically more stable
as it avoids subtracting two
almost identical objects.

\subsection{Noisy acceptance step method I}

The idea behind this noisy method 
\cite{lba_exact} is to generate
a new configuration $(U',\phi')$
with the standard L\"uscher action
and then to accept it according to a correction probability
that is a function of a noise vector $\eta$
\begin{eqnarray}
P^C_{(U,\phi)\to(U',\phi')}(\eta)
\end{eqnarray}
satisfying detailed balance on average over the $\eta$
distribution
\begin{eqnarray}
{\int [d\eta]\; P^C_{(U,\phi)\to(U',\phi')} (\eta)
\over
\int [d\tilde\eta]\; P^C_{(U',\phi')\to(U,\phi)}(\tilde\eta) }
&=&
{\det  [1-R']
\over
\det  [1-R]}
\; .
\end{eqnarray}

In order to do this
one generates random numbers according to the distribution
$P^{HB}$ 
which is constructed to 
give the required determinant factor.
The acceptance of the Metropolis
$P^A$
is fixed
so that the total acceptance
\begin{eqnarray}
P^C_{(U,\phi)\to(U',\phi')}(\eta)
=
P^{HB}_{\eta} P^A_{(U,\phi)\to(U',\phi')}(\eta)
\end{eqnarray}
satisfies detailed balance.
The almost obvious choice for
the heatbath distribution
\begin{eqnarray}
P^{HB}_{\eta} 
= 
{N_{\rm n} \over \det[1-R] }
e^{-\eta^\dagger [1-R]^{-1} \eta }
\end{eqnarray}
with $N_{\rm n}$ a normalisation constant
requires an acceptance
\begin{eqnarray}
P^A_{(U,\phi)\to(U',\phi')}(\eta)
=
{\rm min} \bigl( 1,e^{- \eta^\dagger [1-R']^{-1} \eta +
\eta^\dagger [1-R]^{-1} \eta } \Bigr)
\; .
\end{eqnarray}
The proof of detailed balance
can be seen from the symmetrical behaviour of the involved integrals
\begin{eqnarray}
\hspace*{-1truecm}
&&
{\int [d\eta]\; P^C_{(U,\phi)\to(U',\phi')}(\eta)
\over
\int [d\tilde\eta]\; P^C_{(U',\phi')\to(U,\phi)}(\tilde\eta)}
=
{\int [d\eta]\; {N_{\rm n} \over \det[1-R] }
e^{-\eta^\dagger [1-R]^{-1} \eta }
{\rm min} \bigl( 1,e^{- \eta^\dagger [1-R']^{-1} \eta +
\eta^\dagger [1-R]^{-1} \eta } \Bigr)
\over
\int [d\tilde\eta]\; {N_{\rm n} \over \det[1-R'] }
e^{-\tilde\eta^\dagger [1-R']^{-1} \tilde\eta }
{\rm min} \bigl( 1,e^{- \tilde\eta^\dagger [1-R]^{-1} \tilde\eta +
\tilde\eta^\dagger [1-R']^{-1} \tilde\eta } \Bigr)}
\nn \\
\hspace*{-1truecm} 
&&
{\det[1-R']\over \det[1-R] }
{\int [d\eta]\; 
{\rm min} \bigl( e^{-\eta^\dagger [1-R]^{-1} \eta },
e^{- \eta^\dagger [1-R']^{-1} \eta} \Bigr)
\over
\int [d\tilde\eta]\; 
{\rm min} \bigl( e^{-\tilde\eta^\dagger [1-R']^{-1} \tilde\eta
}
,e^{- \tilde\eta^\dagger [1-R]^{-1} \tilde\eta } \Bigr)}
=
{\det[1-R']
\over
\det[1-R]} 
\; .
\end{eqnarray}

In order to make the algorithm practically working,
we transform the
heatbath distribution
$P^{HB}$
into a simple Gaussian
using the substitution
\begin{eqnarray}
\eta = B \chi
\end{eqnarray}
with $B$ and $B^\dagger$ given by the constraint
\begin{eqnarray}
B^\dagger B = 1-R 
\; .
\end{eqnarray} 
We remark that while this 
transformation
can be done trivially
in the non-Hermitean case \cite{lba_exact},
we here need a little trick.
The decomposition can be done easily
using the product formula
\begin{eqnarray}
1-R
= N_{\rm norm} Q^2 
\prod_k^{n/2} (Q^2-z_k) (Q^2-\bar z_k)
\end{eqnarray}
taking one root of each complex conjugate pair
to yield
\begin{eqnarray}
B = \sqrt{N_{\rm norm}} Q \prod_k^{n/2} (Q^2-z_k)
\; .
\end{eqnarray}
Finally we
obtain
\begin{eqnarray}
P_{\chi} 
= 
N_{\rm n}
e^{-\chi^\dagger B^\dagger[B B^\dagger]^{-1} B \chi }
= 
N_{\rm n}
e^{-\chi^\dagger \chi }
\end{eqnarray}
and
\begin{eqnarray}
\label{e_acceptance}
P^A_{(U,\phi)\to(U',\phi')}
&=&
{\rm min} \bigl( 1,e^{- \chi^\dagger B^\dagger [1-R']^{-1} B \chi +
\chi^\dagger \chi } \Bigr)
\; .
\end{eqnarray}

\subsection{Noisy acceptance step method I with adapted precision}

This method
makes use of a possible
optimisation 
for Metropolis-type
acceptance step schemes
using a restartable solver like
the Conjugent Gradient chosen 
for this study.

As the random number governing the
Metropolis decision is known 
before the solver is applied,
it is in principle possible
to interrupt the solver iterations
and check whether the 
quality of the solver solution is already good enough
to fulfil the requirements of a
clear-cut decision,
i.e. to distinguish the result
from the chosen random number.
If not, the solver is restarted and 
this procedure iterated.

We implemented this idea
in a simplified way.
We demand a very limited
solver precision of $10^{-2}$ in the first step
and
check if this quality is good enough.
If not, we in a second step
demand full solver precision of $10^{-6}$.

\subsection{Noisy acceptance step method II}

Knowing of the
idea of using Gegenbauer polynomials \cite{gegenbauer}
to solve equation of the type
\begin{eqnarray}
A^{1\over 2} x = y
\end{eqnarray}
as explained in App.~\ref{c_gegenbauer},
we want to rewrite
the formulae
of the 
noisy Metropolis update section,
resulting
not
in the acceptance Eq.~\ref{e_acceptance}
but in the
corresponding
\begin{eqnarray}
P^A_{(U,\phi)\to(U',\phi')}
&=&
{\rm min} 
\bigl( 1,e^{
- \chi^\dagger [1-R']^{-{1\over 2}} (1-R) [1-R']^{-{1\over 2}}\chi 
+\chi^\dagger \chi
} \Bigr)
\; ,
\end{eqnarray}
where now 
we effectively do not need the square root of $Q^2P(Q^2)$
but the inverse square root.

It is again easy to see
the idea.
One starts with
a heatbath distribution
including a determinant factor (but now in the numerator)
\begin{eqnarray}
P^{HB}_{\eta} 
= 
N_{\rm n} \det[1-R']
e^{-\eta^\dagger (1-R') \eta }
\; .
\end{eqnarray}
The acceptance needed in this case is
\begin{eqnarray}
P^A_{(U,\phi)\to(U',\phi')}(\eta)
=
{\rm min} \bigl( 1,e^{- \eta^\dagger (1-R) \eta +
\eta^\dagger (1-R') \eta } \Bigr)
\; .
\end{eqnarray}
We skip the proof of detailed balance
as it is identical to the method I case.

We apply the same
transformation trick
eliminating the Gaussian vector $\eta$
via the substitution
\begin{eqnarray}
\eta = [1-R']^{-{1\over 2}} \chi
\end{eqnarray}
given by
the application of the
Gegenbauer
solver mentioned above.

Thus we again constructed
a purely Gaussian distribution
for the $\chi$ variables
\begin{eqnarray}
P_{\chi} 
= 
N_{\rm n}
e^{
- \chi^\dagger [1-R']^{-{1\over 2}} (1-R') [1-R']^{-{1\over 2}}\chi 
}
= 
N_{\rm n}
e^{-\chi^\dagger \chi }
\; ,
\end{eqnarray}
yet the acceptance is now
\begin{eqnarray}
P^A_{(U,\phi)\to(U',\phi')}
&=&
{\rm min} \bigl( 1,e^{ \eta^\dagger (R-R') \eta } \bigr)
\; .
\end{eqnarray}

This formula 
makes it evident that
in this scheme
the main work is done by the Gegenbauer
solver constructing the $\eta$ vectors.
Thus its usefulness
relies on the fact 
that the convergence rate of the
Gegenbauer solver is equivalent to that
of the Conjugent Gradient \cite{gegenbauer}.
In Ch.~\ref{c_cost}
we will give results of the first
simulations with the Gegenbauer solver.
Details of polynomials and 
solver scheme are given in App.~\ref{c_gegenbauer}.

We would like to remark that
an optimisation of this method analogous to that
of the acceptance step method I
with adapted precision is possible.
The Gegenbauer inverter is not restartable,
but a slight alteration storing
some shift vectors in the iteration
could be implemented 
without creating too much overhead.
We thus could
interrupt the solver iteration,
check whether the solver quality
is already good enough for the requested Metropolis decision,
and continue if this is not yet the case.
For reasons of 
limited computer resources
this was not included in this study.

\section{Counting in $Q$ operations}

Our implementation of the 
L\"uscher local bosonic algorithm
applies in the corrected measurement case
\begin{itemize}
\item
$N_{\rm refl}$ reflections
\item
$1$ heatbath step 
\end{itemize}
for the $n$ bosonic fields
and the link field,
while
in the acceptance step cases 
twice that amount is necessary
to symmetrize the trajectory.
We use an abbreviation
\begin{itemize}
\item
$N_{\rm sym}=1$ for
corrected measurement and
\item
$N_{\rm sym}=2$ for
acceptance  step schemes.
\end{itemize}
For the boson field correction part,
both ways have to
invert the
approximation polynomial
$Q^2 P(Q^2)$
which takes
\begin{eqnarray}
2 N_{\rm CG} (n+1)
\end{eqnarray}
$Q$ operations.
Heatbath and reflections for the boson fields both require
\begin{itemize}
\item
1 $Q$ operation to initialise the stored auxiliary fields
\item
1 $Q$ operation in the force calculation routine
\item
1.5  $Q$ operations in the update routine
\end{itemize}
so that we obtain
\begin{eqnarray}
3.5 \cdot n N_{\rm sym} ( N_{\rm refl} +1)
+
2 N_{\rm CG} (n+1)
\; .
\end{eqnarray}
The link updates take in 2 dimensions
for heatbath and reflection each
\begin{itemize}
\item
2 $Q$ operations for the staples
\item
2 $Q$ operations for the bosonic force parts
\end{itemize}
so that we end up with
the total number of matrix multiplications necessary
for one update step of
\begin{eqnarray}
N_{\rm Q \; ops / update}
=
[3.5  n + 4 ] \cdot N_{\rm sym}  ( N_{\rm refl} +1)
+
2 (n+1) N_{\rm CG} 
\; .
\end{eqnarray}

\chapter{The problems I: Instabilities}

In the update algorithms of Ch.~\ref{c_lba}
we encounter
the  basic numerical problem
of evaluating a matrix-valued polynomial of high order.
To be specific,
we will in the following
consider
the problem evaluating a polynomial approximation
of the inverse determinant
\begin{equation} \label{p}
{\rm det} A \approx \left[ {\rm det} P_{n}(A) \right]^{-1}
\; ,
\end{equation}
using
the Chebyshev polynomials
as defined in Ch.~\ref{c_chebyshev}.

The naive idea would be to use the factorized form
Eq.~\ref{e_fact}. 
However, 
the numerical construction 
of a polynomial using the
product representation, can
-- due to rounding errors -- 
easily lead to a loss of precision or even to numerical instabilities.
This holds in particular if
computers with 32-bit floating point 
precision
are used. 

Often, as in the case of
Chebyshev polynomials, 
 numerically stable recursion
relations are available (viz. Ch.~\ref{c_chebyshev})
\cite{chebyshev_book}. 
However, 
exact versions of the 
LBA
\cite{lba_exact} or related approaches 
like the Polynomial Hybrid Monte Carlo (PHMC) algorithm
\cite{fast_fermion,phmc2}  
often need the factorized 
form of the polynomial $P_n$. 
Especially a decomposition
of the polynomial into two or more (e.g. complex conjugate)
parts can in general not be done without
recursion to the factorized form.
This is the numerical problem 
we are faced with in the
Hermitean LBA variant using a Metropolis acceptance step.
It is our intention to investigate in these cases 
several ordering schemes for the complex roots.

\section{Factorized Chebyshev polynomial} 

Let us consider the 
Chebyshev
approximation of a function $f(s)$
depending on a real variable
$s$
by a polynomial $P_n(s)$ of degree $n$.                       
The motivation 
to initially study
a single degree of freedom is 
that we might think of the matrix $A$ as being diagonalized.
Then the problem, Eq.~\ref{p}, reduces to finding
 a polynomial that approximates
each $\lambda^{-1}(A)$ separately, where $\lambda(A)$ is a real eigenvalue of $A$. 
We therefore expect that studying a single degree of freedom can 
provide information also about the
qualitative behaviour of rounding error effects when the matrix valued polynomial $P_n(A)$
is numerically computed. 

In principle 
evaluating 
the partial product $P_q(s)$ 
\begin{equation} \label{partialfact}
 P_q(s)= N^q_{\rm norm} \prod_{k=1}^q (s-z_k)
=
\prod_{k=1}^q n_k (s-z_k)
\;, 
\end{equation} 
one has to define a normalisation $N^q_{\rm norm}$
for the partial products,
effectively
distributing the global normalisation constant
$N_{\rm norm}$
to $n$
$k$-dependent normalisation constants $n_k$.

Considering
precision losses or numerical instabilities
can be understood 
by the following argument:
The absolute values of the, in general complex, 
subsequent partial products $|P_q(s)|$
and $|P_{q+1}(s)|$ can be different by orders of magnitude
if $s \approx z_q$.
If now $|P_{q+1}(s)| \ll |P_q(s)|$ then this must have been achieved by subtracting
two large numbers, which bears the danger of a significant loss
of precision. 
The problem itself suggests, however, its solution: The monomial factors
in Eq.~\ref{e_fact} or equivalently the roots
$z_k$ should be ordered, if possible, in such a way that the absolute values 
of all partial 
products $|P^q(s)|$ have the same order of magnitude.
Regarding an application to vectors (where $s$ is a priori unknown),
this has to be achieved in an $s$-independent way. 

To investigate the effect of reordering,
we propose to use 
a simple
criterion to determine the effects of these rounding errors.
The idea is to evaluate
in a first step for given $q$ the maximal and the minimal value 
of $|P_{q,\epsilon}(s)|$ over the spectral interval $0\le s \le 1$. 
The ratio of the maximum to the minimum value,
i.e. $\tilde{R}_q=\max_{s \in [0,1]} |P_{q,\epsilon} (s)|/
\min_{s \in [0,1]} |P_{q,\epsilon} (s)|$, is then a measure
of how large the fluctuations of the absolute values of the partial
products can become. 
If these fluctuations are very large, 
the polynomial can not be constructed in a safe way.
Building the maximum of $\tilde{R}_q$ with respect to $q$, 
we 
arrive at the final quantitative measure
\begin{equation} \label{rmax} 
R_{\rm max} 
=
\max_{ q \in \{1, \dots, n\} } \left\{
{ \max_{s \in [0,1]} |P_{q,\epsilon} (s)|
\over
 \min_{s \in [0,1]} |P_{q,\epsilon} (s)|}
\right\}
\; .
\end{equation}
It is clear that 
$R_{\rm max}$  
has to be smaller
than the inverse relative accuracy on a given computer 
as a necessary condition for the stability of the evaluation of the full polynomial.  

Another quantity of interest is the maximum value of
the partial products itself
\begin{equation} \label{M}
M_{\rm max}
=
\max_{s \in [0,1], q \in \{1, \dots, n\} } |P_{q,\epsilon}(s)|
\;.
\end{equation}
This has to be 
be smaller 
than the largest representable number in order not to run into
overflow. 
To avoid underflow 
one should thus also study
$M_{\rm min}$ by replacing the maximum in Eq.~\ref{M} by the 
corresponding minimum. 
We, however, restrict ourselves to the maximum quantity.

Note that $R_{\rm max}$ and $M_{\rm max}$
are computed for $s \in [0,1]$, whereas the
Chebyshev polynomial has an exponential convergence only in the
interval $s \in [\epsilon,1]$. 
However, 
as will be explicitly demonstrated below, our results for 
$R_{\rm max}$ and $M_{\rm max}$ do not depend very much on the
choice of the lower end of the interval. 

\section{Ordering schemes}

In this section we
introduce the different ordering schemes
used
for the roots $z_k$ Eq.~\ref{e_roots}. 
In principle,
one could try
to also distribute the normalisation constants
$n_k$ introduced above 
in a $k$-dependent way 
to reduce rounding errors. 
However, 
we found no improvement over 
the naive
homogeneous distribution
\begin{eqnarray}
\label{e_ck}
n_k = (N_{\rm norm})^{1 \over n}
\end{eqnarray}
which we therefore 
adopt throughout the rest of this work.

\begin{figure}[t]
\vspace{-0mm}
\centerline{ \epsfysize=17.5cm
             \epsfxsize=17.5cm
             \epsfbox{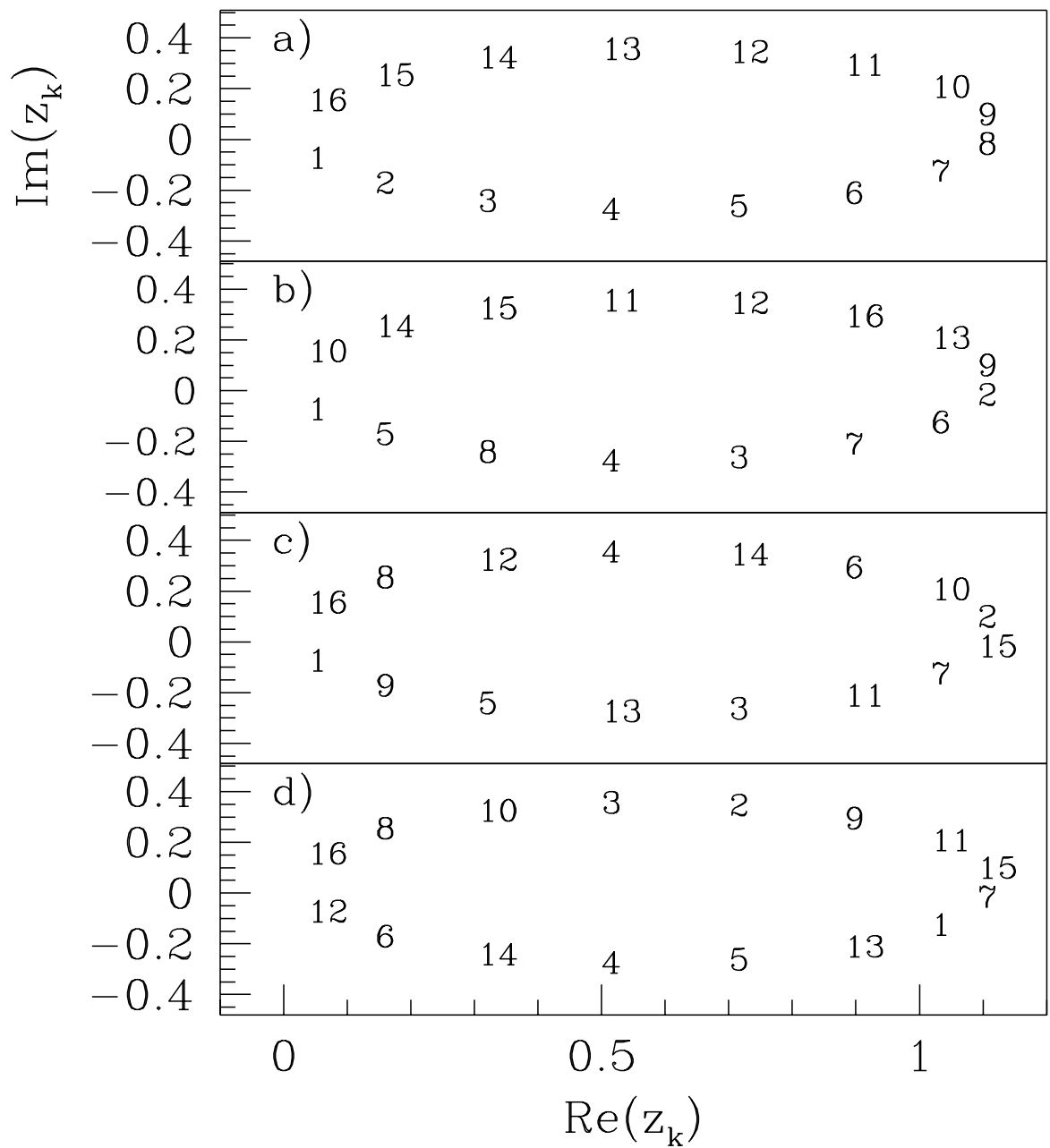}}
\begin{center}
\parbox{12.5cm}{\caption{ \label{f_index}
The roots $z_k$ with $k=1,\dots, 16$ 
and $\epsilon =0.1$ are shown in the complex plane. 
Labels of roots indicate in which order they
are used for the numerical evaluation of the
Chebyshev polynomial
within each ordering scheme. 
We show in a) the naive, b) the pairing, c) 
the bitreversal and d) the Montvay root ordering scheme. 
}}
\end{center}
\end{figure}

\vspace{0.3cm} 
\noindent {\bf Naive ordering}
\\
As naive we regard the ordering 
given by
\begin{equation} \label{znaive}
z^{\rm naive}_k
=
z_k\;, k=1,\cdots,n
\end{equation}
with the $z_k$ given in Eq.~\ref{e_roots}. 
The roots $z_k$ form an ellipse in the complex plane 
and in the naive ordering the roots are  
selected from this ellipse by 
starting at the origin and
moving around anti-clockwise.
This is indicated in Fig.~\ref{f_index}a, where the roots are shown
labelled according to the order in which they are used in the evaluation of
the Chebyshev polynomial Eq.~\ref{e_fact}. 
As we will see later, ordering the roots in this naive way
gives rise to substantial rounding error effects,
even leading to numerical overflow.

\vspace{0.3cm} 
\noindent {\bf Z\"urich group scheme}
\\
Recently,
De Forcrand
brought to our attention
the scheme used for 
the simulations of the Z\"urich group \cite{galli_ordering},
which consists of 
using
the complex pairs
$s_k,s_{(n+1)-k}$
as the $2k-1$ and $2k$-th roots.
Because of results
generally worse than the naive scheme,
we decided to simply state that
this scheme is not to be recommended,
and not to include it in the plots.
We would further like to stress that
the studies undertaken by the Z\"urich group
were of a kind not influenced by
the quality of the ordering scheme.

\vspace{0.3cm}
\noindent {\bf Pairing scheme}
\\
A first improvement over the naive ordering is to 
use a simple
pairing scheme, reordering  
the roots 
\begin{eqnarray}
z_k^{\rm pair}
=
z_{j(k)}\; ,\; k=1,\dots,n
\; .
\end{eqnarray}
We give the 
reorder index $j(k)$ for the example of
$n/2$ being a multiple of $4$ and $n'=n/8$.
In the lower half plane, ${\rm Im}\; z_k < 0$, 
the pairing scheme is achieved by 
\begin{eqnarray}
j 
&=& 
\Bigl\{
1, {n\over 2}, {n\over 4}+1, {n \over 4},
\nn \\&& \phantom{\Bigl\{}
2, {n\over 2}-1, {n\over 4}+2, {n \over 4}-1,
\nn \\&& \phantom{\Bigl\{}
\dots \nn \\
&& \phantom{\Bigl\{}
 n',{n\over 2}-n'+1, {n\over 4}-n', {n \over 4}+n'-1 \Bigr\}
\end{eqnarray}
and for ${\rm Im}\; z_k > 0$  
correspondingly.
An illustration of the ordering in the pairing scheme
is shown in Fig.~\ref{f_index}b. The label of the roots indicates in which order they     
are used in the numerical construction of the polynomial.

In case of $n/2$ not divisible by 4, 
we search for the next integer $m$ smaller than $n/2$ 
and divisible by 4. We then repeat the above described procedure on these 
$m$ roots and simply multiply the remaining roots $z_{m+1} \cdots z_{n/2}$ 
at the end. 
To make our procedure explicit,
we include the Fortran
code to reorder the first half of the roots
in App.~\ref{PSa}.
The code
for the second half of the roots
is constructed analogously.
The Fortran code also contains the case of $n/8$ being odd, where
the above described procedure has to be slightly modified. 

\vspace{0.3cm}
\noindent {\bf Subpolynomial scheme}
\\
The problem with rounding errors in computing the polynomial 
in the product representation arises most severely for a high degree $n$ of the 
polynomial. In order to decrease the effects of rounding errors
one may therefore be guided by the 
following intuitive observation.
Let us consider the polynomial $P_{n,\epsilon}(s)$ with roots $z_k$ and $n\gg 1$. 
If $m$ is an integer divisor of $n$, the roots $z_{1}, z_{1+m}, z_{1+2m},
\dots z_{1+(n/m-1)m}$ turn out to be close to the roots
characterising the polynomial
$P_{n',\epsilon}(s)$ of degree $n'=n/m$ 
(note that we keep the same $\epsilon$). Moreover, the 
normalisation constants 
$c_{k}= \left( N_{\rm norm} (n) \right)^{1 \over n}$ 
and 
$n'_{k}= \left( N_{\rm norm} (n') \right)^{1 \over n'}$ 
are of the same order (the
dependence on $n$ of $n_k$  turns out to be negligible for large $n$). 
Then 
the product
\begin{equation} \label{u} 
u= \prod_{j=0}^{n/m -1} [c_{j+1}(s-z_{1+jm})] 
\end{equation} 
is a good approximation of $P_{n',\epsilon}(s)$, $|u-P_{n',\epsilon}(s)|\ll 1$
for all $\epsilon < s \le1$, and $|u|\le\epsilon$. 
The same argument
may be repeated for the other similar sequences of roots, like
$z_{2}, z_{2+m}, z_{2+2m}, \dots z_{2+(n/m-1)m}$, \dots ,
$z_{m}, z_{2m}, z_{3m}, \dots z_{n}$. 

This means that the product Eq.~\ref{e_fact}
may be split in a product of $m$ subproducts, in such a way that each of
them approximates 
the factorized form of a
polynomial $P_{n',\epsilon}(s)$ of {\it lower} degree $n'=n/m$.
Because of the lower degree of the polynomial given by the products
such as 
Eq.~\ref{u}, one
may expect that much smaller fluctuations occur in the intermediate
steps of the evaluation of each of these subproducts. 

The reordering of the subpolynomial scheme 

\begin{eqnarray}
z_k^{\rm sp}
=
z_{j(k)}
; k=1,\dots,n 
\end{eqnarray}
can be represented by 
\begin{eqnarray}
j 
&=& 
\Bigl\{
1, 1+m, 2+2m, \dots, 1 + ({n\over m}-1)m,
\nn \\&& \phantom{\Bigl\{}
2, 2+m, 2+2m, \dots, 2 + ({n\over m}-1)m,
\nn \\&& \phantom{\Bigl\{}
\dots 
\nn \\&& \phantom{\Bigl\{}
m, m+m, m+2m, \dots, m + ({n\over m}-1)m \Bigr\}
\; ,
\end{eqnarray}
where $m$
is an integer divisor of $n$.

We found that
$m$ has to be chosen $m \approx \sqrt{n}$
to avoid severe loss of precision in the construction of the polynomial
and to reduce rounding error effects to a tolerable level.
We remark that the naive ordering 
is reproduced by the two extreme
 choices $m=1$ and
$m=n$. 
The Fortran code which generates the subpolynomial ordering of the
roots may be found 
in App.~\ref{SPa}.

\vspace{0.3cm}
\noindent {\bf Bitreversal scheme}
\label{s_bit}
\\
The subpolynomial scheme can be generalised, leading to what
we will call the bitreversal scheme. To illustrate how this
scheme works, let us assume that the degree $n$ of the polynomial
is a power of two. 
One now starts with the $n$  
monomial factors in Eq.~\ref{e_fact}, chooses $m=n/2$ and
applies the subpolynomial scheme resulting in $m$ binomial
factors. We then proceed in choosing a $m'=m/2$ and again applying
the subpolynomial scheme to these $m$ binomial factors which leaves us 
with $m'/2$ subpolynomials each of degree four. The procedure can be iterated
until we are left with only one subpolynomial having the degree of
the polynomial itself. 
The above sketched procedure can be realized in practise by 
first representing the integer label (counting from $0$ to $n-1$) 
of the roots in the naive order
by its bit representation. The desired order is then obtained
by simply reversing the bits in this representation. 
The resulting reordering of the roots is shown in
Fig.~\ref{f_index}c with $n=16$ as an example. 

For $n$ not a power of 2,
we pad with dummy roots, chosen to be zero for instance, until the 
artificial  
number of roots is a power of 2.
The bitreversal 
procedure can then be applied as described above.
Afterwards, the dummy roots have to be eliminated from the sequence.

To make the procedure of reordering the roots 
explicit,
we again give the Fortran code used
to generate the bitreversal ordering in App.~\ref{BRa}.

\vspace{0.3cm}
\noindent{\bf Montvay scheme}
\\
Recently, Montvay \cite{ordering_optimization}
suggested to order the roots according to an    
optimisation procedure which can be implemented numerically, e.g.
using algebraic manipulation programs such as Maple. 
Let us shortly sketch how Montvay's ordering scheme works and
refer to \cite{ordering_optimization} for details. 
We assume that we have 
already the optimised order of
the roots for the partial product $P_{q<n}(s)$, Eq.~\ref{partialfact}. 
Then the values of 
$|sP_q(s)(s-z)|$ are computed for all $z$ taken from the set of roots not already used.
The values of $s$ are 
taken from a discrete set of points, 
 $\left\{s_1,\dots,s_N\right\}$ which are all
in the interval $\left[\epsilon,1\right]$.               
Now, the maximal ratio over $s\in \left\{s_1,\dots,s_N\right\}$
of all values 
$|sP_q(s)(s-z)|$ is computed for each root $z$ separately. 
Finally, that root is taken which gives the lowest of
these maximal values. Starting with the trivial polynomial $P_0(s)=1$, this procedure
obviously defines a scheme with which the roots can be ordered iteratively. 
We show in Fig.~\ref{f_index}d 
the resulting order of the roots using Montvay's scheme
by again labelling the roots in the order as they are used to compute the 
Chebyshev polynomial Eq.~\ref{e_fact}. 

\clearpage
\noindent{\bf Clenshaw recursion}
\\
We repeat the statement that
the evaluation of a Chebyshev polynomial normally does, of course,
not rely on the product representation, Eq.~\ref{e_fact}. A 
numerically safe way to construct a Chebyshev polynomial is to use
a recursion relation, such as the Clenshaw recursion
Eq.~\ref{e_chebyshev_recursion}
described in Sec.~\ref{c_chebyshev}
\cite{chebyshev_book}.
The  recursion is known to be numerically very stable 
and will serve
us in the following as a reference procedure for the numerical evaluation
of a Chebyshev polynomial.

\section{Scalar numerical tests}

In this section we will give our numerical results 
for the quantities $R_{\rm max}$, Eq.~\ref{rmax}, and $M_{\rm max}$,
Eq.~\ref{M}. 
When these quantities assume large values, the numerical evaluation of the
Chebyshev polynomial Eq.~\ref{e_fact} is affected by
rounding errors and precision losses or numerical instabilities can
easily be 
encountered. 
All results presented in this section are obtained on computers
using 64-bit arithmetic. 
Montvay's ordering scheme is constructed
applying a Maple program \cite{montvay_program}
requiring 
40 digit precision. 
We want to emphasise at this point that the quantities
$R_{\rm max}$ and $M_{\rm max}$,
which
we are using to test the different ordering schemes,
 do not directly
correspond to the values $|sP_q(s)(s-z)|$, which serve to optimise the root
ordering in the Montvay scheme. 

\begin{figure}[t]
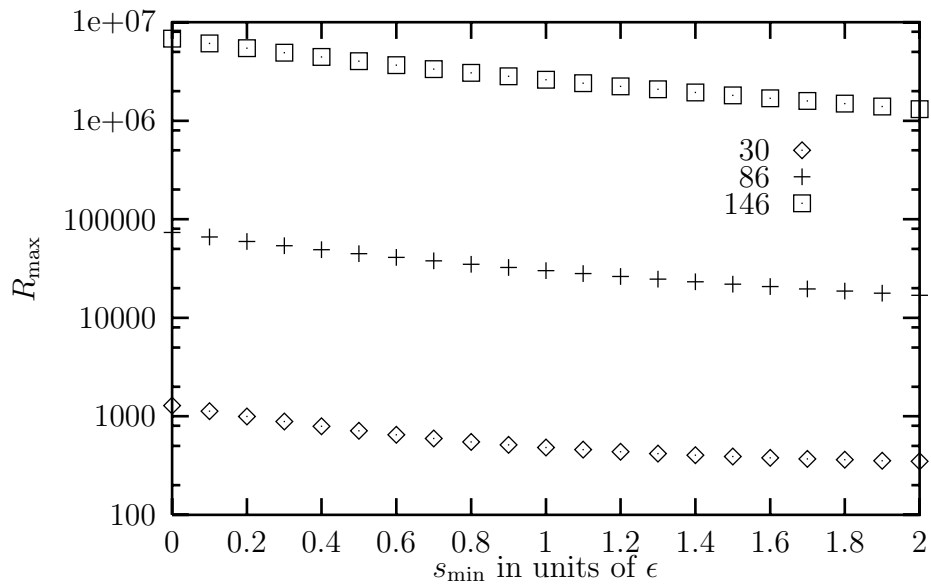


\include{data_test_scale_all_br}

\caption{\label{f_scale}
The ratio $R_{\rm max}$
is shown as a function of the
lower end $s_{\rm min}$ of the interval $[s_{\rm min},1]$ from which
the values of $s$ are taken to compute $R_{\rm max}$. 
$s_{\rm min}$ is measured in units of 
the parameter $\epsilon$.
We show data for three degrees of the Chebyshev polynomial 
$n=30$, $n=86$ and $n=146$
at fixed approximation quality
$\delta=0.001$, using the bitreversal scheme. 
Although different in magnitude, the flat behaviour of
$R_{\rm max}$ as a function of $s_{\rm min}$ is very
similar for the other schemes.}
\end{figure}

In order to compute the values for $R_{\rm max}$, Eq.~\ref{rmax}, and $M_{\rm max}$ 
Eq.~\ref{M} we take $5000$ values of $s$, equally spaced in the
interval $[0,1]$. 
We check explicitly that the values of $R_{\rm max}$ do not depend
very much on the lower end of  the interval $[s_{\rm min},1]$
from which $s$ is taken.
In Fig.~\ref{f_scale}
we show  
$R_{\rm max}$ 
as a function of the
lower end of the interval  $s_{\rm min}$ measured in units of 
the parameter $\epsilon$.
We present data for three polynomial degrees
$n=30$, $n=86$ and $n=146$
at fixed approximation quality
$\delta=0.001$
using the bitreversal ordering scheme.               
As Fig.~\ref{f_scale} shows, the dependence of $R_{\rm max}$ on 
$s_{\rm min}$ is very weak. For the other root ordering schemes 
we find
a very similar behaviour of $R_{\rm max}$ as a function of $s_{\rm min}$.
This justifies the use of $s_{\rm min}=0$ which
we have used for the numerical tests
described in this section. 

\begin{figure}[tbp]
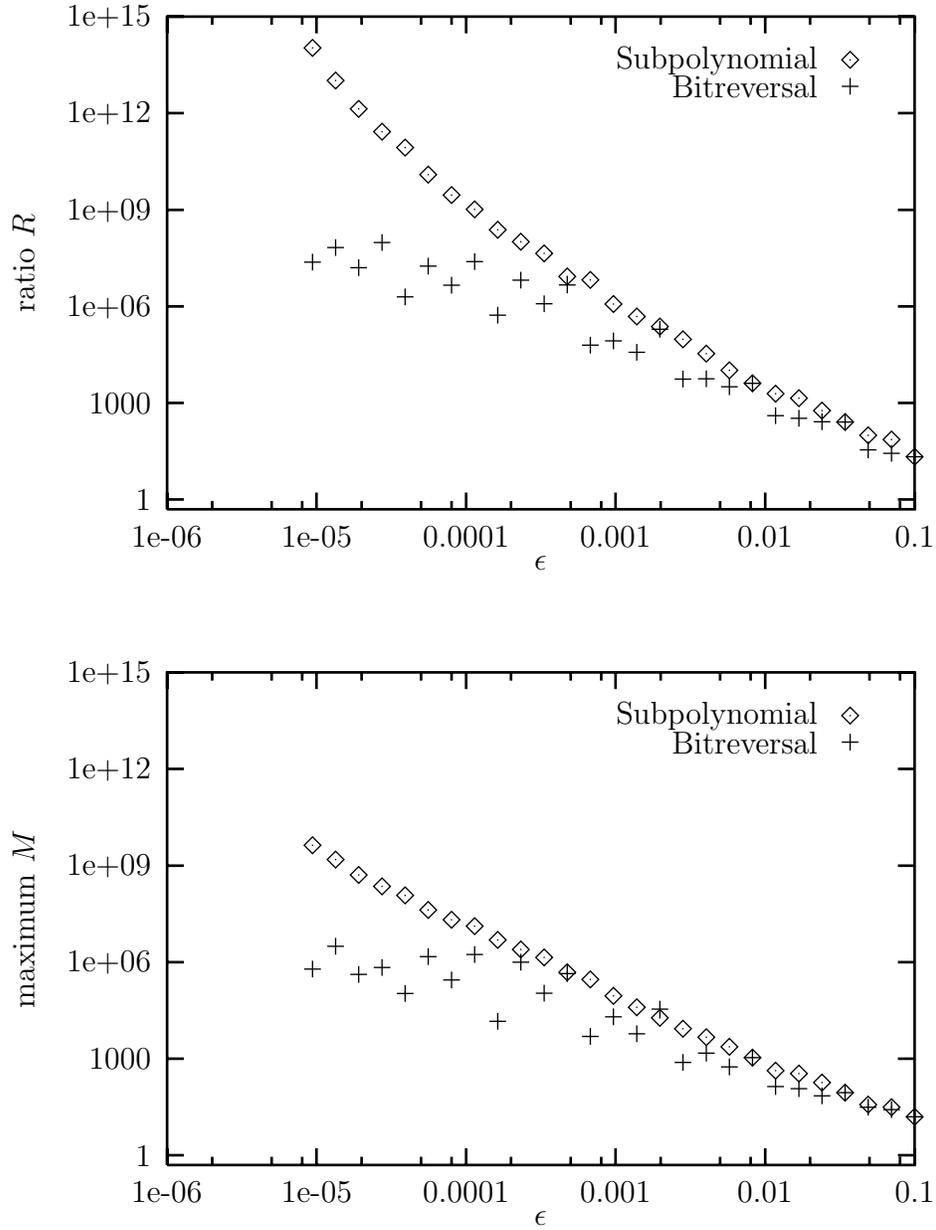


\include{data-3-d0.1-r}
\include{data-3-d0.1-m}
\caption{\label{f_d0.1}
Ratio $R_{\rm max}$
and maximum factor $M_{\rm max}$
are shown as a function of the
degree of the polynomial
at fixed approximation quality
$\delta=0.1$.
We compare subpolynomial and bitreversal
ordering schemes.}
\end{figure}

We start by comparing the subpolynomial and the bitreversal schemes, as they
are closely related to each other. 
In Fig.~\ref{f_d0.1} we show $R_{\rm max}$ and $M_{\rm max}$
as a function of the degree $n$ of the Chebyshev polynomial, 
keeping the maximal fit accuracy
$\delta =0.1$ constant by adjusting the parameter 
$\epsilon$ accordingly. 
For the subpolynomial scheme, the divisor $m$ is chosen to be
$m\approx\sqrt{n}$. Fig.~\ref{f_d0.1} clearly confirms 
the expectation that the bitreversal 
scheme, considered as a generalisation of the 
subpolynomial ordering scheme,
gives smaller values of $R_{\rm max}$ and $M_{\rm max}$. For degrees of the 
Chebyshev polynomial $n>40$ 
rounding error effects
are substantially suppressed in the bitreversal scheme 
compared to the subpolynomial scheme. 

In Fig.~\ref{f_d0.001} 
we show the values of $R_{\rm max}$ and $M_{\rm max}$ 
for the bitreversal, the naive, 
the pairing and the Montvay schemes for the case of $\delta =0.001$
as a function of $n$. 
We have also performed numerical tests 
at $\delta = 0.1$ and $\delta= 0.01$ where 
we found the same qualitative behaviour of 
$R_{\rm max}$ and $M_{\rm max}$ as a function of $n$ for
the different ordering schemes. The first striking
observation in Fig.~\ref{f_d0.001} is that with 
the naive ordering one obtains already for 
moderate degrees $n\approx 30$ of the polynomial large values of $R_{\rm max}$ 
and $M_{\rm max}$, indicating that rounding error effects are becoming
severely problematic. 
Clearly, in the naive ordering scheme
rounding errors can lead to very large
ratios of particular partial products. 
Using the naive scheme, especially on 
machines with only 32-bit precision, 
a safe evaluation of the Chebyshev polynomial
in the product representation can certainly not be guaranteed. 

The behaviour of the values of $R_{\rm max}$ and $M_{\rm max}$ 
obtained by using 
the naive scheme 
demonstrates the necessity of finding better root ordering schemes
which are able to reduce rounding error effects in evaluating
the factorized Chebyshev polynomial. 
That such ordering schemes 
do exist is also demonstrated in Fig.~\ref{f_d0.001}. 
For $n < 100$, the values for 
$R_{\rm max}$ and $M_{\rm max}$ obtained from 
the pairing, bitreversal and Montvay 
schemes are close to each other and many orders of magnitude below the ones
of the naive scheme. However, for $n>120$, the values of 
$R_{\rm max}$ from these ordering schemes also start to 
deviate from each other. 
Taking $R_{\rm max}$ as a measure of the effects of rounding errors,
it seems that the bitreversal scheme 
can reduce the rounding error effects most
efficiently among the ordering schemes investigated here.

\section{Matrix valued numerical tests}

We report 
on a
direct test of the
ordering schemes described above
for
matrix-valued
polynomials.
As these tests were done using 
an existing program running on the powerful 
massively parallel Alenia Quadrics (APE) machines,
we briefly 
detour to 4-dimensional
lattice QCD \cite{qcd_book},
with
details given in App.~\ref{c_qcd}.

The task is to construct the matrix valued Chebyshev polynomial
of the operator $\hat{Q}^2$
\begin{equation} \label{matrixpol}
P_{n}(\hat{Q}^2) = \prod_{k=1}^n n_k(\hat{Q}^2 - z_k) 
\; ,
\end{equation} 
where the roots $z_k$ and the normalisation constant $n_k$ are given by
Eq.~\ref{e_roots} and Eq.~\ref{e_ck}, respectively. 
We evaluated the polynomial, Eq.~\ref{matrixpol}, 
using the Clenshaw recurrence
as well as using the different ordering of roots 
described above. 

The numerical tests 
are performed on thermalized configurations on $8^3\times 16$ lattices using 
the APE computers,
which have only 32-bit precision. 
Simulation parameters were chosen to be $\beta=6.8$, 
$\kappa = 0.1343$ and $c_{\rm sw} = 1.42511$. They correspond 
to realistic parameter values as actually used in  simulations
to determine values of $c_{\rm sw}$ non-perturbatively \cite{improvement2}. 
We adopt the same (Sch\"odinger functional) boundary conditions
as described in \cite{improvement} for the evaluation of $c_{\rm sw}$. 
For the above choice of parameters and setting $c_M=0.735$, 
the lowest eigenvalue of $\hat{Q}^2$ is 
$\lambda_{\rm min} = 0.00114(4)$ and the largest is 
$\lambda_{\rm max} = 0.8721(3)$. 
Investigations are performed at values of $(n,\epsilon)$  
$(16,0.003)$, $(32,0.003)$, $(64,0.0022)$ and $(100,0.0022)$.
At each of these values of $(n,\epsilon)$ we have $O(50)$ 
configurations. 

We apply the matrix $\hat{Q}^2P_{n,\epsilon}(\hat{Q}^2)$, which should be
close to the unit matrix for our choices of $n$ and $\epsilon$, to a 
random Gaussian vector $G_{alpha,s}(x)$, which is a complex vector, located at a
lattice point $(x)$ and carrying
colour $\alpha=1,2,3$ and spinor $s=1,\dots,4$ indices.  
We then compute the vectors
\begin{equation} \label{clenvector}
\Phi_{\rm Clenshaw} = \hat{Q}^2P_{n,\epsilon}(\hat{Q}^2) G
\; ,
\end{equation} 
where $P_{n,\epsilon}(\hat{Q}^2)$ is constructed via Clenshaw's recurrence
relation and 
\begin{equation} \label{rootvector}
\Phi_{\rm order} = \hat{Q}^2P_{n,\epsilon}(\hat{Q}^2) G
\; ,
\end{equation} 
where now $P_{n,\epsilon}(\hat{Q}^2)$ is evaluated using different root
ordering schemes, and order stands for naive, pairing, bitreversal and Montvay. 
On a given configuration and for given $G$ we finally determine
\begin{equation} \label{Delta}
\Delta = \|\Phi_{\rm Clenshaw} - \Phi_{\rm order}\|^2\; .
\end{equation} 

\begin{figure}[t]

\vspace{-0mm}
\centerline{ \epsfysize=13.5cm
             \epsfxsize=13.5cm
             \epsfbox{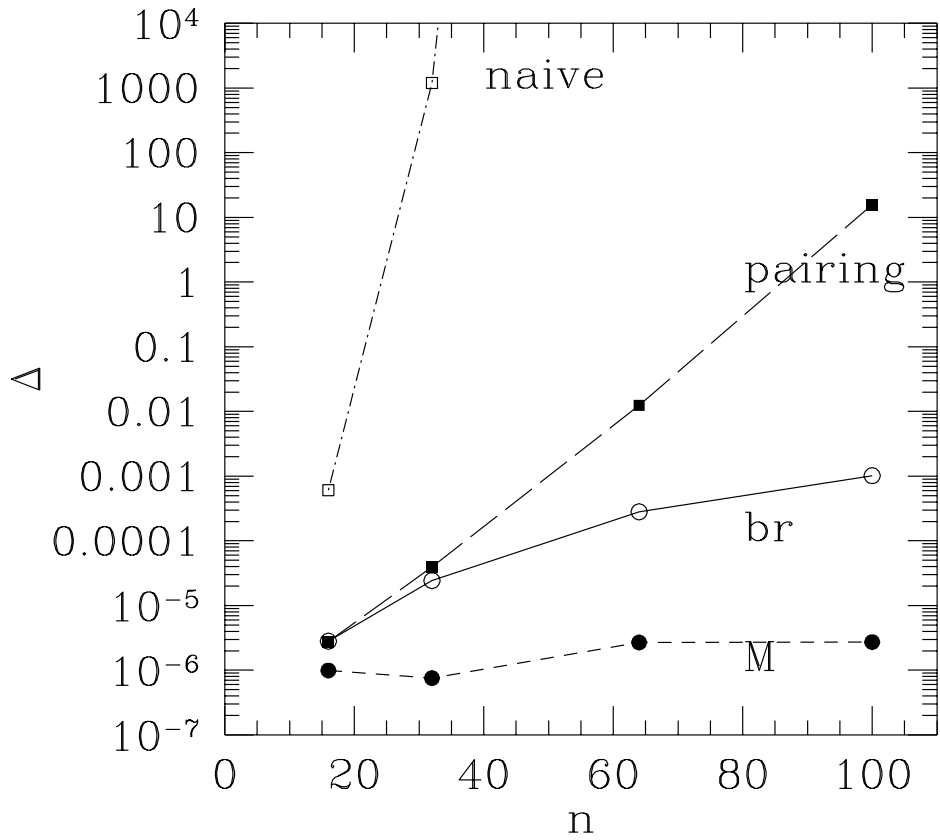}}
\begin{center}
\parbox{12.5cm}{\caption{ \label{f_delta}
The quantity $\Delta$ is shown as a function of the
degree of the Chebyshev polynomial $n$. 
We compare 
the naive, pairing, bitreversal (br) and Montvay
(M) ordering schemes. 
}}
\end{center}
\end{figure}

Since the Clenshaw recurrence is the numerically most stable
method to evaluate the Chebyshev polynomial, 
the values of $\Delta$ provide a measure for the effects 
of rounding errors.
The result for $\Delta$ as a function of $n$ is shown in Fig.~\ref{f_delta}. 
Using the naive ordering scheme, we could not run the cases of $n=64$
and $n=100$ 
as we hit 
numerical overflows. For the pairing scheme 
we find large values for $\Delta$
at $n=64$ and $n=100$.
The bitreversal scheme gives small, but non-negligible values of
$\Delta$ for all values of $n$ used. Finally, Montvay's scheme
gives $\Delta \approx 10^{-6}$ for all values of $n$. Surprisingly,
using roots ordered by the Montvay scheme, the construction
of the Chebyshev polynomial can be done with a stability that
is comparable to the one using the Clenshaw recurrence. 

\clearpage

\begin{figure}[tbp]

\include{data-3-d0.001-r}
\include{data-3-d0.001-m}
\caption{\label{f_d0.001}
Ratio $R_{\rm max}$
and maximum factor $M_{\rm max}$
are shown as a function of the
degree of the polynomial
at fixed approximation quality
$\delta=0.001$. 
We compare naive, pairing, bitreversal and Montvay
ordering schemes.}
\end{figure}

\chapter{The test: Physical results}

\section{Consistency tests}
\label{s_consistency_tests}

In general,
the results of valid simulation algorithms
must be independent of
technical simulation parameters,
random number generator types,
irrelevant boundary conditions
and measurement details like
source or noise types used for the correlators
or fermion densities.

To
check whether
we can rely on the
results of the program package
used to generate the CPU cost
data of Ch.~\ref{c_cost},
we present various
tests
comparing observables
from different measurement strategies
to analytical results,
those from HMC code or other groups.

\vspace{0.3cm}
\noindent{\bf Critical kappa}
\label{critical}

\begin{figure}[tbp]
\caption{\label{f_cg1}
{\bf Number of CG iterations.} 
We show as a function of $\kappa$
the number of 
iterations necessary to invert $Q$
to a precision of $10^{-6}$
using a Conjugate Gradient inverter
on $16 \times 16$ lattices for $\beta=2.0$.
The estimated $\kappa_c$ is indicated by a vertical line.
}

\vspace{0.2cm}
\centering
\makebox[8.8cm]{
\epsfxsize=8.5cm
\epsffile{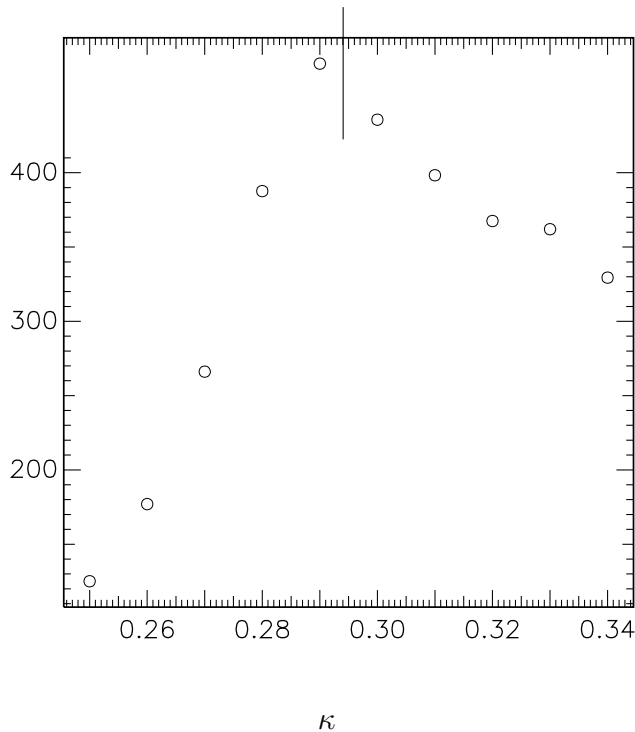}
}
\vspace{0.2cm}
\begin{center}
$\kappa$
\end{center}
\vspace{-0.3cm}
\end{figure}

The critical kappa 
was estimated
using
the peak in the number
of Conjugent Gradient solver iterations needed to invert
$Q$.
We perform dynamical simulations 
using the LBA with acceptance step method I
on $16\times 16$ lattices
with
values of $\beta=2$, $\beta=6$ and $\beta=10$.
Results 
shown in Fig.~\ref{f_cg1} to \ref{f_cg3}
lead to estimated $\kappa_c$
of $0.287$ for $\beta=2.0$,
$0.268$ for $\beta=6.0$ and $0.264$ for $\beta=10.0$
with quite large errors of about $0.02$.
The estimated values are indicated 
via a vertical line in the figures. 
We thus obtain
the expected behaviour
of $\kappa_c \approx {1\over 4}$ for high $\beta$
and an increase in $\kappa_c$ 
as one decreases the value of $\beta$
described in Sec.~\ref{s_kappa_c}.

\begin{figure}[tbp]
\caption{\label{f_dynmasses}
{\bf Meson mass spectrum.}
We show dynamical meson masses as a function of $\kappa$
calculated 
on a $16\times 32$ lattice at $\beta=10.0$.
Included are pion masses from the operators $\bar\psi \gamma^5 \tau \psi$
and $\bar\psi \gamma^5 \gamma^0 \tau \psi$,
eta masses from $\bar\psi \gamma^5 \psi$ and $\bar\psi \gamma^5
\gamma^0 \psi$ and a0 masses from $\bar\psi \tau \psi$.
We give just one pion mass if results are
within errorbars.
}

\vspace{0.2cm}
\centering
\makebox[8.8cm]{
\epsfxsize=8.5cm
\epsffile{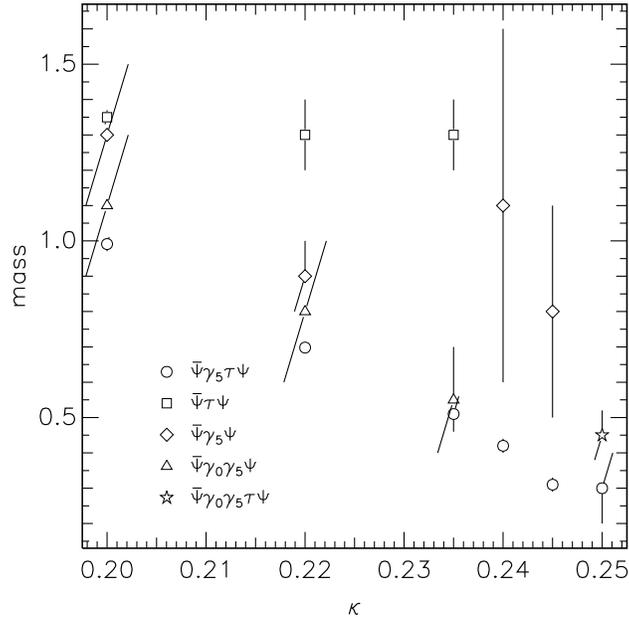}
}
\end{figure}

An independent and usually better way to estimate
the critical kappa
is the extra\-polation to
the kappa value at which the pion mass is vanishing.
Using dynamical data from $16 \times 40$ 
and $16 \times 32$ lattices respectively,
we estimate from Fig.~\ref{f_16_mass6}
a value of $0.266$ for $\beta=6.0$
and
from 
Fig.~\ref{f_dynmasses}
a value of
$0.262$ at $\beta=10.0$
with errors of about the same order of magnitude as above.

Both independent measurement schemes thus yield the same
numbers within errors.
Moreover, results also agree with expectations.

\vspace{0.3cm}
\noindent{\bf Noisy scheme tests}

We
simulate 
full dynamical fermions
on $4\times 4$ lattices
and
$\beta=1.0$
using the LBA
with
different approximation qualities
and approximation 
regions.
The acceptance step method I acceptances
or respectively the reweighting factors
are calculated both in a noisy and exact way.

\ul{Plaquette.}
We compare
in Tab.~\ref{t_plaquette1}
to 
\ref{t_plaquette4}
plaquette averages
to the analytic values
obtained using
a hopping parameter expansion
and the exact plaquette for pure $U(1)$ gauge theory
as described in App.~\ref{s_hopping} and \ref{s_exact_plaquette}.

All calculations
compare simulations 
with parameter choices 
$n=0,2,10$
and $\epsilon=0.5, 0.01$
to the leading order hopping parameter expansion result
which is believed to be reasonable up to $\kappa \approx 0.15$.
We generally accumulated statistics of about
$10^5$ sweeps.
Only in the case of the acceptance step method I 
used with $n=0$, where the Metropolis acceptance step
is correcting for the full fermion determinant 
estimated stochastically,
this was clearly not enough.
We accumulated in this case statistics of up to
$10^7$ sweeps with integrated autocorrelations of $O(1000)$.

The 
agreement of the simulation results
is evident in all cases. 
Analytical results are reproduced up to 
$\kappa=0.15$. 
For larger values of $\kappa$ systematic deviations 
from the analytic results are
encountered as expected.

\ul{Correction factor.}
Checking the influence of the noisy scheme
on the determination of the correction factor
we compare
in Tab.~\ref{t_correction}
factors
calculated from eigenvalues
to those estimated.
We show results for the 
correction factor
from the eigenvalues of $Q^2$
and from 1000 noisy estimation steps.
All calculations
were done on
a $4 \times 4$ lattice
with 
$\beta=1.0, \kappa=0.1,\epsilon=0.01$
and different $n$.

The results
show the 
compatibility
of the exact and 
the noisy estimation scheme.
The $n=2$ results 
nicely demonstrate
the importance of importance sampling,
i.e.
that 
simulating
an almost flat distribution
is impossible
for dynamical fermions
as corrections become too large.
They also show that the correction factor
can in fact deviate decisively from 1.

\ul{Different noises.}
We compare 
in Tab.~\ref{t_ising}
$Z_2$ and $Z_4$ 
complex Ising variables 
used in the noisy estimators for fermion densities
as described in App.~\ref{s_mesons}.
We 
calculate the fermion condensate $\langle \bar\psi \psi \rangle$
and 
pseudo-scalar density $\langle \bar\psi \gamma^5 \psi \rangle$
in the quenched case 
on $16 \times 16$ lattices for $\beta=2.0$
and $\kappa=0.26$.

Results nicely agree with each other.
We remark that the 
imaginary part of the pseudoscalar density
was found to be zero to about $10^{-10}$.

\vspace{0.3cm}
\noindent{\bf $U=1$ tests}

For $U=1$
the condensate can be calculated analytically
as described in App.~\ref{s_free}.
We show results 
for this basic test
from 
runs with links fixed to $U=1$  
on $16 \times 16$ lattices
for a range of $\kappa$ values 
in Tab.~\ref{t_u1_cond}.
They obviously
agree with the analytic predictions.

\vspace{0.3cm}
\noindent{\bf Condensate tests}

In order to check the routines, 
a comparison 
to 
completely 
independent results is desirable.

For the fermion
condensate in dynamical simulations
the group in Graz 
obtained data for the
2 flavour Schwinger model with 
Wilson fermions
using a HMC
update algorithm
on $16 \times 16$ lattices
at $\beta=2.0$
\cite{lang}.

\begin{figure}[tbp]
\caption{\label{f_dyncond}
{\bf Fermion condensate.}
We show as a function of $\kappa$ 
the dynamical fermion condensate,
calculated
on $16 \times 16$ lattices
at $\beta=2.0$.
}

\vspace{0.6cm}
\centering
\makebox[8.8cm]{
\epsfxsize=8.5cm
\epsffile{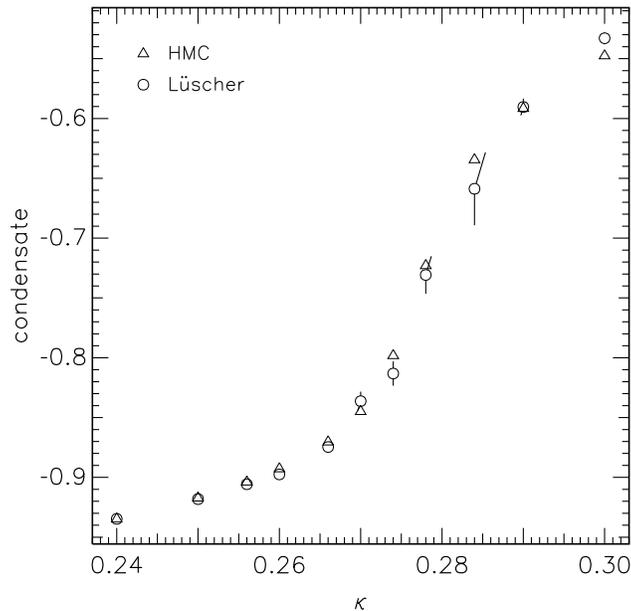}
}

\end{figure}

In Fig.~\ref{f_dyncond}
we show the comparison
of LBA with acceptance step method I data
to these independent results.
The agreement is evident.

\vspace{0.3cm}
\noindent{\bf Random generator tests}
\label{s_RG}

As mentioned in Ch.~\ref{c_hmc},
we
throughout this study
used the high-quality
random number generator (RG)
by M. L\"uscher \cite{random_generator}
to ensure that results 
are independent
from 
random numbers.
To further 
verify this,
we compare 
quenched 
results 
on $16 \times 16$ lattices at $\beta=10$
obtained using 
a simple vectorized XOR
random number generator (XOR RG)
\cite{numerical_recipes}
and the vectorized L\"uscher generator (L\"uscher RG).
In both cases periodic boundary conditions (BCs) are used.

In
Tab~\ref{t_random_condensate}
and
Tab~\ref{t_random_psd}
we compare the
condensate 
and 
pseudo-scalar density 
results to each other.
These results clearly show that 
agreement is almost perfect up to the
critical kappa $\kappa_c \approx {1\over 4}$.
Both
the condensate and pseudoscalar 
density values above $\kappa_c$ are almost meaningless,
as errorbars are of the order of the results themselves.
Still, even those results are consistent with each other.

We 
remark that the pseudoscalar density results
for large $\kappa$ do not agree with zero as was expected.
This effect is due to topological effects
discussed in Ch.~\ref{c_topology}.
Still, pseudoscalar results
are smaller than 
the fermion condensate
by orders of magnitude.

\vspace{0.3cm}
\noindent{\bf Boundary conditions tests}

In a gauge $U(1)$
theory
we expect no 
influence of periodic or anti-periodic
boundary conditions
as $-1$ is in  the centre of the group.
To check this,
we repeat 
in Tab.~\ref{t_random_psd_apbc}
and
\ref{t_random_condensate_apbc}
the periodic boundary condition data 
discussed in the paragraph above
with anti-periodic boundary conditions.

Agreement of results 
with those of the 
periodic boundary conditions runs
is evident in all cases.

\vspace{0.3cm}
\noindent{\bf Pion mass}

We check 
the implementation of the dynamical fermion update
comparing mass results
with dynamical fermions
against an independent calculation.
Using a Hybrid Monte Carlo code,
Irving \cite{pion_mass} 
gives a pseudoscalar vector (pion) 
mass of
$m_\pi=0.369(3)$ 
for a $32 \times 32$ lattice with $\beta=2.29$, $\kappa=0.26$.
We obtain
$m_\pi = 0.377(4)$ from about 2000 measurements 
using acceptance step method I,
thus agreeing with high precision
with the independent value.

\vspace{0.3cm}
\noindent{\bf Symmetry tests}

We test that the
alternative meson operators
in the various channels
give compatible results.

\begin{figure}[tbp]
\caption{\label{f_qu_masses}
{\bf Meson mass spectrum.}
We show quenched meson masses as a function of $\kappa$
calculated 
on a $16\times 32$ lattice at $\beta=6.0$.
Included are pion masses from the operators $\bar\psi \gamma^5 \tau \psi$
and $\bar\psi \gamma^5 \gamma^0 \tau \psi$,
eta masses from $\bar\psi \gamma^5 \psi$ and $\bar\psi \gamma^5
\gamma^0 \psi$ and a0 masses from $\bar\psi \tau \psi$.
Both pion operators give identical results.
}

\vspace{0.3cm}
\centering
\makebox[8.8cm]{
\epsfxsize=8.5cm
\epsffile{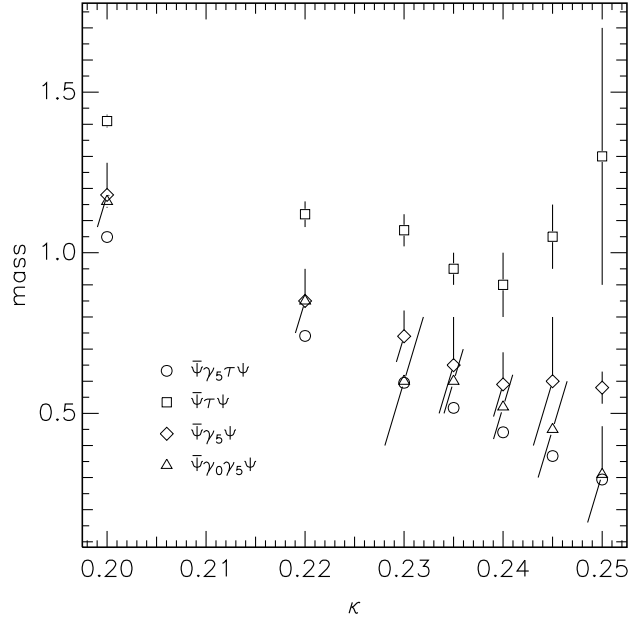}
}
\end{figure}

\ul{Quenched.}
Meson masses in the quenched approximation were 
obtained on $16\times 32$ lattices for
$\beta = 6.0$ and $\kappa=0.20 \dots 0.275$
performing high-statistics runs with about
2000 independent measurements each. 
Fig.~\ref{f_qu_masses} shows the clear signal for the pion mass decreasing as
$\kappa$ is increased, 
its alternative operator gives masses which agree within
errors and is therefore not depicted in the plot. 
For the $\eta$ and its variant, consistency is also verified with
larger errors. The higher states unfortunately 
give noisy results.

\ul{Dynamical.}
Masses were determined on $16\times 32$ lattices for
$\beta = 10.0 $, $\kappa=0.20 \dots 0.25$ and
are shown in Fig.~\ref{f_dynmasses}.
We observe the same symmetry
behaviour as in the quenched case,
with 
the alternative operators for the 
eta and pion masses again agreeing within
errorbars.
Due to the increased complexity
of the computations,
the data at large $\kappa$ and for higher states 
is very noisy compared to the quenched case.

\section{Choice of parameters}
\label{s_parameter_choice}

We want 
to simulate at relevant 
physical parameter choices,
in optimal cases
understanding the 
scaling towards
the
continuum and chiral limits.
To justify the choice
of parameters for the CPU cost studies,
we show that
finite size effects
do not affect the mass results
at the chosen lattice sizes,
$\beta$ and $\kappa$ values.
In a second step,
we demonstrate that 
the simulation parameters
chosen on our larger lattices
correspond to the same physical situation
with a lattice spacing which is smaller by a factor
of about $1.6$.

\vspace{0.3cm}
\noindent{\bf Finite size effects}

In order to check the influence
of the finite
lattice
extent,
we regard the 
meson mass spectrum
on small $8 \times 20$ lattices.
Naively,
one would expect
that
masses can be obtained 
in the
region
$0.5 < m < 2$.

We simulate on
$8\times 20$ lattices with a very conservative
$\beta=3.0$ because of the topological metastability problem
for high $\beta$ values as mentioned in
Sec.~\ref{s_lattice}
and a 
run length of generally $>1000 \tau$.
The topological problems are
discussed in more detail
in Ch.~\ref{c_topology}.

As can be seen in Fig.~\ref{f_8_mass3},
finite size
effects are small up to a $\kappa$ value of about 0.24.
For larger $\kappa$,
a deviation
from the approximately linear behaviour of
the pion mass is detectable.

To justify the linear fits
for the pion mass,
we included in
Fig.~\ref{f_8_mass3}
also a fit assuming the $m^{2\over 3}$ behaviour
suggested by perturbation theory.
It is evident
that this describes the data far worse
than the linear fit.
We are obviously not yet in the regime
where the leading order result is applicable.

This results
in a minimal
pion mass 
$m_\pi(8\times 20)$ = 0.629 possible on this lattices
and 
a mass ratio 
${m_\pi \over m_\eta}= 0.807$.
These parameter choices
are used for the CPU cost tests in Ch.~\ref{c_cost}.

The scaling procedure
described in Sec.~\ref{s_lattice}
relies on the possibility
to 
determine 
for a certain $\beta$
the appropriate pion mass or 
$\kappa$ value
corresponding 
to a fixed ratio
 ${m_\pi \over m_\eta}$.
This is illustrated
in Fig.~\ref{f_8_ratio3}.
The data shows a clear dependency on the pion mass
even taking the errorbars into account.
A linear fit and 
an inversion of the functional dependency is thus
feasible.

\newpage
\noindent{\bf Scaling to larger lattices}
\label{s_scaling}

To achieve scaling towards the continuum limit,
we aim to
to reduce 
the lattice spacing 
going from $\beta=3.0$
and lattices of size $8 \times 20 $
to $\beta=5.0$
and 
appropriate larger lattices
as described in Sec.~\ref{s_lattice}.

We limit ourselves
to these values of $\beta$,
as for higher $\beta$ values
topological metastabilities
contaminate the data
as discussed in Ch.~\ref{c_topology}.
Mass spectra and fits
for $\beta=4,5,6$
on $16\times 40$ lattices
are shown
in Fig.~\ref{f_16_mass4}
to \ref{f_16_mass6}.
This sequence of $\beta$ values is included to show the
deterioration of the data
going to higher $\beta$.

\begin{figure}[tbp]
\caption{\label{f_16_ratio5}
{\bf Mass ratio.}
We show the dynamical 
pi/eta mass ratio 
as a function of the pion mass,
calculated 
on a $16\times 40$ lattice at $\beta=5$.
}

\vspace{0.5cm}
\centering
\makebox[9.8cm]{
\epsfxsize=8.5cm
\epsfig{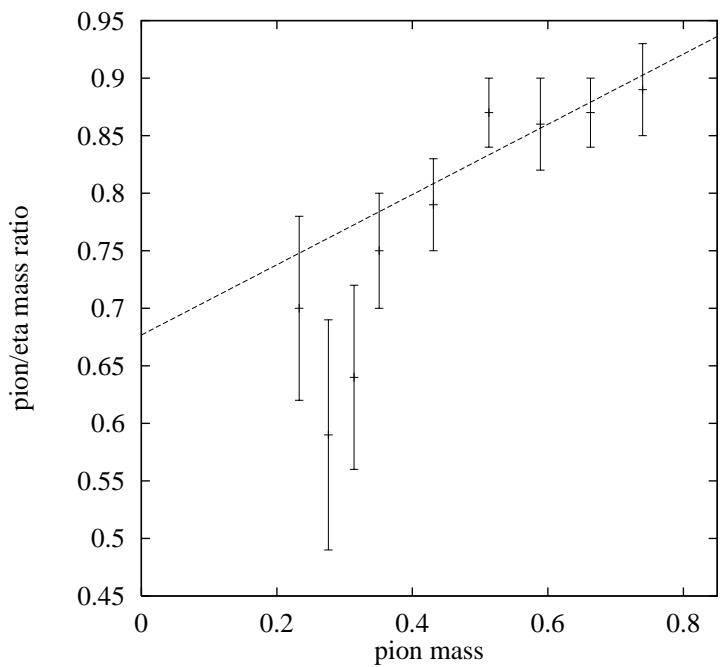}
}
\end{figure}

We have to 
determine the pion mass or respectively
the $\kappa$ value
corresponding to the
same mass ratio ${m_\pi \over m_\eta}$
as described above.
To illustrate this,
we include 
in Fig.~\ref{f_16_ratio5}
a plot of
 ${m_\pi \over m_\eta}$
versus the pion mass.
The lower quality
compared to the $8 \times 20$
results of Fig.~\ref{f_8_ratio3}
is an indication how fast
CPU costs become prohibitively large
for high-precision studies
at larger lattices.

We fix the parameters
for the
optimal CPU cost search runs at
$\beta=5.0$ and
$\kappa=0.245$,
yielding a pion mass
of $m_\pi(16 \times 40)=0.384$
and a ratio of 
{ ${m_\pi \over m_\eta}= 0.826$}.
As this ratio is compatible with the
one obtained on the smaller $8 \times 20$ lattices,
we thus achieve a scaling
by a factor of  ${m_\pi(8 \times 20) \over m_\pi(16 \times 40)} = 1.64$.
Finite size effects are again small
on the lattices we used
as the pion correlation length is about 3.

\clearpage

\begin{figure}[tbp]
\caption{\label{f_cg2} 
{\bf Number of CG iterations.} 
We show as a function of $\kappa$
the number of 
iterations necessary to invert $Q$
to a precision of $10^{-6}$
using a Conjugate Gradient inverter
on $16 \times 16$ lattices for $\beta=6.0$.
The estimated $\kappa_c$ is indicated by a vertical line.
}

\vspace{0.2cm}
\centering
\makebox[8.8cm]{
\epsfxsize=8.5cm
\epsffile{data_kappa_6.0_ps}
}
\vspace{0.2cm}
\begin{center}
$\kappa$
\end{center}
\vspace{-0.3cm}
\end{figure}

\begin{figure}[tbp]
\caption{\label{f_cg3}
{\bf Number of CG iterations.} 
We show as a function of $\kappa$
the number of 
iterations necessary to invert $Q$
to a precision of $10^{-6}$
using a Conjugate Gradient inverter
on $16 \times 16$ lattices for $\beta=10.0$.
The estimated $\kappa_c$ is indicated by a vertical line.
}

\vspace{0.2cm}
\centering
\makebox[8.8cm]{
\epsfxsize=8.5cm
\epsffile{data_kappa_10.0_ps}
}
\vspace{0.2cm}
\begin{center}
$\kappa$
\end{center}
\vspace{-0.3cm}
\end{figure}


\begin{table}[tbp]
\caption
{\label{t_plaquette1}
{\bf Plaquette with exact correction factor.}
We show 
the dynamical average plaquette value
on $4 \times 4$ lattices for $\beta=1.0$.
Columns 4 and 5 
calculated with $\epsilon=0.5$, columns 6 and 7 with $\epsilon=0.01$.
}
\vspace{0.2cm}

\begin{center}
\begin{tabular}{|c|c|cc|cc|cc|cc|cc|}

\hline 
$\kappa$ & hopp & $n=0$ & $\pm$ & $n=2$ & $\pm$ & $n=10$ & $\pm$ & 
$n=2$ & $\pm$ & $n=10$ & $\pm$ \\
\hline\hline  
0.10 & .4470 &.4455 & 11 &.4470&7 &.4410&52 &.4457&10 &.4402&60\\
0.15 & .4493 &.4493 & 48 &.4495&7 &.4494&7  &.4492&14 &.4514&12\\
0.18 & .4523 &.4529 & 13 &.4545&7 &.4547&8  &.4538&12 &.4560&12\\
0.20 & .4554 &.4601 & 23 &.4604&6 &.4607&8  &.4605&17 &.4591&13\\
\hline

\end{tabular}
\end{center}
\end{table}

\begin{table}[tbp]
\caption
{\label{t_plaquette2}
{\bf Plaquette with noisy correction factor.}
We show 
the dynamical average plaquette value
on $4 \times 4$ lattices for $\beta=1.0$.
Columns 4 and 5 
calculated with $\epsilon=0.5$, columns 6 and 7 with $\epsilon=0.01$.
}
\vspace{0.2cm}

\begin{center}
\begin{tabular}{|c|c|cc|cc|cc|cc|cc|}

\hline 
$\kappa$ & hopp & $n=0$ & $\pm$ & $n=2$ & $\pm$ & $n=10$ & $\pm$ & 
$n=2$ & $\pm$ & $n=10$ & $\pm$ \\
\hline\hline
0.10 & .4470 &.4501 &  55 & .4471&8 & .4479&7 & .3575&398& .4474&14\\
0.15 & .4493 &.4729 & 313 & .4502&19& .4497&7 & .4598&159& .4462&18\\
0.18 & .4523 &.5026 & 427 & .4482&43& .4542&9 & .3780&434& .4545&15\\
0.20 & .4554 &.5218 & 402 & .4639&79& .4604&10& .4721&310& .4604&15\\
\hline

\end{tabular}
\end{center}
\end{table}

\begin{table}[tbp]
\caption
{\label{t_plaquette3}
{\bf Plaquette with exact acceptance step.}
We show 
the dynamical average plaquette value
on $4 \times 4$ lattices for $\beta=1.0$.
Columns 4 and 5 
calculated with $\epsilon=0.5$, columns 6 and 7 with $\epsilon=0.01$.
}
\vspace{0.2cm}

\begin{center}
\begin{tabular}{|c|c|cc|cc|cc|cc|cc|}

\hline 
$\kappa$ & hopp & $n=0$ & $\pm$ & $n=2$ & $\pm$ & $n=10$ & $\pm$ & 
$n=2$ & $\pm$ & $n=10$ & $\pm$ \\
\hline\hline
0.10 & .4470 & .4457&51& .4465&7& .4465&7& .4483&13& .4460&11\\
0.15 & .4493 & .4487&56& .4501&6& .4492&7& .4497&16& .4502&19\\
0.18 & .4523 & .4532&52& .4549&8& .4529&8& .4528&16& .4532&12\\
0.20 & .4554 & .4643&42& .4600&8& .4596&8& .4589&17& .4596&9\\
\hline

\end{tabular}
\end{center}
\end{table}

\begin{table}[tbp]
\caption
{\label{t_plaquette4}
{\bf Plaquette with noisy acceptance step method I.}
We show 
the dynamical average plaquette value
on $4 \times 4$ lattices for $\beta=1.0$.
Columns 4 and 5 
calculated with $\epsilon=0.5$, columns 6 and 7 with $\epsilon=0.01$.
}
\vspace{0.2cm}

\begin{center}
\begin{tabular}{|c|c|cc|cc|cc|cc|cc|}

\hline 
$\kappa$ & hopp & $n=0$ & $\pm$ & $n=2$ & $\pm$ & $n=10$ & $\pm$ & 
$n=2$ & $\pm$ & $n=10$ & $\pm$ \\
\hline\hline
0.10 & .4470 & .4472 &4 & .4465&9 & .4466&7 & .4071&345& .4491&14\\
0.15 & .4493 & .4488 &7 & .4496&20& .4515&8 & .4736&117& .4462&18\\
0.18 & .4523 & .4558 &21& .4518&37& .4520&11& .4301&312& .4549&14\\
0.20 & .4554 & .4557 &32& .4654&42& .4612&14& .4954&190& .4619&14\\
\hline

\end{tabular}
\end{center}
\end{table}

\begin{table}[tbp]
\caption
{\label{t_correction}
{\bf Correction factor.}
We show 
the average correction factor $\det [1-R]$ 
calculated exactly from the eigenvalues of $Q^2$ 
and stochastically estimated
from dynamical simulations 
on $4 \times 4$ lattices for $\beta=1.0$ and $\kappa=0.1$
using
generally $\epsilon=0.01$ and different $n$.
}

\vspace{0.5cm}
\begin{center}
\begin{tabular}{|c|l|lc|}
\hline

$n$ & from $Q^2$     &  noisy    & $\pm$ \\
\hline\hline
40 & .99967 &  .99966  & 7 \\
10 & .40885 &  .408    & 12  \\
8  & .73572 &  .69     & 3 \\
6  & .43990 &  .44     & 5 \\
4  & .00164 &  .0033   & 16 \\
2  & .00010 &  .002    & 2 \\
\hline

\end{tabular}
\end{center}
\end{table}

\begin{table}[tbp]
\caption
{\label{t_ising}
{\bf Ising Random Number Test.}
We show 
the quenched average fermion condensate
and pseudoscalar density
on $16 \times 16$ lattices for
$\beta=2.0$ and  $\kappa=0.26$
using different
Ising noises with
$Z_4$ or $Z_2$ symmetry.
}

\vspace{0.5cm}
\begin{center}

\begin{tabular}{|c|c|cc|cc|}
\hline 

&& \rule[-0.3cm]{0cm}{0.8cm} $Z_4$ & $\pm$ & $Z_2$ & $\pm$ \\
\hline\hline

real part &
\rule[-0.3cm]{0cm}{0.8cm}
$\bar\psi\psi $ & 
-.9084 & .0013 &
-.9098 & .0016 \\

\hline

&
\rule[-0.3cm]{0cm}{0.8cm}
$\bar\psi\gamma^5 \psi $ &
 .0018 & .0027 &
-.0035 & .0028 \\
\hline\hline

Imaginary part &
\rule[-0.3cm]{0cm}{0.8cm}
$\bar\psi\psi $ & 
-.0008 & .0019 & 
-.0004 & .0015 \\
\hline

&
\rule[-0.3cm]{0cm}{0.8cm}
$\bar\psi\gamma^5 \psi $ &
 .0000 & .0000 &
 .0000 & .0000 \\
\hline

\end{tabular}
\end{center}
\vspace{-0.5cm}
\end{table}

\begin{table}[tbp]

\caption
{\label{t_u1_cond}
{\bf $U=1$ Condensate.}
We show 
the average fermion condensate
from $U=1$ simulations
on $16 \times 16$ lattices,
compared to analytic results.
}

\vspace{0.5cm}
\begin{center}
\begin{tabular}{|c|c|cc|}
\hline 

$\kappa$ & analytic result & \rule[-0.3cm]{0cm}{0.8cm} $\bar\psi \psi$ & $\pm$ \\
\hline\hline

0.000   &  -1.0000  & -1.0000 &.0000\\
\hline	              	            
0.100   &   -.9982  &  -.9982 &.0007\\
\hline	            	            
0.200   &   -.9554  &  -.9543 &.0018\\
\hline	            	            
0.300   &   -.5253  &  -.5278 &.0020\\
\hline	            	            
0.400   &   -.3491  &  -.3505 &.0016\\
\hline	            	            
0.500   &   -.2587  &  -.2596 &.0014\\
\hline	            	            
0.600   &   -.2029  &  -.2034 &.0012\\  
\hline

\end{tabular}
\end{center}
\vspace{-0.5cm}
\end{table}

\begin{table}[tbp]

\caption
{\label{t_random_condensate}
{\bf Condensate -- periodic BCs.}
We show 
the quenched fermion condensate
on $16 \times 16$ lattices
at $\beta=10.0$
comparing
results from L\"uscher RG to those of XOR RG.
}

\vspace{0.5cm}
\begin{center}
\begin{tabular}{|c|cc|cc|}
\hline 

$\kappa$ & L\"uscher: \rule[-0.3cm]{0cm}{0.8cm} $\bar\psi \psi$ & $\pm$    & XOR: $\bar\psi \psi$ & $\pm$    \\
\hline\hline			 			 
				 			 
0.05     & -.99998  & .00010 & -.99983  & .00010 \\
\hline	                         			 
0.10     & -.99842  & .00019 & -.99816  & .00020 \\
\hline	                         			 
0.15     & -.98999  & .00031 & -.98965  & .00031 \\
\hline	   
0.20     & -.95808  & .00010 & -.95808  & .00010 \\
\hline	                         			 
0.22     & -.92873  & .00012 & -.92873  & .00012 \\
\hline	                         			 
0.24     & -.88164  & .00017 & -.88122  & .00017 \\
\hline	                         			 
0.26     & -.77180  & .10871 & -.59028  & .09895 \\
\hline

\end{tabular}
\end{center}
\end{table}

\begin{table}[tbp]
\caption
{\label{t_random_psd}
{\bf Pseudoscalar density -- periodic BCs.}
We show 
the quenched pseudoscalar density
on $16 \times 16$ lattices
at $\beta=10.0$,
comparing
results from L\"uscher RG to those of XOR RG.
}

\vspace{0.5cm}
\begin{center}
\begin{tabular}{|c|cc|cc|}
\hline 

$\kappa$ & L\"uscher: \rule[-0.3cm]{0cm}{0.8cm}$\bar\psi \gamma^5 \psi$ & $\pm$ & XOR: $\bar\psi \gamma^5 \psi$ & $\pm$    \\
\hline\hline

0.05     & -.00021  & .00014 & .00002  & .00013 \\
\hline	                         			 
0.10     & -.00050  & .00030 & .00015  & .00027 \\
\hline	                         			 
0.15     & -.00090  & .00051 & .00083  & .00045 \\
\hline	   
0.20     &  .00241  & .00015 & .00241  & .00015 \\
\hline	                      	   
0.22     &  .00566  & .00019 & .00566  & .00019 \\
\hline	                      	   
0.24     &  .01935  & .00035 & .01920  & .00031 \\
\hline	                      	   
0.26     &  .02001  & .09420 & .16371  & .09485 \\
\hline

\end{tabular}
\end{center}
\end{table}

\begin{table}[tbp]
\caption
{\label{t_random_condensate_apbc}
{\bf Condensate -- anti-periodic BCs.}
We show 
the quenched fermion condensate
on $16 \times 16$ lattices
at $\beta=10.0$,
comparing
results from L\"uscher RG to those of XOR RG.
}

\vspace{0.5cm}
\begin{center}
\begin{tabular}{|c|cc|cc|}
\hline 

$\kappa$ & L\"uscher: \rule[-0.3cm]{0cm}{0.8cm}$\bar\psi \psi$ & $\pm$    & XOR: $\bar\psi \psi$ & $\pm$   \\
\hline\hline

0.05     & -.99998  & .00010 &  -.99982  & .00010 \\
\hline	                         			 
0.10     & -.99843  & .00020 &  -.99817  & .00020 \\
\hline	                         			 
0.15     & -.98998  & .00032 &  -.98969  & .00031 \\
\hline	   
0.20     & -.95802  & .00006 &  -.95800  & .00005 \\
\hline	                         			 
0.22     & -.92883  & .00007 &  -.92862  & .00012 \\
\hline	                         			 
0.24     & -.88137  & .00012 &  -.88087  & .00018 \\
\hline	                         			 
0.26     & -.55999  & .11331 & -1.44166  & .85625 \\
\hline

\end{tabular}
\end{center}
\end{table}

\begin{table}[tbp]
\caption
{\label{t_random_psd_apbc}
{\bf Pseudoscalar densities -- anti-periodic BCs.}
We show 
the quenched pseudoscalar density
on $16 \times 16$ lattices
at $\beta=10.0$,
comparing
results from L\"uscher RG to those of XOR RG.
}

\vspace{0.5cm}
\begin{center}
\begin{tabular}{|c|cc|cc|}
\hline 

$\kappa$ & L\"uscher: \rule[-0.3cm]{0cm}{0.8cm}$\bar\psi \gamma^5 \psi$ & $\pm$    & XOR: $\bar\psi \gamma^5 \psi$ & $\pm$    \\
\hline\hline

0.05     & -.00012 & .00015 & .00001 & .00014 \\
\hline	                         			 
0.10     & -.00029 & .00032 & .00016 & .00029 \\
\hline	                         			 
0.15     & -.00051 & .00053 & .00096 & .00047 \\
\hline	   
0.20     &  .00220 & .00009 & .00219 & .00008 \\
\hline	                      	  
0.22     &  .00536 & .00014 & .00565 & .00019 \\
\hline	                      	  
0.24     &  .01882 & .00024 & .01922 & .00030 \\
\hline	                      	  
0.26     & -.23836 & .13051 & -1.14357 & 1.29132 \\
\hline

\end{tabular}
\end{center}
\end{table}

\clearpage

\begin{figure}
\caption{\label{f_8_mass3}
{\bf Meson mass spectrum.}
We show dynamical meson masses 
as a function of $\kappa$
calculated
on a $8\times 20$ lattice at $\beta=3$.
}

\vspace{0.2cm}
\centering
\makebox[8.8cm]{
\epsfig{file=notes_060797_search.data8_1_ps, width=14cm}
}
\vspace{-1.5cm}
\end{figure}

\begin{figure}
\caption{\label{f_8_ratio3}
{\bf Mass ratio.}
We show the dynamical 
pi/eta mass ratio 
as a function of the pion mass,
calculated
on a $8\times 20$ lattice at $\beta=3$.
}

\vspace{0.5cm}
\centering
\makebox[8.8cm]{
\epsfig{file=notes_060797_search.data8_2_ps, width=13cm}
}
\vspace{-1.5cm}
\end{figure}

\begin{figure}[tbp]
\caption{\label{f_16_mass4}
{\bf Meson mass spectrum.}
We show dynamical meson masses 
as a function of $\kappa$,
calculated
on a $16\times 40$ lattice at $\beta=4$.
}

\vspace{0.3cm}
\centering
\makebox[9.8cm]{
\epsfxsize=8.5cm
\epsfig{file=notes_060797_search.data16_2_4_ps, width=13cm}
}
\end{figure}

\begin{figure}[tbp]
\caption{\label{f_16_mass5}
{\bf Meson mass spectrum.}
We show dynamical meson masses 
as a function of $\kappa$,
calculated
on a $16\times 40$ lattice at $\beta=5$.}

\vspace{0.3cm}
\centering
\makebox[8.8cm]{
\epsfxsize=8.5cm
\epsfig{file=notes_060797_search.data16_2_5_ps, width=14cm}
}
\end{figure}

\begin{figure}[tbp]
\caption{\label{f_16_mass6}
{\bf Meson mass spectrum.}
We show dynamical meson masses 
as a function of $\kappa$,
calculated on a $16\times 40$ lattice at $\beta=6$.
}

\vspace{0.3cm}
\centering
\makebox[8.8cm]{
\epsfxsize=8.5cm
\epsfig{file=notes_060797_search.data16_2_6_ps, width=14cm}
}
\end{figure}

\chapter{The problems II: Topology}
\label{c_topology}

Constructing decorrelated configurations
in lattice simulations
can be difficult if large energy barriers 
exist 
between
regions of the configuration space
to be sampled.
Unfortunately,
one such possibility
are topological sectors
for the
$U(1)$ model \cite{topology_gattringer_u1},
part of the 
Schwinger model \cite{topology_joos,topology_gattringer_sm,topology_dilger}
we are considering in this study.

While local updates do not work well,
we can use in two dimensions
a global heatbath update 
to obtain reasonably high tunnelling rates.
This trick obviously works also in
the quenched case,
while for full dynamical simulations 
global heatbath updates 
are not known.

A possible way of
dealing with such systems
which has been suggested \cite{aachen}
is to
stay in one fixed topological
charge sector and define
quantities in that way.
In order to study this for relevant observables,
we measure
pion correlators
in fixed (low)
topological sectors.
In the quenched case
we compare local and global update schemes.
For dynamical fermions
we compare low and high $\beta$ results,
using both the
local bosonic algorithm (LBA) \cite{lba}
and Hybrid Monte Carlo (HMC) \cite{hmc}.

\section{Algorithm}
\label{s_algorithm}

\ul{Quenched case: local update.}
We use 
a local link update 
consisting of
one exact heatbath \cite{u1_heatbath}
and three over-relaxation
steps per trajectory.

\ul{Quenched case: global update.}
Alternatively,
it is possible
to use
a global heatbath for the plaquettes.
As the new configuration
is constructed without recursion
to the old one,
there are no problems with metastabilities.

The global update
uses the fact that 
nearly all plaquettes are independent
even on a finite lattice.
The plaquettes only have to satisfy
the constraint
\begin{eqnarray}
\label{e_condition}
\prod_P U_P  = 1
\; .
\end{eqnarray}
We are therefore able to
update $LT-1$ plaquettes with
a heatbath algorithm.
The last plaquette is then determined by the 
condition Eq.~\ref{e_condition}
and the configuration has to be accepted
or rejected according to a Metropolis decision
to ensure 
the correct distribution.
Finally, this plaquette configuration
has to be translated into a valid link
configuration.
To achieve this, 
we utilise the freedom
to choose a gauge.
We use a maximal tree prescription,
setting $LT-2$ links to 1.
Then $LT$ links can be recursively 
determined from the plaquettes.
The unconstrained two 
remaining links
correspond to the free
Polyakov loops $P_T$ and $P_L$ 
in 2 dimensions.

We want to state
that the problem
of slow topological charge fluctuations
could also be solved
by explicit topological updates
\cite{topology_dilger}.
Unfortunately,
this trick is not applicable
in the presence of dynamical fermions.
Improvements claimed to increase
tunnelling for staggered fermions \cite{topology_dilger}
did not work for the Wilson fermions used in this study.
Whether a
recently suggested reweighting method
to enhance topological updates \cite{reweighting_topology}
really reduces the problem is still under discussion.

\ul{Full dynamical simulations.}
In this case
we use 
the Hermitean version of
L\"u\-scher's local 
bosonic algorithm
with
noisy acceptance
step method I
as described in Ch.~\ref{c_lba}.
To have an independent check,
we compare to a Hybrid Monte Carlo algorithm
implemented as described in Ch.~\ref{c_hmc}.

\section{Topological sectors}
\label{topsec}

The integer-valued topological charge functional 
\begin{eqnarray}
{e\over 4 \pi} \int d^2x \epsilon_{\mu\nu} F_{\mu\nu}
\end{eqnarray}
can be represented
on the lattice
by
\begin{eqnarray}
Q 
= 
{1 \over 2\pi} \sum_{P} \phi_P
\end{eqnarray}
with plaquette angle 
$\phi_P={\rm Im} \ln (U_P) \in (-\pi,\pi)$ \cite{top_charge}.
We denote by
tunnel events
all updates resulting in a 
change of the topological charge
and as tunnel probability
the number of tunnel events
divided by the total number
of updates.
The
topological susceptibility
is defined via
\begin{eqnarray}
\chi_{\rm top}
=
{1\over N_P}
[<Q^2>-<Q>^2]
\; .
\end{eqnarray}

\begin{figure}[tbp]
\caption{\label{tunnel}
{\bf Tunnel probability.}
Topological tunnel probability
of pure $U(1)$ as a function of $\beta$
using local and global updates on a $16 \times 16$ lattice.
}

\centering
\makebox[8.8cm]{
\epsfxsize=8.5cm
\epsffile{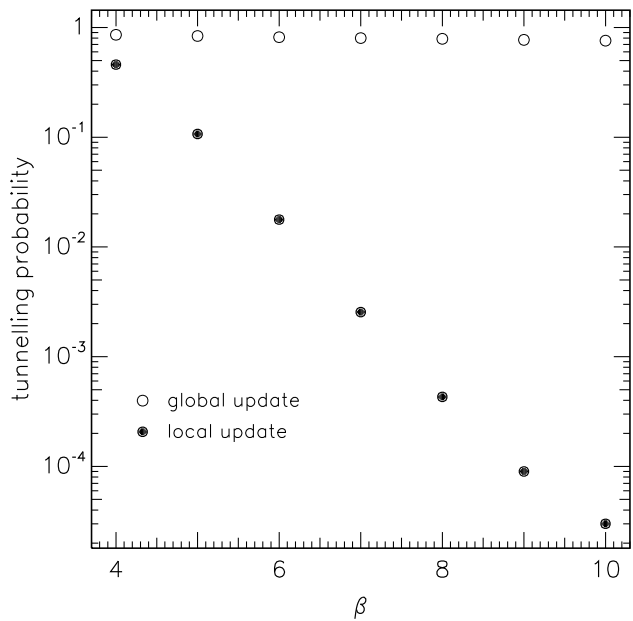}
}
\end{figure}

We demonstrate the relevance
of topological sectors on a $16\times32$ lattice 
showing in Fig.~\ref{tunnel} the
tunnel rate plotted
against $\beta$
for local and global updates.
The exponential decrease of the tunnel probability
for this local update algorithm
gives rise to metastabilities
in simulations at large $\beta$.
We therefore consider it worthwhile
to investigate observables within fixed topological sectors.

\section{Simulations}
\label{sim}

\ul{Simulation parameters.}
We simulate on
$8\times20$ and $16\times40$
lattices
at a beta value of $\beta=12$
generating 10000 configurations.
We only show data for the larger $16\times40$ lattices.
For the fermion part,
we choose
$\kappa= 0.24$,
 where 
the pion correlation length (in the quenched case)
is found to be around 3
and finite size effects can be expected to be
small.

\ul{Local updates.}
We monitor the topological charge
during the simulations.
Due to metastability,
no tunnelling of $Q$ is observed
in the runs
with local updates.
At $\beta = 12$ 
we therefore are able 
to perform a simulation in a given topological sector
by using an initial configuration with this particular charge.
This is done
generating a classical homogeneous
plaquette configuration
of the desired charge
and
converting this to the links
as described in Sec.~\ref{s_algorithm}.
The two Polyakov loops $P_T, P_L$
are chosen according to a flat
random distribution.

\ul{Global updates.}
In the limit of independent plaquettes
the topological susceptibility
can be calculated analytically
as shown in App.~\ref{s_susceptibility}.
The result in the large $\beta$ approximation
for $\beta=12$ is
\begin{eqnarray}
\chi_{\rm top} |_{\beta=12}
\approx
{1\over 4 \pi^2 \beta} |_{\beta=12}
=
0.0022
\; .
\end{eqnarray}
To check the global update,
we simulate
on a $16 \times 32$ lattice
obtaining
$
\chi_{\rm top}
=
0.0021(1)
$.
This shows that the global update works well
without metastabilities in the topological charge.

\ul{Observables.}
We generally 
measure the pion correlator 
using the prescription
detailed in App.~\ref{s_mesons}.

\vspace{0.3cm}
\noindent{\bf Quenched results}

The results of the quenched runs
are shown in Fig.~\ref{quenched12}.
Obviously,
for $Q=0,4$ we are not able to find
a plateau in the effective mass plots.
We find a valley-like structure
in the mass
for the low $Q$ cases.
For high $Q$
this turns into 
a hill-like structure with the peak situated at
half of the temporal lattice extent.
To show 
that in the intermediate region
the valley and hill structures can approximately cancel
and suggest a fake plateau,
we
include the $Q=1$ plot.

In the quenched case,
we are able to compare to the
results using a global update scheme.
For the global update we find a tunnel probability
of $P=0.76$ and therefore do not
expect any influence of topology.
The effective mass
is shown in the lower right plot of Fig.~\ref{quenched12}.
A plateau is clearly 
more reasonable than in the fixed $Q$ cases.

\vspace{0.3cm}
\noindent{\bf Dynamical fermion results}

Effective masses from
full dynamical simulations
are 
shown in Fig.~\ref{dyn12}.
We do not expect quantitatively the same results
as in the quenched case,
yet the behaviour is
qualitatively similar.

To establish
that dynamical results
are not influenced by
the chosen parameters of the local bosonic algorithm,
we repeat the calculation
with topological charge $Q=4$
using a standard HMC algorithm.
This is also shown in Fig.~\ref{dyn12}.
The results agree nicely within errorbars.

From these results,
we conclude
that we need 
to average over the topological sectors
to obtain a plateau in the effective mass.

\section{Projections to topological sectors}
\label{project}
\vspace{-0.05cm}

To gain further insight,
we now use a slightly different approach.
In principle, we
could also restrict ourselves
to definite topological sectors
by selecting measurements
with fixed topological charge 
from a 
simulation,
i.e. effectively simulating the path integral
given by
\begin{eqnarray}
Z[q] = \int D[U] D[\bar\psi] D[\psi] \delta_{Q,q} e^{-S}
\; .
\end{eqnarray}
To this end
we need simulations with a reasonably high fluctuation.
Such simulations can be done e.g. 
with dynamical simulations at low $\beta$
or quenched simulations using global updates.

\begin{figure}[tbp]
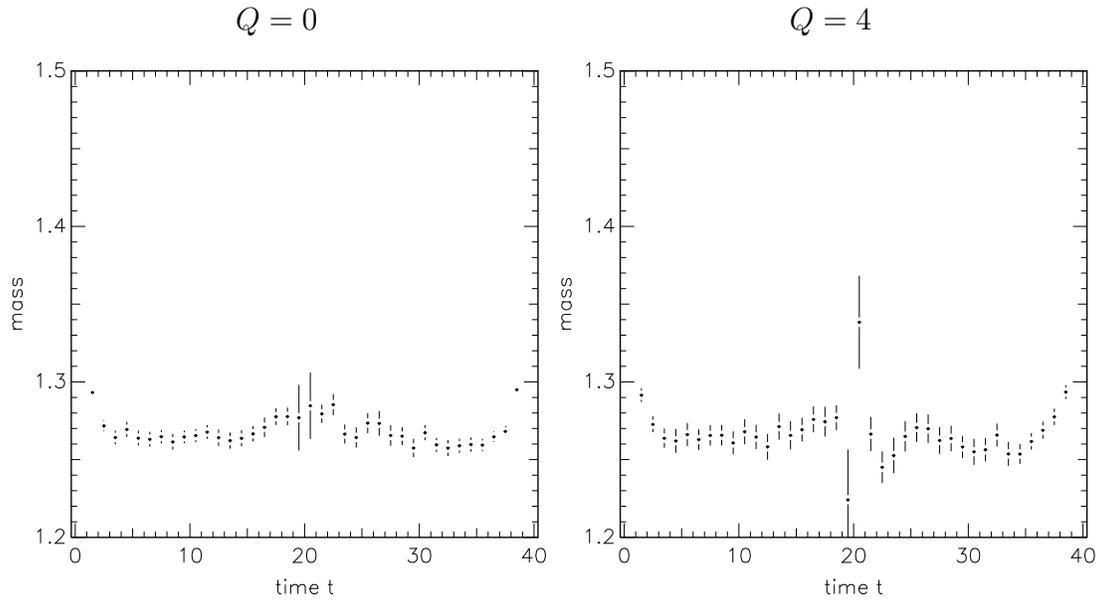

\caption{\label{dyn1-fixedQ}
{\bf Effective pion mass.}
Dynamical projected effective pion mass
as a function of time for
$\beta=1.0$, $\kappa=0.22$.
}

\unitlength1cm

\vspace{0.5cm}
\hspace{3cm}
$Q=0$
\hspace{6cm}
$Q=4$
\vspace{-0.5cm}

\begin{picture}(20,8)
\epsfxsize=7.3cm
\epsffile{p-dyn1-fixedQ_1_ps}
\epsfxsize=7.3cm
\epsffile{p-dyn1-fixedQ_2_ps}
\end{picture}

\end{figure}

\ul{Full dynamical case.}
We work at low $\beta=1$
with a slightly smaller $\kappa=0.22$.
Effective masses are depicted in Fig.~\ref{dyn1-fixedQ}.
We can detect no discrepancy between
masses calculated in different topological sectors.
This result
was also reported by a group working with staggered fermions,
which concluded that topological sectors 
are of no importance to mass estimates in the Schwinger model
\cite{aachen}.

\ul{Quenched case.}
Here we exploit the opportunity
to use the same parameters $\beta=12,\kappa=0.24$
as in Sec.~\ref{sim}.
The results are shown in  Fig.~\ref{q12-fixedQ}.
At this $\beta$, we do not find 
agreement.
Rather the effective masses
are
nearly the same 
as in the simulations without tunnelling
presented in Fig.~\ref{dyn12}.
We would like to point out
that they do not agree within errorbars.
On the other hand, we remark that the statistical sample
was very much smaller for the projected data
due to the fact that only a part of the generated
configurations
is projected into
the appropriate sectors.

The striking difference between low and high $\beta$ results
makes a sound understanding of this
behaviour
highly desirable.
As can be seen from the
quenched results here,
it is evidently not just the
averaging over the topological sectors (as found in Sec.~\ref{sim})
which is lacking in high $\beta$ simulations.
There seems to be some more subtle dynamical 
effect involved.

\vspace{-0.15cm}
\section{External gauge configurations}
\label{external}

In order to investigate 
how much averaging is necessary,
we plot in Fig.~\ref{config1} effective 
masses for external
configurations with 
fixed topological charge
$Q=0$ and $Q=4$.
These are generated
from homogeneous
plaquette configurations
in the way described in Sec.~\ref{s_algorithm}.
For the fermions we use
$\kappa=0.24$. 

We stress that 
we use random Polyakov loops $P_T, P_L$
in both cases,
so that one should not expect free fermion behaviour in the $Q=0$ case.
For $Q=4$ a completely irregular behaviour is observed. 
It is thus not possible to measure
a meaningful pion correlator from one (even very smooth)
configuration alone. The translation invariance is
manifestly broken
even for that smooth configuration.

\begin{figure}[tbp]
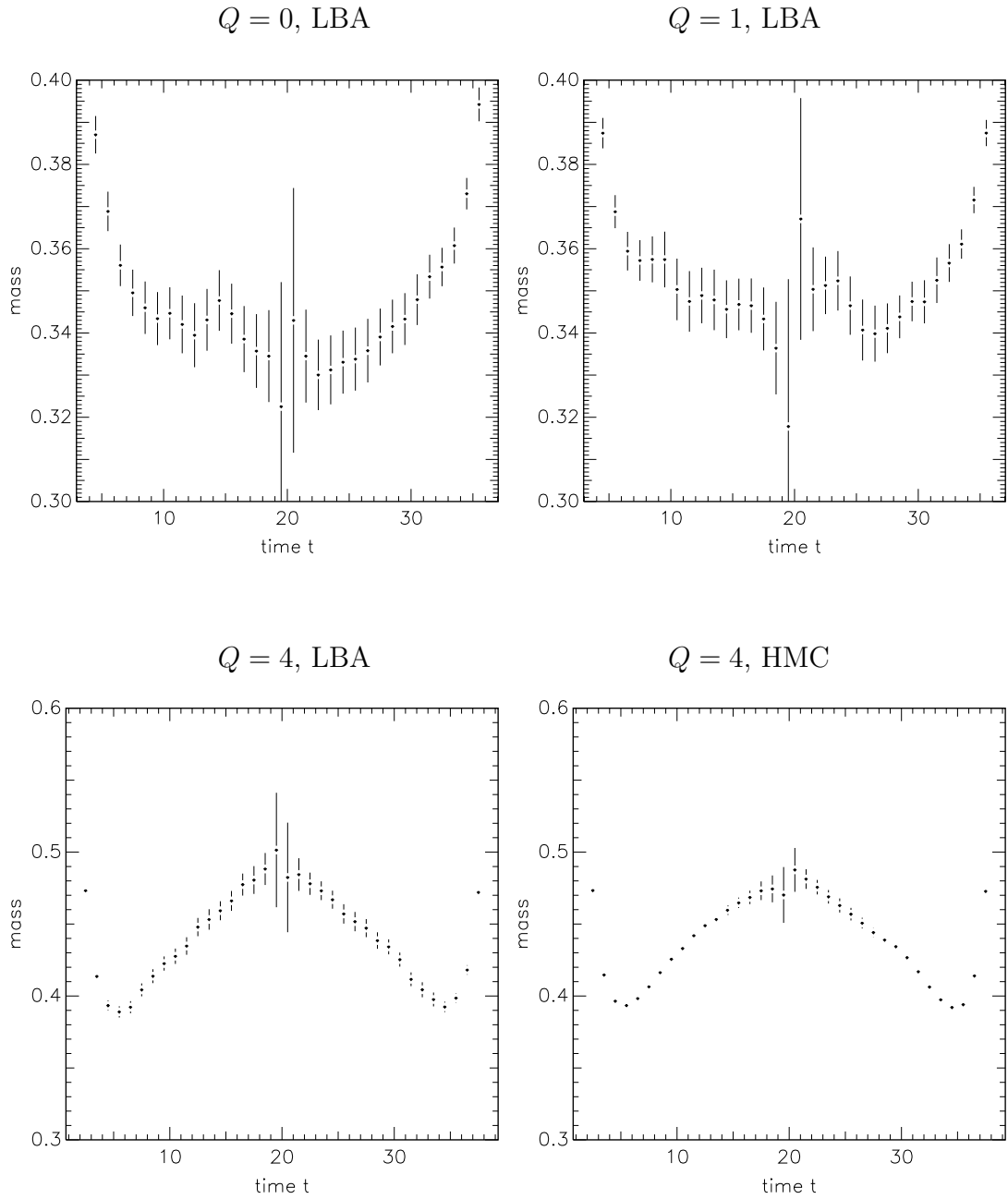

\caption{\label{dyn12}
{\bf Effective pion mass.}
Dynamical effective pion mass as a function of time for 
$\beta=12.0$, $\kappa=0.24$.
}

\unitlength1cm

\vspace{0.5cm}
\hspace{3cm}
$Q=0$, LBA
\hspace{4cm}
$Q=1$, LBA
\vspace{-0.45cm}

\begin{picture}(20,8)
\epsfxsize=7.3cm
\epsffile{p6_ps}
\epsfxsize=7.3cm
\epsffile{p7_ps}
\end{picture}

\vspace{1cm}
\hspace{3cm}
$Q=4$, LBA
\hspace{4cm}
$Q=4$, HMC
\vspace{-0.45cm}

\begin{picture}(20,8)
\epsfxsize=7.3cm
\epsffile{p8_ps}
\epsfxsize=7.3cm
\epsffile{p10_ps}
\end{picture}

\end{figure}

\begin{figure}[tbp]
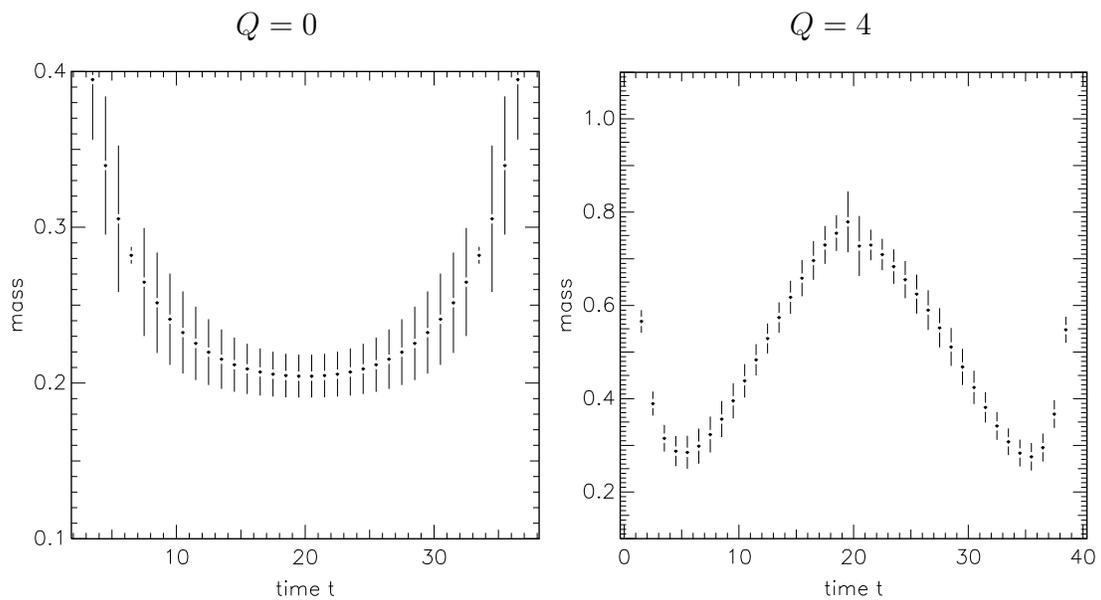

\caption{\label{config10}
{\bf Effective pion mass.} 
Shown
as a function of time 
from 10 external configurations at $\kappa=0.24$.
}

\unitlength1cm

\vspace{0.45cm}
\hspace{3cm}
$Q=0$
\hspace{6cm}
$Q=4$
\vspace{-0.55cm}

\begin{picture}(20,8)
\epsfxsize=7.3cm
\epsffile{p-config10_1_ps}
\epsfxsize=7.3cm
\epsffile{p-config10_2_ps}
\end{picture}

\end{figure}

The result of 
averaging over 10 values of  $P_T$ and $P_L$
is shown in Fig.~\ref{config10}.
We clearly regain the qualitatively
expected regular valley and hill structure 
observed in Fig.~\ref{dyn12}
in both the
$Q=0$ and $Q=4$ case.


\begin{figure}[tbp]
\caption{\label{quenched12}
{\bf Effective pion mass.}
Quenched effective pion mass
as a function of time for $\beta=12.0$, $\kappa=0.24$.
}

\unitlength1cm

\vspace{0.5cm}
\hspace{3cm}
$Q=0$, local
\hspace{4cm}
$Q=1$, local
\vspace{-0.5cm}

\begin{picture}(20,8)
\epsfxsize=7.3cm
\epsffile{p1_ps}
\epsfxsize=7.3cm
\epsffile{p2_ps}
\end{picture}

\vspace{1cm}
\hspace{3cm}
$Q=4$, local
\hspace{5cm}
global
\vspace{-0.5cm}

\begin{picture}(20,8)
\epsfxsize=7.3cm
\epsffile{p3_ps}
\epsfxsize=7.3cm
\epsffile{p5_ps}
\end{picture}

\end{figure}

\begin{figure}[tbp]
\caption{\label{q12-fixedQ}
{\bf Effective pion mass.}
Quenched projected effective pion mass
as a function of time for 
 $\beta=12.0$, $\kappa=0.24$.
}

\unitlength1cm

\vspace{0.45cm}
\hspace{3cm}
$Q=0$
\hspace{6cm}
$Q=4$
\vspace{-0.5cm}

\begin{picture}(20,8)
\epsfxsize=7.3cm
\epsffile{p-q12-fixedQ_1_ps}
\epsfxsize=7.3cm
\epsffile{p-q12-fixedQ_2_ps}
\end{picture}

\end{figure}

\begin{figure}[tbp]
\caption{\label{config1}
{\bf Effective pion mass.} 
Shown as a function of time 
from one external configuration at $\kappa=0.24$.
 }

\unitlength1cm

\vspace{0.45cm}
\hspace{3cm}
$Q=0$
\hspace{6cm}
$Q=4$
\vspace{-0.55cm}

\begin{picture}(20,8)
\epsfxsize=7.3cm
\epsffile{p-config1_1_ps}
\epsfxsize=7.3cm
\epsffile{p-config1_2_ps}
\end{picture}

\end{figure}

\chapter{The aim: CPU cost optimisation}
\label{c_cost}

We remind the reader of the 
schemes included in our investigation.
Besides
\\
\hspace*{0.5cm}$\bullet$ Hybrid Monte Carlo (HMC) to set the scale,
\\
we compare
\\
\hspace*{0.5cm}$\bullet$  LBA with { reweighting},
\\
\hspace*{0.5cm}$\bullet$  LBA with { acceptance step method I} ,
\\
\hspace*{0.5cm}$\bullet$  LBA with { acceptance step method I with adapted precision} and
\\
\hspace*{0.5cm}$\bullet$  LBA with { acceptance step method II}
\\
as described in Ch.~\ref{c_lba},
investigating the CPU cost behaviour
changing $n,\epsilon$
and the number of reflections for the simulation runs.

We would  like to point out again
that
reweighting and acceptance step algorithms
result in
two different ensembles,
so that
one has to use the effective CPU cost
$
C_{\rm eff}
=
N_{\rm total \; Q \; ops} \cdot {\sigma_{\rm tot}^2 (A) \over <A>^2}
$
defined in Sec.~\ref{s_correction}.

\section{$8\times 20$ lattices}
\label{s_cost8}

Our search for optimal
CPU cost behaviour
is conducted
on $8 \times 20$ lattices with
parameters
$\beta=3.0$ and
$\kappa=0.24$
measuring costs
for the average plaquette
and the pion correlator at distance $\Delta t =3$.

We generally use 
anti-periodic boundary conditions
and 
5 hits for the link heatbath algorithm
to provide high acceptance rates.
To ensure numerical stability,
we apply the 
bitreversal scheme for ordering of roots
as described in Ch.~\ref{s_bit}
whenever a Chebyshev polynomial
in the factorized form was necessary.
The rescaling factor
$c_M$ (viz. App.~\ref{s_fermion_matrix}, \cite{lba})
is set to a conservative
$c_M = 1.02$,
so that there is no slowing down from the large eigenvalues of $Q^2$.
The precision of the 
inverters demanded
is generally
$10^{-6}$,
with
$10^{-2}$ for the 
reduced precision
of the adapted precision scheme.
We accumulated statistics
of 10000 calculation sweeps,
applying 1000 sweeps thermalization
with integrated autocorrelations generally
below 50.

\begin{figure}
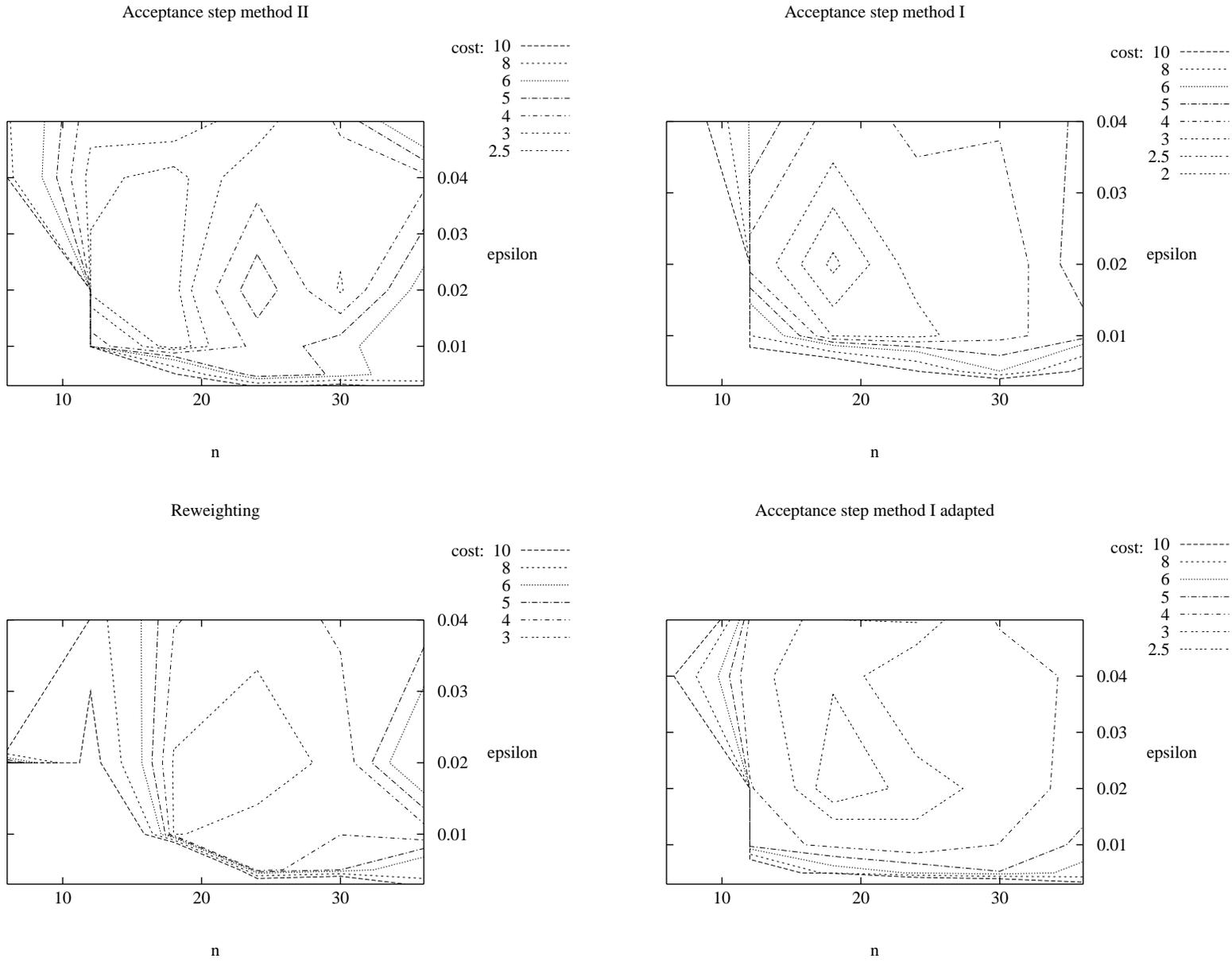

\caption
{\label{f_search8_1}
{\bf Plaquette CPU cost.}
Contour plots of
the plaquette CPU cost as a function of
$n$ and $\epsilon$
calculated
 on $8\times 20$ lattices
at $\beta=3.0, \kappa=0.24$.
}

\unitlength1cm
\begin{picture}(20,14)
\hspace{-1.2cm}
\epsfxsize=9.2cm
\epsffile{notes_060697_8x20_CPU.data120_ps}
\hspace{-1cm}
\epsfxsize=9.2cm
\epsffile{notes_060697_8x20_CPU.data130_ps}
\end{picture}

\begin{picture}(20,11)
\hspace{-1.2cm}
\epsfxsize=9.2cm
\epsffile{notes_060697_8x20_CPU.data140_ps}
\epsfxsize=9.2cm
\hspace{-1cm}
\epsffile{notes_060697_8x20_CPU.data220_ps}
\end{picture}

\end{figure}

To illustrate our 
search for optimal parameters,
we depict in Fig.~\ref{f_search8_1}
and
Fig.~\ref{f_search8_2}
the 
CPU cost
in the 
$n-\epsilon$ plane,
where
the
number of reflections is optimised
for each $n,\epsilon$ pair.
The figure clearly shows that we obtain 
flat
optima.

In Tab.~\ref{t_cost8_hmc}
we show the CPU cost 
for the Hybrid Monte Carlo algorithm
run with
different acceptance probability. 
This illustrates that the cost is not very dependent
on the acceptance.
To set the scale,
we take the optimal values
for an acceptance of $0.84$.

We in Tab.~\ref{t_cost8}
compare the CPU cost 
of only the optimal
parameter sets 
of the various LBA variants 
with the HMC scale.

We find that the 
LBA is doing 
better than the HMC
by a factor of about 3 for the plaquette and
about 2 for the pion correlator at distance $\Delta t =3$.
Thus the gain in the plaquette is typically better than
that in the pion correlator. 
The reweighting method is a bit worse than
the optimised acceptance step methods,
losing for both observables by 
about a factor of 1.3.
The
gain using the adapted precision
trick
is quite considerable
for the pion correlator (1.6),
while it is even a slight detoriation for the plaquette
as the overhead is too large.
We remark that the configurations found for
the optimised and unoptimised 
acceptance step method I scheme runs are not identical
as one might expect in the case of no numerical errors.
Overall,
the number of reflections is 
important to the optimisation.
A fixed number of 1 would not
have reproduced the 
real optima.
This leaves room for improvement
in simulations
where this has not yet been utilised.

\section{$16\times 40$ lattices}
\label{s_cost16}

For the larger lattices we choose
the parameter values 
$\beta=5.0$ and $\kappa=0.245$
on $16 \times 40$ lattices
as
detailed in Sec.~\ref{s_scaling}.
Technical simulation
parameters
are the same as described in Sec.~\ref{s_cost8}.

We again illustrate our 
search for optimal parameters
in Fig.~\ref{f_search16_1}
and
Fig.~\ref{f_search16_2},
showing
the 
CPU cost
in the 
$n-\epsilon$ plane,
where
the
number of reflections is optimised
for each $n,\epsilon$ pair.
The figure shows that 
although optima are fairly flat,
we 
do not find
as 
flat
behaviour
as for the smaller lattices.
We admit, though,
that somewhat larger $n$ should have been included in the study.

In Tab.~\ref{t_cost16_hmc}
we give the CPU costs 
for the Hybrid Monte Carlo algorithm
run with
different acceptance probability. 
As we did not study this as
exhaustively as in the smaller lattices case,
we take the optimal values
from all HMC runs
to set the scale.

We in table~\ref{t_cost16}
again
compare CPU cost
of the 
optimal
parameter sets of the various LBA variants
with the HMC scale.

We find 
that the 
LBA is again doing 
better than the HMC
by a factor of about 3 for the plaquette and
about 2.5
 for the pion correlator at distance $\Delta t =3$.
Thus the gain in the plaquette is again
better than
that in the pion correlator,
but not by as large a margin as for the smaller lattices. 
The reweighting method is
again less efficient than
the optimised acceptance step methods,
this time even more so than
for the smaller lattices,
losing for both observables by 
about a factor of 1.8.
The
gain using the adapted precision
trick
is negligible
for both observables.
Overall,
the number of reflections is
not as  
important to the optimisation
as for the small lattices.

\clearpage



\clearpage

\begin{figure}
\caption
{\label{f_search8_2}
{\bf Pion CPU cost.}
Contour plots of
the pion correlator CPU cost as a function of
$n$ and $\epsilon$
calculated
 on $8\times 20$ lattices
at $\beta=3.0, \kappa=0.24$.
}

\unitlength1cm
\begin{picture}(20,14)
\hspace{-1.2cm}
\epsfxsize=9.2cm
\epsffile{notes_060697_8x20_CPU.data120_2_ps}
\hspace{-1cm}
\epsfxsize=9.2cm
\epsffile{notes_060697_8x20_CPU.data130_2_ps}
\end{picture}

\begin{picture}(20,11)
\hspace{-1.2cm}
\epsfxsize=9.2cm
\epsffile{notes_060697_8x20_CPU.data140_2_ps}
\hspace{-1cm}
\epsfxsize=9.2cm
\epsffile{notes_060697_8x20_CPU.data220_2_ps}
\end{picture}

\end{figure}

\clearpage 


\begin{table}[tbp]
\caption
{\label{t_cost8_hmc}
{\bf HMC cost.}
Simulations at $\beta=3.0, \kappa=0.24$
on $8\times 20$ lattices.
We vary the number of trajectory steps
$n_{tr}$,
while holding the trajectory length $n_{tr} \cdot \Delta \tau$
constant.
}

\vspace{0.3cm}
\begin{center}
\begin{tabular}{|r|r|r|r|r|r|}
\hline 

$n_{tr}$ & $\Delta \tau$ & Accept & 
Plaq. & Cond. & Pion  
\\
\hline\hline
4 &0.209   &  .3910&    7.49&      14.68&   846.88
\\ \hline 		      		  	
5 &0.169   &  .6015&    6.69&       7.61&   759.65
\\ \hline 		      		  	
7 &0.137   &  .7170&    6.33&       8.85&   577.20
\\ \hline 		      		  	
9 &0.103   &  .8441&    6.10&       7.77&   573.66
\\ \hline 		      		  	
12&0.082   &  .8996&    6.61&      11.81&   813.38
\\ \hline

\end{tabular}
\end{center}
\end{table}

\begin{table}
\caption{\label{t_cost8}
{\bf LBA CPU cost minina.}
Simulations at $\beta=3.0, \kappa=0.24$
on $8\times 20$ lattices.
}

\vspace{0.5cm}
\begin{center}
\begin{tabular}{|c|c|c|c|c|c|c|c|c|c|}
\hline\hline
& \multicolumn{4}{c|}{Plaquette} && \multicolumn{4}{c|}{Pion correlator} \\
\hline
algorithm & $n$ & $\epsilon$ & refl. & cost && $n$ & $\epsilon$ & refl. & cost 
\\ \hline
HMC               &    &      &   & 6.1  &&    &      &   & 574           
\\ \hline\hline			                         
Accept I          & 18 & 0.02 & 2 & 1.9  && 24 & 0.01 & 2 & 501
\\ \hline         		                         
Accept I adapted  & 18 & 0.02 & 3 & 2.1  && 18 & 0.04 & 1 & 320
\\ \hline         
Accept II         & 18 & 0.01 & 2 & 2.0  && 12 & 0.04 & 1 & 304
\\ \hline         		                         
Reweighting       & 24 & 0.02 & 4 & 2.6  && 18 & 0.02 & 4 & 414
\\

\hline\hline
\end{tabular}
\end{center}

\end{table}



\clearpage 

\begin{figure}
\caption
{\label{f_search16_1}
{\bf Plaquette CPU cost.}
Contour plots of
the plaquette CPU cost as a function of
$n$ and $\epsilon$
calculated
 on $16\times 40$ lattices
at $\beta=5.0, \kappa=0.245$.
}

\unitlength1cm
\begin{picture}(20,14)
\hspace{-1.2cm}
\epsfxsize=9.2cm
\epsffile{notes_060697_16x40_CPU.data120_ps}
\hspace{-1cm}
\epsfxsize=9.2cm
\epsffile{notes_060697_16x40_CPU.data130_ps}
\end{picture}

\begin{picture}(20,11)
\hspace{-1.2cm}
\epsfxsize=9.2cm
\epsffile{notes_060697_16x40_CPU.data140_ps}
\hspace{-1cm}
\epsfxsize=9.2cm
\epsffile{notes_060697_16x40_CPU.data220_ps}
\end{picture}

\end{figure}

\clearpage

\begin{figure}
\caption
{\label{f_search16_2}
{\bf Pion CPU cost.}
Contour plots of
the pion correlator CPU cost as a function of
$n$ and $\epsilon$
calculated
 on $16\times 40$ lattices
at $\beta=5.0, \kappa=0.245$.
}

\unitlength1cm
\begin{picture}(20,14)
\hspace{-1.2cm}
\epsfxsize=9.2cm
\epsffile{notes_060697_16x40_CPU.data120_2_ps}
\hspace{-1cm}
\epsfxsize=9.2cm
\epsffile{notes_060697_16x40_CPU.data130_2_ps}
\end{picture}

\begin{picture}(20,11)
\hspace{-1.2cm}
\epsfxsize=9.2cm
\epsffile{notes_060697_16x40_CPU.data140_2_ps}
\hspace{-1.2cm}
\epsfxsize=9.2cm
\epsffile{notes_060697_16x40_CPU.data220_2_ps}
\end{picture}

\end{figure}

\clearpage 


\begin{table}[tbp]
\caption
{\label{t_cost16_hmc}
{\bf HMC cost.}
Simulations 
at $\beta=5.0, \kappa=0.245$ on $16\times 40$ lattices.
We vary the number of trajectory steps
$n_{tr}$,
while holding the trajectory length $n_{tr} \cdot \Delta \tau$
constant.
}

\vspace{0.3cm}
\begin{center}
\begin{tabular}{|r|r|r|r|r|r|}
\hline 

$n_{tr}$ & $\Delta \tau$ & Accept & 
Plaq. & Cond. & Pion  
\\
\hline\hline
11 & 0.0847 & .6232 &  1.15&   14.08&    856.97 \\
11 & 0.0847 & .6172 &  1.57&   11.73&    671.16 \\
17 & 0.0556 & .8410 &   .93&   14.28&   1017.99
\\ \hline

\end{tabular}
\end{center}
\end{table}

\begin{table}
\caption
{\label{t_cost16}
{\bf LBA CPU cost minina.}
Simulations 
at $\beta=5.0, \kappa=0.245$ on $16\times 40$ lattices.
}

\vspace{0.5cm}
\begin{center}

\begin{tabular}{|c|c|c|c|c|c|c|c|c|c|}
\hline\hline
& \multicolumn{4}{c|}{Plaquette} && \multicolumn{4}{c|}{Pion correlator} \\
\hline
algorithm & $n$ & $\epsilon$ & refl. & cost && $n$ & $\epsilon$ & refl. & cost 
\\ \hline
HMC                &    &       &   & .93 &&    &       &   & 671
\\ \hline\hline			                      
Accept I          &  18 & 0.01  & 1 & .33 && 30 & 0.005 & 1 & 272
\\ \hline                           
Accept I adapted  &  36 & 0.005 & 4 & .36 && 36 & 0.02  & 1 & 273
\\ \hline
Accept II         &  18 & 0.01  & 1 & .33 && 30 & 0.005 & 1 & 272
\\ \hline
Reweighting       &  30 & 0.005 & 2 & .61 && 36 & 0.005 & 2 & 476
\\
\hline\hline
\end{tabular}
\end{center}

\end{table}

\chapter{Conclusions}
\vspace{-0.5cm}

The massive 
two-flavour Schwinger
model has physical properties similar to QCD in four dimensions. 
It
is much easier to simulate even with dynamical fermions and 
allows to determine observables with high precision
as
the
cost will be lower by orders of magnitude as compared to the case of
QCD in four dimensions. This makes the Schwinger model a reasonable
testing ground for dynamical fermion algorithms.
Still, in the case of the local bosonic algorithm,
this study is the first application using Wilson fermions.

On the other hand,
due to problems with
topological sectors (as discussed below)
and the
scaling properties towards the continuum limit,
it is definitely
not that easy a toy model
as one might expect.

\vspace{-0.22cm}
\section{Instabilities}

In a class of fermion simulation algorithms
 relying on the
local bosonic algorithm (LBA)
a matrix valued Chebyshev polynomial is involved.
Recursion relations
allow the evaluation of these
polynomials in a numerically stable way.
Yet in a number of cases,
like our implementation of the Acceptance step method I in the
 Hermitean LBA or
the Polynomial Hybrid Monte Carlo (PHMC) algorithm \cite{phmc2},
the polynomial is needed in 
the factorized form.
Then
rounding errors can easily lead
to significant precision losses and even numerical instabilities,
especially if simulations are 
done on machines having only 32-bit floating point arithmetic
precision. 

We investigate the 
effects of using various ordering schemes of monomial factors,
or equivalently the complex roots, 
on the numerical construction of the Chebyshev polynomials
now commonly used for the LBA. 
We find
that different ordering schemes for the
roots can lead to rounding error effects ranging from numerical overflow
to retaining a precision 
comparable to the one 
numerically stable
recurrence relations can provide.
 
In the case of a Chebyshev polynomial of a single real variable $s$
approximating the function $1/s$, 
we find that the bitreversal scheme
and a scheme suggested by Montvay can keep rounding error effects to a low
level for degrees of the Chebyshev polynomial up to $n \approx 220$.

Applying 
these reordering schemes
to the evaluation of 
a matrix valued polynomial,
we study  numerical simulations of 
4-dimensional lattice QCD. 
We find that Montvay's ordering scheme of the roots seems to be particularly
suited for this problem. The rounding errors
could be kept on a level which is comparable to the one that is reached
when using the stable Clenshaw recurrence
relation.

We conclude
that the precision with which the numerical evaluation of the Chebyshev polynomial
can be performed depends strongly on the chosen ordering of the roots.
We expect also severe 
consequences for the dynamics of the simulation algorithms
where Chebyshev polynomials in the product representation are used,
depending on the root ordering scheme employed.
 
As the most important outcome of this 
investigation, 
we consider that there
exist orderings of the roots which allow a numerically very stable evaluation
of a Chebyshev polynomial, even up to degrees $n$ of the polynomial of
about 200.
Since these values of $n$ correspond to degrees of the Chebyshev polynomials
commonly used in simulations, 
we consider our findings as promising 
for future applications of the local bosonic algorithm.

\vspace{-0.22cm}
\section{Topology}

During these studies
we encountered
simulations at high $\beta$ ($\approx 8$)
exhibiting metastabilities
in the topological charge.
For dynamical simulations,
no cure to this problem is known.

Motivated by this,
we studied
simulations
at extreme $\beta$ values ($\gsim 12$)
which are 
effectively at fixed charge
as no tunnelling occurs
within these runs.
Results for effective pion masses
show that a
definition of a 
pion mass 
from a plateau is not possible 
for either quenched or dynamical simulations
even for vanishing topological charge.

A comparison with quenched 
global updates exhibiting no metastabilities
demonstrates
that a plateau can be found
in the correct path integral sample.
Furthermore, 
cross-checks against the Hybrid Monte Carlo algorithm
(HMC)
indicate that this problem
is not an artifact
stemming from the fermion algorithm.

To gain insight into
the mechanism,
we studied
effective pion masses
calculated from 
projections
to fixed topological sectors. 
Results
from
fluctuating
ensembles
show no dependence on the
topological charge.
On the other hand,
those projected from fluctuating quenched 
ensembles at high $\beta$
show approximately
the same problematic behaviour
as simulations completely without tunnelling.
This suggests a rather subtle dynamical
effect we do not understand yet.

We find that
external homogeneous plaquette configurations
with 
fixed Polyakov loop values $P_T$ and $P_L$
exhibit completely irregular behaviour.
After averaging over $P_T$ and $P_L$
we regain the 
effective mass results
characteristic for the
topological charge sectors these 
configurations lie in.

We conclude that
there is a 
need to obtain a better understanding
of the interplay between
the dynamical mass generation of mesonic states and 
topological sectors.
We would like to point out that
higher statistics runs
could reveal similar phenomena
in other models with non-trivial topological structure.

\vspace{-0.22cm}
\section{CPU cost optimisation}
\vspace{-0.12cm}

We optimised
the CPU cost 
of the local bosonic algorithm
varying
3 technical parameters of the 
algorithm,
namely
the number of over-relaxation steps
in each trajectory,
the order of the approximating Chebyshev polynomial $n$
and the lower cut of the approximation region $\epsilon$.
We simulated
at two different lattice sizes $8 \times 20$ and $16 \times 40$
keeping the physical mass ratio
${m_\pi\over  m_\eta} = 0.81$ approximately constant.
This resulted in pion masses of about $m_\pi = 0.629$
and $m_\pi = 0.384$, or equivalently
in a scaling of the lattice spacing of about $1.6$.

The tuning of the LBA is
demonstrated 
to be surprisingly easy.
It became more difficult for the
larger lattices, though.
Technically,
the number of reflections
per heatbath update,
usually fixed in other studies of the algorithm,
is found to be an important optimisation tool.

The CPU cost can be
lower than for HMC,
but not by a large factor
with present techniques.
We find a
gain for the plaquette
of approximately 3
for both lattice sizes
and for the pion propagator 
of $\approx 2$
on the smaller and $\approx 2.5$ on the larger lattices.
The gain thus
differs 
and estimates from plaquette-like
observables 
are too optimistic.
Still, our main point is that the gain is
detectable and consistent for both lattice sizes.

We also demonstrate
that using a
noisy Metropolis acceptance step scheme
to make the LBA exact is
also possible
for the Hermitean case.
The use of the Gegenbauer solver,
which
avoids instabilities
in the evaluation of the polynomial,
in the method labelled as acceptance step method II,
is shown to be 
competitive to CG
in the first real simulation.
Costs are virtually identical
to those of the acceptance step method I
for both lattice sizes.

In general,
all investigated acceptance 
variants
performed similarly.
The reweighting method, though,
had decisively higher costs.
Especially for the larger lattices
it performed worse than
the acceptance step methods 
for both observables studied
by a factor of 1.8
(1.3 for the smaller lattices).
Still, as 
the difference is not really large,
the possible advantages of this method
regarding exceptional configurations \cite{phmc1} 
make further study worthwhile.

The optimisation of the
acceptance step method I,
interrupting the solver iterations and checking
at intermediate steps whether the solution quality was already
good enough for the requested Metropolis decision,
did result in a
significant gain of a factor of 1.6 
compared to the unoptimised version
regarding the cost of the pion correlator
on the smaller lattices.
On the larger lattices, though,
we did not find any improvement.
As this is nevertheless
an optimisation
offering possible gain,
future studies
should have this trick in mind as the gain could be different
for larger inverter trajectories. 

We  conclude that these results
give further evidence
that the
LBA is a competitive 
algorithm for the simulation of dynamical fermions.

\begin{appendix} 

\chapter{Conventions}
\label{c_conventions}

\section{General conventions}
\label{s_basicdefs}

\ul{Dimensions.}
We use the standard conventions
\begin{eqnarray}
\hbar =c=1
\; .
\end{eqnarray}
In $d=2$ dimensions then follow
\begin{eqnarray}
[p]=[e]
&=&
[m]
\nn \\  
\; [l] = [t] 
&=&
[m]^{-1} 
\; .
\end{eqnarray}
The inverse lattice coupling $\beta={1\over e^2 a^2}$ 
and the Wilson parameter $\kappa={1\over 2(ma+d)}$ are
dimensionless.
Gauge links $U_{x,\mu}$ are dimensionless, while 
Grassmann fermion fields $\psi$ have the dimension $[m]^{\1,2 }$.
In the electromagnetic tensor 
\begin{eqnarray}
F^{\mu\nu} = \ma[0,-E,E,0] 
\end{eqnarray}
in 1+1 dimensions 
only the electric field $E$ appears,
as because of the missing transverse
directions no magnetic field is existing.
The electric field has the same dimension as the charge.
\\
\ul{Indices.} We denote \\
flavour indices
with Latin lower case letters
usually starting with $a$,
\\
lattice sites with Latin lower case letters
usually starting with $x$,
\\
spinor indices 
with Latin lower case letters
usually starting with $s$.
\\
Directions
we abbreviate with 
Greek lower case letters
$
\mu, \nu, 1, 2, \dots
$
\\
and use barred letters
for the
orthogonal direction (only one in 2D)
$
\bar\mu,\bar\nu, \bar{1},\bar{2}, \dots \; .
$
\\
Generally, 1 signifies the spatial direction,
2 the temporal.
\\
Unit vectors are denoted by a hat
$
\hat\mu,\hat{\bar\mu},\hat{1},\hat{\bar{1}},\dots
$.
\\
\ul{Trace conventions.} The different traces are
\begin{eqnarray}
{\rm TR} &=& \sum_{x,s,a} 
\quad ; \quad \mbox{i.e. sum over sites, Dirac and flavour indices,}
\nn \\
{\rm Tr} &=& \sum_{x,s}
\quad ; \quad \mbox{i.e. sum over sites and Dirac indices,}
\nn \\
{\rm tr} &=& \sum_{s} 
\quad ; \quad \mbox{i.e. sum over Dirac indices,}
\nn \\
{\rm tr_f} &=& \sum_{a} 
\quad ; \quad \mbox{i.e. sum over flavour indices.}
\; .
\end{eqnarray}
\\
In an update trajectory,
new variables and configurations are primed,
the old ones unprimed.
\\
\ul{Links.} 
We use compact $U(1)$ link variables
$U_{x,\mu}$.
Plaquettes are
defined starting at the lower left corner 
going anti-clockwise
\begin{eqnarray}
\label{e_plaquette2}
U_{P\;x}
=
U^\dagger_{x,\mu} \;
U^\dagger_{x+\hat{\mu},\bar{\mu}}\;
U        _{x+\hat{\bar{\mu}},\mu}\;
U        _{x,\bar{\mu}}              
\; .
\end{eqnarray}
Plaquette angles $\phi_x \in [-\pi,\pi]$
are defined via
\begin{eqnarray}
\label{e_angle}
e^{i\phi_x} = U_{P\;x} \; .
\end{eqnarray}
Polyakov loops $P_T,P_L$ are
used averaged over the orthogonal direction,
i.e.
\begin{eqnarray}
\label{e_polyakov}
P_L 
= 
{1\over T} \sum_{x_2=1}^T \prod_{x_1=1}^L U_{x, 1}
\quad {\rm and} \quad
P_T 
= 
{1\over L} \sum_{x_1=1}^L \prod_{x_2=1}^T U_{x, 2} 
\; .
\end{eqnarray}
\\
\ul{Fermions}.
Spinors are denoted with Greek letters
$\psi_{x,s}$ throughout.

\section{Gamma matrices}

We use Hermitean gamma matrices
\begin{eqnarray}
\gamma^0
= 
\ma[1,0,0,-1] \;,
\quad\quad
\gamma^1
= 
\ma[0,1,1,0] 
\end{eqnarray}
and a Hermitean $\gamma^5$
\begin{eqnarray}
\gamma^5
=
i\gamma^0\gamma^1
=
\ma[0,i,-i,0]
\end{eqnarray}
with the relations
\begin{eqnarray}
\Bigl\{ \gamma^\mu,\gamma^\nu \Bigr\}
=
2 \delta^{\mu,\nu},
\quad\quad
\Bigl\{
\gamma^5,\gamma^\mu
\Bigr\}
=
0,
\quad
\quad
\gamma^5 \gamma^\mu \gamma^5 
=
-\gamma^\mu 
\; .
\end{eqnarray}
Note that
\begin{eqnarray}
\gamma^5 (1 \pm \gamma^\mu) \gamma^5 (1 \pm \gamma^\mu)
=
1 \pm \gamma^5 \gamma^\mu \gamma^5 \pm \gamma^\mu
\pm \gamma^5 \gamma^\mu \gamma^5 \gamma^\mu 
=
0 
\; ,
\end{eqnarray}
i.e. that the
$Q^2$ matrix (Eq.~\ref{e_q2}) does not
connect sites with two straight links in-between.
For general purposes we introduce the set
\begin{eqnarray}
\Gamma &\in & \tilde\Gamma=\{ \id, (i\gamma^5), \gamma^0, 
(i\gamma^1) \}
\end{eqnarray}
chosen so that
\begin{eqnarray}
(\bar\psi \Gamma \psi)^\dagger
=
(\bar\psi \Gamma \psi) \quad \forall \Gamma \; .
\end{eqnarray}

\section{Flavour matrices}

The flavour matrices set is 
\begin{eqnarray}
T &\in & \tilde{T} = \{ \tau^+,\tau^-, \id, \tau^0 \} 
\; ,
\end{eqnarray}
where the tau matrices are 
derived from the  Pauli matrices given by
\begin{eqnarray}
\tau^0 
&=&
\ma[1,0,0,-1] ,
\quad\quad
\tau^1
=
\ma[0,1,1,0] ,
\quad\quad
\tau^2 
=
\ma[0,-i,i,0]
\end{eqnarray}
via
\begin{eqnarray}
\tau^+
&=&
{1 \over \sqrt{2}} (\tau^1+i\tau^2),
\quad\quad
\tau^- 
=
{1 \over \sqrt{2}} (\tau^1-i\tau^2)
\; .
\end{eqnarray}
The Hermiticity relations
\begin{eqnarray}
\id^\dagger=\id , 
\quad\quad
 \tau^{0\dagger}=\tau^0 ,
\quad\quad
\tau^{+\dagger}=\tau^-
\end{eqnarray}
hold.
Further useful relations are
\begin{eqnarray}
{\rm tr_f} (\tau^a) 
=
0 \; ,
\quad\quad
{\rm tr_f} (\tau^{a\dagger} \tau^a) 
= 
2 \; ,
\quad\quad
{\rm tr_f} (\id) 
= 
2 
\end{eqnarray}
and
\begin{eqnarray}
\label{e_triplettrace}
\sum_a {\rm tr_f} (\tau^{a\dagger} \tau^a) 
= 3 
{\rm tr_f} (\tau^{b\dagger} \tau^b) \quad \forall b 
\; . 
\end{eqnarray}

\section{Fermion matrix}
\label{s_fermion_matrix}

Starting from the standard Wilson fermion matrix
\begin{eqnarray}
M
=
\delta_{x,y}
-\kappa  \sum_\mu
\Bigl(
\delta_{x-\hat\mu,y} (1+\gamma^\mu) U_{x-\hat\mu,\mu}
+
\delta_{x+\hat\mu,y} (1-\gamma^\mu) U^\dagger_{x,\mu}
\Bigr)
\end{eqnarray}
with the Wilson kappa definition
\begin{eqnarray}
\kappa
&=&
{1\over 2(m+d)}
\; ,
\end{eqnarray}
we define via
\begin{eqnarray}
\label{e_hopping}
M
=
\id - \kappa H
\end{eqnarray}
the hopping matrix $H$.
The abbreviation $G$ for 
the Greens function $M^{-1} \delta_{\rm flavour} $
is used throughout.
We construct a Hermitean matrix $Q$
scaled such that eigenvalues lay
in [-1,1] \cite{lba} via 
\begin{eqnarray}
\label{e_Q2}
Q
=
c_o \gamma^5 \delta_{x,y}
-
c_o \kappa \sum_\mu
\Bigl(
\delta_{x-\hat\mu,y} \gamma^5 (1+\gamma^\mu) U_{x-\hat\mu,\mu}
+
\delta_{x+\hat\mu,y} \gamma^5 (1-\gamma^\mu) U^\dagger_{x,\mu}
\Bigr)
\end{eqnarray}
with the scaling factor
\begin{eqnarray}
c_0 
&=& {m+d \over m+2d} {1\over c_M} 
=
{1 \over 1+2d\kappa} {1\over c_M} 
\quad;\quad c_M \ge 1 
\; .
\end{eqnarray}
Inserting explicit gamma matrices,
this results in the programmed formulae for $Q$
\begin{eqnarray}
\left( Q\phi \right)_{x,1}
=
&+&
i c_0 \phi_{x,2}
\nn \\ &-&
2i c_0 \kappa U^\dagger_{x,0} \phi_{x+\hat{0},2}
\nn \\ &-&
i c_0 \kappa U^\dagger_{x,1} (\phi_{x+\hat{1},2}-\phi_{x+\hat{1},1})
\nn \\ &-&
i c_0 \kappa U_{x-\hat{1},1} (\phi_{x-\hat{1},1}+\phi_{x-\hat{1},2})
\nn \\
\left( Q\phi \right)_{x,2}
=
&-&
i c_0 \phi_{x,1}
\nn \\ &-&
i c_0 \kappa U^\dagger_{x,1} (\phi_{x+\hat{1},2}-\phi_{x+\hat{1},1})
\nn \\ &+&
2i c_0 \kappa U_{x-\hat{0},0} \phi_{x-\hat{0},1}
\nn \\ &+&
i c_0 \kappa U_{x-\hat{1},1} (\phi_{x-\hat{1},1}+\phi_{x-\hat{1},2})
\end{eqnarray}
and $Q^2$
\begin{eqnarray}
\label{e_q2}
Q^2_{x,z}
&=& 
c_0^2 \delta_{x,z}
\nn \\
&-&
2 c_0^2 \kappa \sum_\mu [
U_{x-\hat\mu,\mu} \delta_{x-\hat\mu,z} +
U^\dagger_{x,\mu} \delta_{x+\hat\mu,z} ]
\nn \\
&+&
c_0^2 \kappa^2 \sum_{\mu,\mu'}
\nn \\
&(&
\delta_{x-\hat\mu-\hat\mu',z} \;\; (1-\gamma^\mu) (1+\gamma^{\mu'})\;\; 
U_{x-\hat\mu,\mu} U_{x-\hat\mu-\hat\mu',\mu'}
\nn \\
&+&
\delta_{x+\hat\mu-\hat\mu',z} \;\; (1+\gamma^\mu) (1+\gamma^{\mu'})\;\; 
U^\dagger_{x,\mu} U_{x+\hat\mu-\hat\mu',\mu'}
\nn \\
&+&
\delta_{x-\hat\mu+\hat\mu',z} \;\; (1-\gamma^\mu) (1-\gamma^{\mu'})\;\; 
U_{x-\hat\mu,\mu} U^\dagger_{x-\hat\mu,\mu'}
\nn \\
&+&
\delta_{x+\hat\mu+\hat\mu',z} \;\; (1+\gamma^\mu) (1-\gamma^{\mu'})\;\; 
U^\dagger_{x,\mu} U^\dagger_{x+\hat\mu,\mu'} )
\; .
\end{eqnarray}
The diagonal parts are
\begin{eqnarray}
Q_{x,x}
&=&
c_0 \gamma^5
=
\ma[0,i c_0,-i c_0,0]
\nn \\
Q^2_{x,x}
&=&
c^2_0 ( 1 + 4 d  \kappa^2)
=
\ma[c^2_0 ( 1 + 4 d  \kappa^2),0,0,c^2_0 ( 1 + 4 d  \kappa^2)]
\; .
\end{eqnarray}

\section{Meson operators, fermion densities and correlators}
\label{s_mesons}

\ul{Operators.} 
As meson operators we
introduce fermion bilinears
\begin{eqnarray}
O^{\Gamma T}_x
=
\bar\psi_x \Gamma T \psi_x
\; , 
\end{eqnarray}
sandwiching both a gamma matrix $\Gamma \in \tilde\Gamma$ and a 
flavour matrix $T \in \tilde{T}$ from the
sets described above.
We show the list of operators in 2 dimensions
and the corresponding mesonic states 
in Tab. (\ref{t_operators}).
\begin{table}[tbp]
\caption[Meson Operators]
{{\bf Meson Operators.}
}
\label{t_operators}

\begin{center}
\begin{tabular}{|c|c|c|c|c|c|c|c|}
\hline 

No.&$O$ & $\Gamma$ & $T$ & $\Gamma'$ & $T'$ & $\gamma^5\Gamma$ & 
meson \\
\hline\hline

\rule[-0.3cm]{0cm}{0.8cm} 1&$\bar\psi \psi$ & 
$\id$ & $\id$ & $\id$ & $\id$ & $\gamma^5$ & $f_0$
\\
\hline
\rule[-0.3cm]{0cm}{0.8cm} &$\bar\psi \gamma^0 \psi$ & 
$\gamma^0$ & $\id$ & $\gamma^0$ & $\id$ & $\gamma^5\gamma^0$ & 
$\omega_0$ 
\\
\hline\hline
\rule[-0.3cm]{0cm}{0.8cm} 2&$\bar\psi (i\gamma^1) \psi$ & 
$i\gamma^1$ & $\id$ & $i\gamma^1$ & $\id$ & $i\gamma^5\gamma^1$ &
$\omega_1$ 
\\
\hline
\rule[-0.3cm]{0cm}{0.8cm} &$\bar\psi (i\gamma^5) \psi$ & 
$i\gamma^5$ & $\id$ & $i\gamma^5$ & $\id$ & $i\id$ & 
$\eta$ 
\\
\hline\hline

\rule[-0.3cm]{0cm}{0.8cm} 3&$\bar\psi \tau \psi$ & 
$\id$ & $\tau$ & $\id$ & $\tau$ & $\gamma^5$ &
$a_0$ 
\\
\hline
\rule[-0.3cm]{0cm}{0.8cm} &$\bar\psi \gamma^0 \tau \psi$ & 
$\gamma^0$ & $\tau$ & $\gamma^0$ & $\tau$ & $\gamma^5\gamma^0$ & $\rho_0$
\\
\hline\hline
\rule[-0.3cm]{0cm}{0.8cm} 4&$\bar\psi (i\gamma^1) \tau \psi$ & 
$i\gamma^1$ & $\tau$ & $i\gamma^1$ & $\tau$ & $i\gamma^5\gamma^1$ &
$\rho_1$ 
\\
\hline
\rule[-0.3cm]{0cm}{0.8cm} &$\bar\psi (i\gamma^5) \tau \psi$ & 
$i\gamma^5$ & $\tau$ & $i\gamma^5$ & $\tau$ & $i\id$ & $\pi$
\\
\hline\hline

\end{tabular}
\end{center}
\end{table}
The 2D peculiarities
\begin{eqnarray}
\gamma^5\gamma^0 
=
-i\gamma^1 
\quad {\rm and} \quad
i\gamma^5\gamma^1 
=
-\gamma^0 
\end{eqnarray}
reduce the possible gamma structures to four.
The trivial action of $\gamma^0$,
if used on momentum zero states,
leads to still further reduction to
2 relevant operators each for the
flavour singlet and triplet.
We remark that these symmetries are 
not used in the program.
Correlations from  all operators 
are explicitly calculated
and the symmetry properties verified.
\\
\ul{Fermion densities.}
To evaluate densities
we need the 
general expression
\begin{eqnarray}
\label{e_wick}
\langle\psi_{x_1}^{a_1} \bar\psi_{y_1}^{b_1}
\dots
\psi_{x_n}^{a_n} \bar\psi_{y_n}^{b_n}
\rangle
=
\sum_{x'_1 \dots x'_n}
\epsilon_{x_1 \dots x_n}^{x'_1 \dots x'_n}
M^{-1}_{x'_1 y_1} \delta^{a_1 b_1} 
\dots M^{-1}_{x'_n y_n} \delta^{a_n b_n}
\end{eqnarray}
specialised to the 2 fermion isosinglet case
\begin{eqnarray}
{1 \over V d n_f} \langle \psi_{x,r}^a \Gamma_{rs} \bar\psi_{x,s}^a \rangle
=
{-1 \over V d} \sum_{rsx}  \Gamma_{rs} M^{-1}_{x,s;x,r} 
\; .
\end{eqnarray}
Corresponding to the
four possible gamma matrices $\Gamma \in \tilde\Gamma$
we obtain 
\begin{itemize}
\item
for $\Gamma=\id$ the fermion condensate,
\item
for $\Gamma=\gamma^5$ the axial condensate or pseudo-scalar density,
\item
for $\Gamma=\gamma^1,\gamma^0$ the $\gamma^0$ and $\gamma^1$ densities.
\end{itemize}
Rewriting this in terms of $Q$ (Eq.~\ref{e_Q2})
and normalising
by a constant
${1 \over V \cdot d }$ to 
suppress trivial volume and dimension behaviour
we arrive at the desired expression
\begin{eqnarray}
{-c_0 \over V d} \sum_{rst x}  \gamma^5_{rt} \Gamma_{ts}
Q^{-1}_{x,s;x,r}
\; .
\end{eqnarray}
\\
\ul{Noise.}
We use noise
vectors
with the properties
\begin{eqnarray}
\label{e_noise}
<\eta^*_x \eta_y >= \delta_{x,y},
\qquad
<\eta_x>=0
\; ,
\end{eqnarray}
where the
components
are either continuous, having
a Gaussian distribution,
or are discrete $Z_2$ ($\pm 1$) or $Z_4$ ($\pm 1, \pm I$) Ising
variables. 
The vectors do not carry spinor indices.
\\
\ul{Noisy scheme.}
Using these noise vectors
we apply the
noisy estimator
of the inverse of $Q$
\begin{eqnarray}
Q^{-1}_{xr,xs}
=
\langle (\eta^*_x \sum_z Q^{-1}_{xr zs} \eta_z) \rangle
\end{eqnarray}
where the inversion
has to be done for each spinor component separately
to obtain the full information.
\\
\ul{Correlators.}
For the meson 2-point functions
(correlators) we specialise Eq.~\ref{e_wick}
to the 4 fermion case
\begin{eqnarray}
\hat{C}_{xy}^{\Gamma T \Gamma' T'}
=
\langle
\bar\psi_x \Gamma T \psi_x
\bar\psi_y \Gamma' T' \psi_y
\rangle
\end{eqnarray}
can then be expressed as
\begin{eqnarray}
\hat{C}_{xy}^{\Gamma T \Gamma' T'}
&=&
\langle
\sum_{abcd,\alpha\beta\gamma\delta}
\bar\psi^a_{x\alpha} \Gamma_{\alpha\beta} T^{ab} \psi^b_{x\beta}
\bar\psi^c_{y\gamma} \Gamma'_{\gamma\delta} T'^{cd} \psi^d_{y\delta}
\rangle
\nn \\
&=&
\sum_{abcd,\alpha\beta\gamma\delta}
-\Gamma_{\alpha\beta} T^{ab}
\Gamma'_{\gamma\delta} T'^{cd}
\langle
\psi^b_{x\beta}
\bar\psi^c_{y\gamma}  \psi^d_{y\delta}
\bar\psi^a_{x\alpha} 
\rangle
\nn \\
&=&
\sum_{abcd,\alpha\beta\gamma\delta}
\Gamma_{\alpha\beta} T^{ab}
\Gamma'_{\gamma\delta} T'^{cd}
\Bigl(
G^{dc}_{y\delta, y\gamma} G^{ba}_{x\beta,  x\alpha}
-
G^{bc}_{x\beta,  y\gamma} G^{da}_{y\delta, x\alpha}
\Bigr)
\; .
\end{eqnarray}
For operators having an expectation value
we have to subtract this value from
the correlator to get the connected Greens
function.
This is not explicitly written out here,
nor is it used in the analysis routine - 
there the constant part is simply fitted.

Using the flavour structure
of the Greens function,
it is possible to write general formulae
for all flavour 
matrices
\begin{eqnarray}
\hat{C}_{xy}^{\Gamma T }
&=&
\sum_{\alpha\beta\gamma\delta}
\Gamma_{\alpha\beta}  
\Gamma_{\gamma\delta} 
\Bigl(
({\rm tr_f} T)^2
G_{y\delta, y\gamma} G_{x\beta,  x\alpha}
-
({\rm tr_f} T^\dagger T)
G_{x\beta,  y\gamma} G_{y\delta, x\alpha}
\Bigr)
\; ,
\end{eqnarray}
where we introduce the simplified notation
\begin{eqnarray}
C_{xy}^{\Gamma T}
&=&
\langle
\bar\psi_x \Gamma T \psi_x
(\bar\psi_y \Gamma T \psi_y)^\dagger
\rangle
=
\hat{C}_{xy}^{\Gamma T;\gamma^0 T^\dagger \Gamma^\dagger \gamma^0}
\; .
\end{eqnarray}
Using the connection to the scaled matrix $Q$
\begin{eqnarray}
G=c_0 Q^{-1} \gamma^5; \quad Q^\dagger=Q 
\end{eqnarray}
we get 
\begin{eqnarray}
C_{xy}^{\Gamma T}
=
\sum_{\alpha\beta\epsilon\iota}
c_0^2
(\gamma^5\Gamma)_{\iota\beta}  
(\gamma^5\Gamma)_{\epsilon\delta} 
\Bigl(
({\rm tr_f} T)^2
Q^{-1}_{y\delta, y\epsilon} Q^{-1}_{x\beta,  x\iota}
-
({\rm tr_f} T^\dagger T)
Q^{-1}_{x\beta,  y\epsilon} Q^{-1}_{y\delta, x\iota}
\Bigr) 
\; ,
\end{eqnarray}
where the matrices
\begin{eqnarray}
A_{\epsilon\delta}
=
\sum_{\gamma}
\gamma^5_{\epsilon\gamma}
\Gamma_{\gamma\delta} 
\end{eqnarray}
can be calculated and coded in the program.

We generally use correlators
with randomly located point source
and summation over the target time slice
\begin{eqnarray}
C^{\Gamma T}(t)
&=&
\sum_x C_{(x+x_0,t+t_0)(x_0,t_0)}^{\Gamma T}
\; ,
\end{eqnarray}
where the index structure should be clear from the context.
\\
\ul{Noisy scheme.}
Using Gaussian or Ising noise
vectors
with the properties Eq.~\ref{e_noise},
we are able
to sum over the 
time slice $S(x_0 t_0)$ including the
initial site $x_0,t_0$.
Using Hermiticity of $Q$
we can derive 
in the triplet case
\begin{eqnarray}
C^\Gamma_{{\rm triplet}} (t) 
= 
-6 c_0^2
\sum_{x abcd}
(\gamma^5\Gamma)_{ab} (\gamma^5\Gamma)_{cd}
\;
Q^{-1}_{x+x_0 t+t_0 b; x_0 t_0 c}
\;
Q^{-1}_{x_0 t_0 d; x+x_0 t+t_0 a}
\nn\\
=
{-6 c_0^2 \over L}
\sum_{xabcd}
(\gamma^5\Gamma)_{ab} (\gamma^5\Gamma)_{cd}
(\sum_z
Q^{-1}_{x+z t+t_0b; z t_0 c}
\eta_{z})
(\sum_{x_0}
(Q^{-1}_{x+x_0 t+t_0 a;x_0 t_0 d}
\eta_{x_0})^\dagger )
\end{eqnarray}
and, correspondingly, for the singlet
\begin{eqnarray}
\hspace*{-1truecm}
C^\Gamma_{{\rm singlet}} (t)
&=&
{1\over 3} C^\Gamma_{{\rm triplet}} (t)
\\
&+&
\sum_{x \alpha\beta\gamma\delta}
c_0^2
(\gamma^5\Gamma)_{\alpha\beta}  
(\gamma^5\Gamma)_{\gamma\delta} 
\Bigl(
4
Q^{-1}_{x_0 t_0\delta, x_0 t_0\gamma} 
Q^{-1}_{x_0+x t_0+t\beta,  x_0+x t_0+t\alpha}
\Bigr)
\; ,
\nn
\end{eqnarray}
where one would estimate the
disconnected part from
the full point source.
\\
\ul{Pion.}
For the pion a special trick is possible.
The gamma structure allows to
use noise vectors including spinor indices with
\begin{eqnarray}
<\eta^*_{xs} \eta_{yt} >= \delta_{x,y} \delta_{s,t},
\qquad
<\eta_{xs}>=0
\; .
\end{eqnarray}
Using these
only one inversion is necessary,
as can be seen from
\begin{eqnarray}
\vspace{-1cm}
C^{\rm pion} (t) 
&=& 
6 c_0^2 
\sum_{x abcd}
\delta_{a,b} \delta_{c,d}
\;\;
Q^{-1}_{x+x_0 t+t_0 b; x_0 t_0 c}
\;\;
Q^{-1}_{x_0 t_0 d; x+x_0 t+t_0 a}
\nn\\
&=&
{6 c_0^2 \over L}
\sum_{ax}
(\sum_{ze}
Q^{-1}_{x+z t+t_0a; z t_0 e}
\eta_{z e} )
\;
(\sum_{x_0 c}
(Q^{-1}_{x+x_0 t+t_0 a;x_0 t_0 c}
\eta_{x_0c})^\dagger )
\; .
\end{eqnarray}

\chapter{Analytic results}

\section{Free fermions}
\label{s_free}

We just give the result for
the fermion condensate using free fermions
where the momenta are restricted
using anti-periodic boundary conditions to
\begin{eqnarray}
k_\mu
=
{2\pi \over L_\mu} (n_\mu+{1 \over 2}) 
; \quad 
n_\mu \in \{0, \dots L_\mu-1\} 
\; .
\end{eqnarray}
We obtain
the condensate
\begin{eqnarray}
{1 \over {n_f V d}} \langle {\rm TR} (\bar\psi \psi) \rangle 
=
{-1 \over V } \sum_k 
{(1-2\kappa 
\sum_\mu 
\cos k_\mu) 
\over
\sum_\mu (2\kappa \sin k_\mu)^2  +(1-2\kappa \sum_\mu \cos k_\mu)^2} 
\; .
\end{eqnarray}

\section{Pure gauge topological susceptibility}
\label{s_susceptibility}

In the limit 
of independent plaquettes
the topological susceptibility
can be calculated analytically.
Introducing the
generating functional with periodic BCs
for $N_P$ independent plaquettes
we obtain
\begin{eqnarray}
Z(\beta,\theta) 
= \left[ \int_{-\pi}^{\pi} {d\phi \over 2\pi}
e^{i\theta Q+\beta \cos \phi} \right]^{N_P}
\; ,
\end{eqnarray}
where $\phi$ denotes the plaquette angle
defined in Eq.~\ref{e_angle}.
From this the topological susceptibility
is given by
\begin{eqnarray}
\chi_{\rm top}
=
{-1 \over N_P}
\partial_\theta^2 \log Z(\beta,\theta ) |_{\theta=0}
\end{eqnarray}
which can be evaluated to
\begin{eqnarray}
\chi_{\rm top}
&=&
-\partial_\theta^2 \log 
\int_{-\pi}^{\pi} {d\phi \over 2\pi}
e^{i\theta Q+\beta  \cos \phi}
|_{\theta=0} 
\nn \\
&=&
{\int_{-\pi}^{\pi} {d\phi \over 2\pi}
({\phi \over 2\pi})^2
e^{\beta \cos \phi}
\over
\int_{-\pi}^{\pi} {d\phi \over 2\pi}
e^{\beta \cos \phi}} 
=
<({\phi \over 2\pi})^2>
\; .
\end{eqnarray}
\\
\ul{Approximations.}
Two convenient approximations are possible.
On the one hand the small $\beta$ limit yields
\begin{eqnarray}
\chi_{\rm top}
=
\int_{-\pi}^{\pi} {d\phi \over 2\pi}
({\phi \over 2\pi})^2
=
{1\over12}
\; .
\end{eqnarray}
On the other 
hand 
the integral can be approximated by a Gaussian
one in the large $\beta$ limit using 
\begin{eqnarray}
\beta \cos \phi = \beta - {\beta \over2}\phi^2 + \dots
\quad \to \quad
<\phi^2> \approx {1\over\beta}
\end{eqnarray}
yielding a topological susceptibility of
\begin{eqnarray}
\chi_{\rm top}
\approx
{1\over 4 \pi^2 \beta}
\; .
\end{eqnarray}

\section{Pure gauge plaquette}
\label{s_exact_plaquette}

For the pure $U(1)$
theory in 2 dimensions it is possible
to derive an analytical expression
for the plaquette even on a finite lattice.
We label
link angles
as
$
\phi_{l t \mu}
$
with $l$ and $t$ the site indices 
from 1 to $L$ and $T$ respectively
and express the path integral in these variables
\begin{eqnarray}
\hspace*{-1truecm}
Z
&=&
e^{-\beta \Omega} 
\int D[\phi_{l t \mu}] e^{\beta\sum_{lt} \cos(
\phi_{l t 1}+\phi_{l+1 t 2}-\phi_{l t+1 1}-\phi_{l t 2} ) }
\\
&=&
e^{-\beta \Omega} 
\prod_{l t}
\int_{-\pi}^{\pi} 
{d\phi_{l t 1}\over 2\pi} {d\phi_{l t 2} \over 2 \pi} 
\sum_{n_{lt}}
I_{n_{lt}}(\beta)
e^{in_{lt}(
\phi_{l t 1}+\phi_{l+1 t 2}-\phi_{l t+1 1}-\phi_{l t 2} ) }
\; ,
\nn
\end{eqnarray}
where
we introduced
the Fourier-transform
of the exponential
\begin{eqnarray}
e^{\beta\cos \phi}
=
\sum_{-\infty}^{\infty} c_n e^{in\phi}
\end{eqnarray}
with coefficients
\begin{eqnarray}
c_n 
=
{1\over 2 \pi} \int_{-\pi}^{\pi}
e^{\beta\cos \phi } e^{-in\phi} d\phi
=
{1\over \pi} \int_{0}^{\pi}
e^{\beta\cos \phi} \cos(n\phi)d\phi
=
I_n(\beta) \; .
\end{eqnarray}
$I_n$ signifies
the modified Bessel function \cite{numerical_recipes}.
As this expression for the path integral
factorizes 
we
can
rewrite $Z$ to
\begin{eqnarray}
e^{-\beta \Omega} 
\sum_{n_{lt}} 
[
\prod_{l t} \left(
\int_{-\pi}^{\pi} 
{d\phi_{l t 1} \over 2\pi} 
e^{in_{lt}
\phi_{l t 1}-in_{l t-1} \phi_{l t 1} } 
\int_{-\pi}^{\pi} 
{d\phi_{l t 2} \over 2 \pi} 
e^{-in_{lt}
\phi_{l t 2}+in_{l-1 t}\phi_{l t 2} }
\right)
\prod_{l t}
I_{n_{lt}}(\beta)
]
\nn \\
=
e^{-\beta \Omega} 
\sum_{n_{lt}} 
\left[
\prod_{l t} 
\left(
\delta_{n_{lt},n_{l t-1}}
\delta_{n_{lt},n_{l-1 t}}
I_{n_{lt}}(\beta) 
\right)
\right] \; .
\qquad\qquad\qquad\qquad\qquad\qquad
\end{eqnarray}
This reduces $Z$ to a simple sum
\begin{eqnarray}
Z
&=&
e^{-\beta \Omega} 
\sum_{n=-\infty}^{\infty}
I_{n}(\beta)^{\Omega}
=
e^{-\beta \Omega} \left[
I_{0}(\beta)^{\Omega}
+
\sum_{n=1}^{\infty}
2 I_{n}(\beta)^{\Omega}
\right]
\; ,
\end{eqnarray}
where we used the symmetry
$
I_n(x) = I_{-n}(x) 
$ valid for all $n$ and $x$.
To finally get the plaquette
we use the relation
\begin{eqnarray}
<P>
&=&
1 + {1\over Z \Omega}
{\partial Z \over \partial \beta}
\end{eqnarray}
and
\begin{eqnarray}
{\partial I_n(\beta) \over \partial \beta}
&=&
{1\over 2} \left[ I_{n-1}(\beta) + I_{n+1}(\beta) \right]
\end{eqnarray}
to get after some algebra to
\begin{eqnarray}
<P>
&=&
{\displaystyle 
{\displaystyle I_1(\beta) \over I_0(\beta)}
+
\displaystyle \sum_{n=1}^{\infty} \left[ 
\left(
{\displaystyle I_n(\beta) \over I_0(\beta)}
\right)^{\Omega-1}
\left(
{\displaystyle I_{n-1}(\beta) \over I_0(\beta)} + 
{\displaystyle I_{n+1}(\beta) \over I_0(\beta)}
\right)
\right]
\over
1 + 2 \displaystyle \sum_{n=1}^{\infty}
\left( {\displaystyle I_n(\beta) \over I_0(\beta)} \right)^{\Omega}
}
\; .
\end{eqnarray}
This expression is then numerically evaluated
using Fortran routines for Bessel functions \cite{numerical_recipes}.

\section{Hopping parameter expansion}
\label{s_hopping}

We start with the distribution
of the links in a dynamic fermion simulation
\begin{eqnarray}
P_{\rm eff}
\propto
(\det [\id- \kappa H])^2 e^{\beta \sum_x {\rm Re} \; U_{P\;x} }
\end{eqnarray}
with $H$ the hopping terms (Eq.~\ref{e_hopping})
and rewrite
using
\begin{eqnarray}
\det[M^2]
=
e^{2 {\rm Tr} \log M}
\; .
\end{eqnarray}
We 
expand the logarithm
\begin{eqnarray}
\log M = \log (1-\kappa H)
=
-\kappa H - {\kappa^2 H^2 \over 2}
- {\kappa^3 H^3\over 3} - {\kappa^4 H^4 \over 4} +\dots
\end{eqnarray}
and observe that the first link-dependent term
appearing in the trace is the
$H^4$ term as the odd powers of $H$ vanish in the trace
$
{\rm Tr} H 
=
{\rm Tr} H^3 =
\dots
=
0
$
and the second order contribution
\begin{eqnarray}
{\rm Tr} H^2
=
{\rm const.}
+ 
{\rm const.} \cdot U_{x-\hat\mu,\mu} \cdot U^\dagger_{x-\hat\mu,\mu}
\dots
=
{\rm const.}
\end{eqnarray}
is an irrelevant constant.
So we end up with
\begin{eqnarray}
\det[M^2]
=
e^{-{1\over2}\kappa^4 {\rm Tr} H^4}
\end{eqnarray}
to leading order.

Writing out the $H^4$ term,
we see that only closed loops of links can contribute to the trace,
i.e. to leading order only plaquettes are contributing.
This results in 
\begin{eqnarray}
{\rm Tr} H^4
\hspace{-0.3cm}
&= \displaystyle \sum_{x,\mu} \big( & 
\hspace{-0.3cm}
{\rm tr}[
(1+\gamma^\mu)         
(1+\gamma^{\bar{\mu}}) 
(1-\gamma^\mu)         
(1-\gamma^{\bar{\mu}}) 
]
U_        {x-\hat{\mu},\mu}                    \;
U_        {x-\hat{\mu}-\hat{\bar{\mu}},\bar{\mu}}\;
U^\dagger_{x-\hat{\mu}-\hat{\bar{\mu}},\mu}\;
U^\dagger_{x-\hat{\bar{\mu}},\bar{\mu}}
\nn \\
\hspace{-0.3cm}
&+&
\hspace{-0.3cm}
{\rm tr}[
(1+\gamma^\mu)         
(1-\gamma^{\bar{\mu}}) 
(1-\gamma^\mu)         
(1+\gamma^{\bar{\mu}}) ]
U_        {x-\hat{\mu},\mu}\;
U^\dagger_{x-\hat{\mu},\bar{\mu}}\;
U^\dagger_{x-\hat{\mu}-\hat{\bar{\mu}},\mu}\;
U_        {x,\bar{\mu}}
\nn \\
\hspace{-0.3cm}
&+&
\hspace{-0.3cm}
{\rm tr}[
(1-\gamma^\mu)         
(1+\gamma^{\bar{\mu}}) 
(1+\gamma^\mu)         
(1-\gamma^{\bar{\mu}}) ]
U^\dagger_{x,\mu}\;
U        _{x+\hat{\mu}-\hat{\bar{\mu}},\bar{\mu}}\;
U        _{x-\hat{\bar{\mu}},\mu}\;
U^\dagger_{x-\hat{\bar{\mu}},\bar{\mu}}
\nn \\
\hspace{-0.3cm}
&+&
\hspace{-0.3cm}
{\rm tr} [(1-\gamma^\mu)         
(1-\gamma^{\bar{\mu}}) 
(1+\gamma^\mu)         
(1+\gamma^{\bar{\mu}})] 
U^\dagger_{x,\mu} \;
U^\dagger_{x+\hat{\mu},\bar{\mu}}\;
U        _{x+\hat{\bar{\mu}},\mu}\;
U        _{x,\bar{\mu}} 
\big)
\; .
\end{eqnarray}
To evaluate the gamma matrix expressions we observe that
\begin{eqnarray}
{\rm tr } \id &=& 2 
\; ,
\quad\quad\quad
{\rm tr } \gamma^i = 0 \quad \forall i
\; ,
\nn \\
{\rm tr } \gamma^i \gamma^j &=& 2 \delta_{i,j}
\; ,
\quad\quad
{\rm tr } \gamma^i \gamma^j \gamma^k = 0 \quad\forall i,j,k
\end{eqnarray}
and that therefore all gamma terms give the same contribution
\begin{eqnarray}
{\rm tr \;[ gamma \; terms]}
=
2 - 2 - 2 + {\rm tr} [\gamma^\mu  \gamma^{\bar \mu}  \gamma^\mu  
\gamma^{\bar \mu}]
=
-4\; .  
\end{eqnarray}
Returning to the full expression
we abbreviate the plaquette as above
and obtain
\begin{eqnarray}
{\rm Tr} H^4
=-4 \sum_{x,\mu} \left(
U_{P\; x-\hat{\mu}-\hat{\bar{\mu}}}
+
U^\dagger_{P\; x-\hat{\mu}}
+
U^\dagger_{P\; x-\hat{\bar{\mu}}}
+
U_{P\; x}
\right)
\; .
\end{eqnarray}
As we sum over the lattice sites,
we can reorder
the contributions
and find that we have each term twice
for each direction.
We further combine the complex conjugate pairs
so that 
the final expression for the determinant is
\begin{eqnarray}
\det[M^2]
=
e^{16\kappa^4 
\sum_{x} \Re \;
U_{P\; x}}
\; .
\end{eqnarray}
This first order result can be combined with the standard
link action term
so that the result is an effective
pure gauge theory with a shifted beta value of
\begin{eqnarray}
\beta'
=
\beta + 16\kappa^4 \; .
\end{eqnarray}

\chapter{Force}
\label{c_force}

\section{Force on bosonic fields}

Starting with the effective action
\begin{eqnarray}
S_L
&=&
\beta \Re \sum_{p} (1- U_p)
+
\sum_{k,x,y} \phi_x^{k\dagger} (Q-\mu_k)^2_{x,y} \phi_y^k
+ 
\sum_{k,x} \nu^2_k \phi_x^{k\dagger} \phi_x^k
\end{eqnarray}
we want the part dependent on 
one specified  field spinor $\phi_x^k$.
Rewriting yields
\begin{eqnarray}
S_L(U,\phi)
&=&
{\rm const.}
+
\phi_x^{k\dagger} \Bigl[ \sum_{y | y\ne x} (Q^2 -2Q \mu_k)_{x,y} \phi_y^k \Bigr]
+
\Bigl[ \sum_{y | y\ne x} \phi_y^{k\dagger} (Q^2 -2Q \mu_k)_{y,x} \Bigr] \phi_x^k
\nn \\
&& \phantom{ {\rm const.}} +
\phi_x^{k\dagger} (\mu^2_k+\nu^2_k+ (Q^2 -2Q \mu_k)_{x,x})  \phi_x^k
\nn \\
&=&
{\rm const.}
+
\phi_x^{k\dagger} B_x^k
+
B_x^{k\dagger} \phi_x^k
+ 
\phi_x^{k\dagger} K_x^k \phi_x^k
\; .
\end{eqnarray}
Mapping this to a Gaussian 
distribution via
\begin{eqnarray}
e^{-S}
&\propto&
e^{ -\phi_x^{k\dagger} B_x^k
-
B_x^{k\dagger} \phi_x^k
-
\phi_x^{k\dagger} K_x^k \phi_x^k
}
\propto
e^{-R^\dagger R}
\end{eqnarray}
we easily see
heatbath updates
\begin{eqnarray}
R = 
(K_x^k)^{{1\over 2}} \phi_x^k + (K_x^k)^{-{1\over2}} B_x^k 
\nn \\
\phi_x^k
=
(K_x^k)^{-{1\over2}} R  - (K_x^k)^{-1} B_x^k 
\end{eqnarray}
and micro-canonical reflections 
steps
\begin{eqnarray}
R \to -R &=& -(K_x^k)^{1\over2} \phi_x^k - (K_x^k)^{-{1\over2}} B_x^k 
\nn \\
\phi_x^{k\prime}
&=&
-\phi_x^k
-2 (K_x^k)^{-1} B_x^k 
\; .
\end{eqnarray}
The necessary 
coefficients
$B_x^k, (K_x^k)^{-1}, (K_x^k)^{-{1\over2}}$ can be calculated using
the formulae in App.~\ref{s_fermion_matrix}
giving explicit expressions for $K_x^k$
\begin{eqnarray}
K_x^k
=
c^2_0 (1 +  4 d \kappa^2)
- 2 \mu_k c_0 \gamma^5
+\nu_k^2+\mu_k^2
= K^{k}
\; ,
\end{eqnarray}
which is not a function of the lattice site (though a Dirac matrix)
and $B_x^k$
\begin{eqnarray}
B_x^k
&=&
\sum_{y\ne x} (Q^2 -2Q \mu_k)_{x,y} \phi_y^k
\nn \\
&=&
\sum_{y} [(Q^2 -2Q \mu_k)_{x,y} \phi_y^k ]
-
[c_0^2 (1 + 4 d \kappa^2 ) -2 \mu_k c_0 \gamma^5] \phi_x^k
\; .
\end{eqnarray}

\section{Force on links}

The force on links is defined by
\begin{eqnarray}
\label{e_force}
-S_L=
{\rm const.} + Re F_{x,\mu} U_{x,\mu} 
\end{eqnarray}
with the
intention 
to apply this to the heatbath formulae 
\begin{eqnarray}
P \propto e^{b \cdot \cos\phi}
\; .
\end{eqnarray}
Observing that the links $U_{x,\mu}$ are 
complex numbers,
we define
\begin{eqnarray}
b 
&=& 
|F_{x,\mu}| 
\nn \\
\cos \phi 
&= &
\cos(\phi_1+\phi_2 )
\quad {\rm with} \quad
e^{i\phi_1} = {F_{x,\mu} \over |F_{x,\mu}| };
\quad
e^{i\phi_2} = U_{x,\mu}
\; .
\end{eqnarray}
Starting with the action one easily sees
\begin{eqnarray}
S_L
&=&
{\rm const.} -
\Re \; \beta U_p
+
\sum_{k,x,y} \phi_x^{k\dagger} (Q^2_{x,y}-2\mu_k Q_{x,y}) \phi_y^k
\; .
\end{eqnarray}
Expanding
and using 
$
2 \Re \; z = z + z^*
$
and the gamma matrix relations
\begin{eqnarray}
\vspace*{-0.6cm}
\left( \gamma^5 (1+\gamma^\mu) \right)^\dagger
=
 \gamma^5 (1-\gamma^\mu)
\; , \quad
\left( (1\pm \gamma^\mu) (1\pm\gamma^{\mu'}) \right)^\dagger
=
(1\pm \gamma^{\mu'}) (1\pm\gamma^\mu)
\end{eqnarray}
to combine the complex conjugate pairs
and also complex conjugating one of the
staple expressions (allowed as only the real part is taken) one  obtains
\begin{eqnarray}
F_{x,\mu} 
&=&
\beta \sum_{\mu' \ne \mu}
( U^\dagger_{x,\mu'} U^\dagger_{x+\hat\mu',\mu} U_{x+\hat\mu,\mu'} +
  U^\dagger_{x+\hat\mu-\hat\mu',\mu'} U^\dagger_{x-\hat\mu',\mu} U_{x-\hat\mu',\mu'} )
\nn \\
&+&
 4 c_0^2 \kappa 
\sum_{k} 
\Bigl[\;
\phi_{x+\hat\mu}^{k\dagger} \phi_x^k  \;\Bigr] 
\nn \\
&-&  2c_0^2 \kappa^2 \sum_{k} 
\Bigl[ \;\;
\sum_{\mu'} \Bigl(
\nn \\
&&\phantom{+}
\phi^\dagger_{x+\hat\mu}
 \;\; (1-\gamma^\mu) (1+\gamma^{\mu'})\;\; \phi_{x-\hat\mu'}
U_{x-\hat\mu',\mu'}  
\nn \\
&&+
\phi^\dagger_{x+\hat\mu}
 \;\; (1-\gamma^\mu) (1-\gamma^{\mu'})\;\; \phi_{x+\hat\mu'}
U^\dagger_{x,\mu'}  
\nn \\
&&+
\phi^\dagger_{x+\hat\mu'+\hat\mu}
 \;\; (1-\gamma^{\mu'}) (1+\gamma^{\mu})\;\; \phi_{x}
U_{x+\hat\mu,\mu'} 
\nn \\
&&+
\phi^\dagger_{x-\hat\mu'+\hat\mu}
 \;\; (1+\gamma^{\mu'}) (1+\gamma^{\mu})\;\; \phi_{x}
U^\dagger_{x-\hat\mu'+\hat\mu,\mu'}  \Bigr) \Bigr] 
\nn \\
&-&
 4 c_0 \kappa
\sum_{k} 
\Bigl[
\mu_k
\phi^{k\dagger}_{x+\hat\mu} \gamma^5 (1+\gamma^\mu) \phi_x^k 
\Bigr]
\; . 
\end{eqnarray}

\chapter{Root Ordering Fortran codes}

In this appendix we want to give the Fortran codes which are used
for the reordering of the roots Eq.~\ref{e_roots}. In the following
$n$ will always denote the degree of the Chebyshev polynomial 
and $m$ is an integer divisor of $n$ as needed for the 
subpolynomial scheme.

\section{Simple pairing scheme}
\label{PSa}

We show the code for the first half of roots.
The second half is constructed ana\-logously.
{\footnotesize
\begin{verbatim}
      if (abs(n/4 - one*n/4) .ge. 0.01) then
        med1 = n/4 + 1                             ! n/4 odd
        med2 = n/4 + 1
      else
        med1 = n/4                                 ! n/4 even
        med2 = n/4 + 1 
      endif

      i=0      
      do k=1,n/8                                   ! take complete
        i = i+1 ; j1(i) = k                        ! groups of 4
        i = i+1 ; j1(i) = n/2-k+1
        i = i+1 ; j1(i) = med1+k                   ! -> hashing table
        i = i+1 ; j1(i) = med2-k
      end do
      
      r = n/2 - n/2 /4*4
      if (r .ge. 2) then                           ! last pair
        k = n/8 + 1
        i = i+1 ; j1(i) = k
        i = i+1 ; j1(i) = n/2-k+1
        r = r-2
      endif     
      if (r .ge. 1) then                           ! last lonely one
        i = i+1 ; j1(i) = med1
      endif
\end{verbatim} 
}

\section{Subpolynomial scheme}
\label{SPa}

{\footnotesize
\begin{verbatim}
      kstep = n / m                                ! \# of subpolynomials
      i =0
      do k=1,m
        do l=1,kstep
          i = i+1 ; j1(i) = (l-1)*m + k            ! reset index
        end do
      end do
\end{verbatim}
}

\section{Bitreversal scheme}
\label{BRa}

{\footnotesize
\begin{verbatim}
      do b=0,20
        if (2**b .ge. n) then                      ! length of array
          bits=b; goto 333
        end if
      end do
 333  continue
      
      do k=1,2*n
        j(k) = 0                                   ! init
      end do

      do k=1,n
        do b=0,bits
          bit(b)=0                                 ! init
        end do
        
        nn = k-1                                   ! shift
        do b=bits-1,0,-1
          if (2**b .le. nn) then
            nn     = nn - 2**b
            bit(b) = 1                             ! extract bit
          end if
        end do

        i = 0
        do b=0,bits-1
          i = i + bit(b) * 2**(bits-1-b)           ! reverse
        end do
        i    = i+1                                 ! reshift
        j(i) = k                                   ! -> hashing table
      end do
      
      i = 0
      do k=1,2*n            
        if (j(k) .ne. 0) then                      ! no dummy root
          i        = i+1
          j1(j(k)) = i                             ! reset index
        end if
      end do
\end{verbatim}
}

\section{Montvay scheme}
\label{MSa}

{\footnotesize
\begin{verbatim}

read(`approxima.txt`);

n   :=  12 ;
eps :=  .1000000044703483 ;
c0  :=  52987.39383322105 ;
Root:=[
6.299918614676031E-02   -.146958373796435 *I,
.2375643902633147   -.2602503551965 *I,
.483704827825326   -.313922116577824 *I,
.7450326909011724   -.295678104222787 *I,
.9616809154023317   -.209697801497745 *I,
1.08401800398973   -7.567825958203078E-02 *I,
1.08401800398973 +  7.567825958203071E-02 *I,
.9616809154023317 +  .2096978014977446 *I,
.7450326909011726 +  .2956781042227866 *I,
.483704827825326 +  .3139221165778238 *I,
.2375643902633154 +  .2602503551965006 *I,
6.299918614676020E-02 +  .146958373796435 *I ];

Optimord(n,Root,c0,1,1.,x,eps,1.,100,40,yes);
\end{verbatim}
}

\chapter{Gegenbauer Solver}
\label{c_gegenbauer}

\section{Gegenbauer Polynomials}

We define
polynomials $C^\gamma_n(z)$ 
of the complex
variable $z$ 
with degree $n$ and a real parameter $\gamma >
0$
via their
generating function
\begin{eqnarray}
\label{e_geg}
(1 + t^2 - 2 t z)^{-\gamma} 
= 
\sum_{n=0}^\infty t^n C^\gamma_n(z)
\; .
\end{eqnarray}
Alternatively, an
integral representation
\begin{eqnarray}
C^\gamma_n(z)
&=& 
\frac{2^{1-2\gamma} \Gamma(2\gamma+n)}{n! \, \Gamma(\gamma)^2}     
\int_0^\pi d\phi \, (\sin\phi)^{2\gamma-1}
(z + \sqrt{z^2-1} \cos\phi)^n
\end{eqnarray}
is possible.
For the many known features
of the thus defined Gegenbauer polynomials
like
trigonometric representation,
coefficients of the highest monomial,
parity, bounds and large $n$ approximations 
we defer the reader to the literature
\cite{gegenbauer,transcendental_functions}.
For completeness, 
we 
mention the
first few polynomials
\begin{eqnarray}
C^\gamma_0(z) 
&=& 1           
\nn \\
C^\gamma_1(z) 
&=& 2\gamma z       
\nn \\
C^\gamma_2(z) 
&=& 2\gamma(\gamma+1)z^2 - \gamma       
\nn \\
C^\gamma_3(z) 
&=& \frac{4}{3}\gamma(\gamma+1)(\gamma+2)z^3 - 2\gamma(\gamma+1)z 
\; .
\end{eqnarray}
The relative error of the partial sums
is given by
\begin{eqnarray}
\label{e_relerror}
R_n(z) 
&=& 
1 -  (1 + t^2 - 2 t z)^\gamma \sum_{k=0}^n t^k C^\gamma_k(z)  
\nn \\
&=& 
(1 + t^2 - 2 t z)^\gamma \sum_{k=n+1}^\infty t^k C^\gamma_k(z)
\; .
\end{eqnarray}
The main
item we
want to stress
is the existence of an
recursion relation
\begin{eqnarray}
\label{e_gegenbauer_recursion}
(n+1) C^\gamma_{n+1}(z) + (n+2\gamma-1) C^\gamma_{n-1}(z) 
= 
2(n+\gamma) \, zC^\gamma_n(z)
\end{eqnarray}
enabling an efficient calculation
of these polynomials.

\section{Solver}

The
generating function
suggests the use of this
polynomial expansion
to build up a solver method.
Assume $M$ a Hermitean matrix
with spectrum
$\mbox{spec}(M) \subset [\lambda_1,\lambda_2]$
and
$0 < \lambda_1 \leq \lambda_2$.
We want to solve for $x$ in
\begin{eqnarray}
M^\gamma x = b
\; .
\end{eqnarray}
To map to the Gegenbauer polynomials
defined in Eq.~\ref{e_geg},
we
transform to a normalised matrix 
$A$ with $\mbox{spec}(A) \subset [-1,1]$
\begin{eqnarray}
M
=
c(1+t^2-2tA)
\quad \leftrightarrow \quad
A 
=
 -{1\over 2tc} M + {1 + t^2 \over 2t}
\; ,
\end{eqnarray}
defining parameters $t$ and $c$
\begin{eqnarray}
\label{e_t}
t
=
{\sqrt{\lambda_2 \over \lambda_1} -1 \over
\sqrt{\lambda_2 \over \lambda_1} +1},
\quad\quad
c
=
{\lambda_2-\lambda_1 \over 4t} 
\; .
\end{eqnarray}
We can then write the solution $x$ as
\begin{eqnarray}
x 
&=& 
M^{-\gamma} b   
= 
c^{-\gamma} (1 + t^2 - 2tA)^{-\gamma} b       
\nn \\
&=& 
c^{-\gamma} \sum_{n=0}^\infty t^n C^\gamma_n(A) b     
=
\sum_{n=0}^\infty t^n s_n      
\end{eqnarray}
with
\begin{eqnarray}
s_n 
&=& 
c^{-\gamma} C^\gamma_n(A) b
\; ,
\end{eqnarray}
i.e. as a sum over shift vectors $s_n$
with exponentially decreasing factors.
Moreover,
the shifts have an easy recursion relation
\begin{eqnarray}
(n+1) s_{n+1} + (n+2\gamma-1) s_{n-1} = 2(n+\gamma) \, A s_n
\end{eqnarray}
with start shifts $s_{-1}$ and $s_0$ defined by
\begin{eqnarray}
s_{-1} 
&=& 
0      
\nn \\
s_0 
&=& 
c^{-\gamma} b
\; .
\end{eqnarray}

To obtain a valid solver,
we have to find a stopping criterion,
i.e we have to calculate (at least bounds for)
the relative error Eq.~\ref{e_relerror} and the 
rest vector
\begin{eqnarray}
r_n 
&=& 
b - M^\gamma x_n    
= 
R_n(A) b
\; .
\end{eqnarray}

\section{Real solver for $\gamma = {1\over 2}$}

For
$\gamma=1$ one obtains a standard inverter algorithm.
Most interesting are
non-standard cases,
like e.g. $\gamma={1\over 4}$, 
which is applicable to SUSY
models \cite{susy}. 
The general cases for $\gamma$ and $z$ are
discussed in more detail in \cite{gegenbauer}.
We for obvious reasons 
at this place only consider the real case
with $\gamma={1\over 2}$.

\vspace{0.3cm}
\noindent{\bf Legendre Polynomials}
\noindent
The Gegenbauer polynomials in the $\gamma={1\over 2}$ case
are the Legendre polynomials
\begin{eqnarray}
C^{1/2}_n(z) 
\equiv 
P_n(z)
\; .
\end{eqnarray}
In this case
the integral representation is given by
\begin{eqnarray}
P_n(\cos\phi) 
&=& 
\int_0^\pi {d\phi \over \pi}
(z + \sqrt{z^2-1} \cos\phi)^n
\end{eqnarray}
with a normalisation
$
P_n(1) = 1
$.
The recursion is identical to Eq.~\ref{e_gegenbauer_recursion}.

The important point is that in this case
the relative error can be
estimated for real $z \in [-1,1]$ to
\begin{eqnarray}
|R_n(t)| 
\le 
|t|^{n+1}
\; .
\end{eqnarray}
For the proof we defer to \cite{gegenbauer}.
More involved is the
estimation 
for 
$|z| \geq 1$, so we briefly sketch the idea.
We regard the Legendre polynomials
with complex argument $z$
parametrised by
\begin{eqnarray}
z = \cosh(\tau + i\varphi)
\; .
\end{eqnarray}
Assume that $z$
an $t$ given by Eq.~\ref{e_t} 
are inside the ellipse given by
\begin{eqnarray}
Z = \cosh(\theta + i\phi)
\end{eqnarray}
with $ \theta \geq 0,\phi \in [0,2\pi]$.
Then we
get a uniform bound for
the relative error of the expansion
\begin{eqnarray}
|R_n(z)|
\leq
(|t|e^\theta)^{n+1}
\end{eqnarray}
and equivalently for the
error of the solver assuming $A$ is normal
with spectrum inside the ellipse.

A priori, $\theta$ is unknown.
An estimate can be deducted from the shifts solving
\begin{eqnarray}
{||s_n|| \over ||s_0||}
=
{||P_n(A)b|| \over ||b||}
\leq
P_n( \cosh\theta )
\end{eqnarray}
e.g. 
by Newton-Raphson iterations.

\newpage
\vspace{0.3cm}
\noindent{\bf Solver implementation}
\noindent
For the Hermitean local bosonic algorithm case
$
M = Q^2 P_{n_\phi} (Q^2)
$
we are able to give an realistic upper bound for the
spectrum
\begin{eqnarray}
\lambda_{max}(M) = 1 + \delta(n,\epsilon)
\; ,
\end{eqnarray}
but we have to use
a guess for the unknown $\lambda_{min}(M) > 0$
\begin{eqnarray}
\lambda_1
&=&
{r_{\rm GB} \epsilon} P_{n_\phi}( {r_{\rm GB} \epsilon})
\approx
r_{\rm GB} (n_\phi+1) \sqrt{\epsilon}
\; ,
\end{eqnarray}
assuming that the spectrum
starts at $r_{\rm GB} \epsilon, r_{\rm GB} \leq 1$.

If $\lambda_{min}(M) > \lambda_1$, the convergence factor is $t$, 
but a smaller
value would be more efficient. 
In case of $\lambda_{min}(M) < \lambda_1$, 
$\cosh\theta > 1$ holds.
The convergence factor is $t e^\theta$. If $t e^\theta \ge 1$, the series
does not converge at all. 
So there exists an optimal choice $t_{opt}$
which maps the extreme value $A = 1$ 
exactly to $\lambda_{min}(M)$
\begin{eqnarray}
\left( 
\frac{1-t_{opt}}{1+t_{opt}} \right)^2 
= 
\frac{\lambda_{min}(M)}{1 + \delta}
\; ,
\end{eqnarray}
where the expansion converges with $t_{opt}$. 
Using the knowledge of $\theta$,
we can determine
$t_{opt}$ from
\begin{eqnarray}
      \left( \frac{1-t_{opt}}{1+t_{opt}} \right)^2 
            = \frac{1 + t^2 - 2t\cosh\theta}{(1 + t)^2}
\; .
\end{eqnarray}

In reality,
this is done by adjusting parameters 
$t$ and $c$ for the next expansion
\begin{eqnarray}
t 
\to
t' = 
\left\{ 
\begin{array}{c @{\quad\mbox{if}\quad} l}
t^{1.1}     & \theta = 0 
\\
t_{opt}     & \theta > 0
\end{array}    
\right. 
\quad {\rm and} \quad
c 
\to
c' = {1 + \delta \over (1 + t')^2} 
\end{eqnarray}
using the information from the last
and the fact that we expect similar behaviour
for configurations close in an updating sequence.

\newpage
\vspace{0.3cm}
\noindent{\bf Computer realization}

\begin{enumerate}

\item 
 start of basic constants
 \begin{enumerate}
 \item
 \mbox{read } t
 \item
 $c={1+\delta(n,\epsilon)\over (1+t)^2}$
 \end{enumerate}

\item
 initialise work scalars and vectors 
 \begin{enumerate}
 \item
 $s_0 = c^{-{1\over 2}}b$
 \item
 $s_{-1} = 0$
 \item
 $x_0 = s_0$
 \item
 $N = || s_0 ||$
 \item
 $n = 0$
 \item
 ${\rm mult} = 1$
 \end{enumerate}

\item
 recursion
 \begin{enumerate}
 \item
 ${\rm Ms} \leftarrow M s_0$
 \item
 ${\rm As} \leftarrow {1 + t^2 \over 2t} s_0 - {1\over 2tc} {\rm Ms}$
 \item
 $ W \leftarrow 
 {2n+1 \over n+1} {\rm As}
 -
 {n \over n+1} s_{-1}$
 \item
 $s_{-1} \leftarrow s_0$
 \item
 $ s_0 \leftarrow W$
 \item
 ${\rm mult} \leftarrow {\rm mult} \cdot t$
 \item
 $ x_0 \leftarrow x_0 + {\rm mult} \cdot s_0 $
 \item
 $n \leftarrow n+1$
 \item
 $\theta$ from 
 $\max \left({||s_0|| \over N},1 \right) = P_n(\cosh \theta)$
 \end{enumerate}

\item
 test
 \begin{enumerate}
 \item
 if $(|t|e^\theta)^n < \delta_{\rm rec}$ 
 \hspace{2.2cm} $\to$ store $t$, exit
 \item
 else 
 \hspace{4.6cm} $\to$ 
 iterate the recursion
 \end{enumerate}

\end{enumerate}

\chapter{4D QCD conventions}
\label{c_qcd}

The theory is established in $d=4$ dimensions
on a Euclidean space-time lattice 
with
size $L^3\times T$. 
A gauge field $U_{\mu}(x)\in SU(3)$ is assigned to the link
pointing from point $x$ to point $(x+\mu)$, where
$\mu=0,1,2,3$ designates the 4 forward directions in space-time.
The matrix defining the interaction of the fermions is
\begin{eqnarray} 
\label{qmatrix}
Q(U)_{xy} \!\!\!&=& \!\!\!
c_0
\gamma^5 [
(1+\sum_{\mu\nu}
[{i \over 2}c_{\rm sw}\kappa\sigma_{\mu\nu}{\cal F}_{\mu\nu}(x)])\delta_{x,y}
 \nonumber \\
&-&\kappa\sum_{\mu} \{
   (1-\gamma^{\mu})U_{\mu}(x)\delta_{x+\mu,y} +
(1+\gamma^{\mu})U^{\dagger}_{\mu}(x-\mu)\delta_{x-\mu,y}\}]  \;\;,
\end{eqnarray} 
where $\kappa$ and $c_{\rm sw}$ are parameters that have to be chosen
according to the physical problem under consideration,  
$c_0=[c_M(1+2d\kappa)]^{-1}$ and 
$c_M$ is a constant serving to optimise 
simulation algorithms. 
Typically $\kappa \approx 1/8$ and both $c_{\rm sw}$ and $c_M$ are 
${\rm O}(1)$. 
We refer to the following notation section 
for the definitions of the matrices $\gamma^\mu$, 
$\gamma^5$, 
$\sigma_{\mu\nu}$ and the anti-Hermitian tensor ${\cal F}_{\mu\nu}(x)$. 

In order to speed up the Monte Carlo simulation,
not the original matrix $Q$ 
but an even-odd preconditioned matrix $\hat{Q}$ is used. 
We rewrite the matrix $Q$ as
\begin{equation}
Q \equiv c_0 \gamma^5  \left( \begin{array}{cc}
                1+T_{ee} & M_{eo} \\
                M_{oe} & 1+T_{oo} \\
                \end{array} \right)\;\;,
\end{equation} 
where we introduce the matrix $T_{ee}$($T_{oo}$) on the
even (odd) sites as
\begin{equation} \label{eq:t}
(T)_{xa\alpha,yb\beta} =
\sum_{\mu\nu}[{i \over 2}c_{sw}\kappa\sigma^{\alpha\beta}_{\mu\nu}
{\cal F}^{ab}_{\mu\nu}(x) \delta_{x,y}]\;\;.
\end{equation}
The off-diagonal parts $M_{eo}$ and $M_{oe}$
connect the even with odd and odd with even lattice
sites respectively.        
Preconditioning is now realized 
by writing the determinant of $Q$,
apart from an irrelevant constant factor, as
\begin{eqnarray}
\det[Q]&\propto&\det[1+T_{ee}]\det[\hat{Q}]
\nonumber \\
\hat{Q}&=&
\hat{c}_0 \gamma^5(1 + T_{oo} - M_{oe}(1+T_{ee})^{-1}M_{eo})\;\;.
\end{eqnarray} 
The constant factor
$\hat{c}_0$ is given by $\hat{c}_0=[c_M(1+64\kappa^2)]^{-1}$,
and the constant $c_M$ 
is chosen such that the eigenvalues of $\hat{Q}$ are
in the interval $[-1,1]$. 
Since for the simulation algorithms the eigenvalues have to be positive,
we finally work with the matrix $\hat{Q}^2$.

\vspace{0.3cm}
\noindent{\bf Notations in 4D} 
\vspace{0.3cm}

\underline{Gamma matrices.} 
The matrices $\sigma_{\mu\nu}$, $\mu,\nu=0,...,3$, are
defined via the commutator of $\gamma$-matrices
\begin{equation} \label{sigmamunu}
\sigma_{\mu\nu}=\frac{i}{2}\left[\gamma^\mu,\gamma^\nu\right]\;.
\end{equation}
The $4\otimes 4$ $\gamma$-matrices are given by
\begin{equation} \label{gamma}
\gamma^\mu = \left( \begin{array}{cc} 
                 0 & e_\mu \\
                 e_\mu^\dagger & 0 
              \end{array} \right)\;,
\end{equation} 
with the $2\otimes 2$ matrices 
\begin{equation}
e_0 = -1 \quad ; \quad e_j = i\sigma_j,\; j=1,2,3
\end{equation} 
and Pauli-matrices $\sigma_j$
\begin{equation}
\sigma_1 = \left( \begin{array}{cc} 0 & 1 \\ 1 & 0 \end{array} \right)\;\;
\sigma_2 = \left( \begin{array}{cc} 0 & -i \\ i & 0 \end{array} \right)\;\;
\sigma_3 = \left( \begin{array}{cc} 1 & 0 \\ 0 & -1 \end{array} \right)\;\; .
\end{equation}
The matrix $\gamma^5 = \gamma^0\gamma^1\gamma^2\gamma^3$
is thus diagonal
\begin{equation} \label{gamma5}
\gamma^5 = \left( \begin{array}{cc}
                 1 & 0 \\
                 0 & -1
              \end{array} \right)\;.
\end{equation}

\underline{${\cal F}_{\mu\nu}(x)$.}
This antisymmetric and anti-Hermitian tensor
is a function of the gauge links and given by
\begin{eqnarray} 
{\cal F}_{\mu\nu}(x) &=& {1 \over 8} \left[
 U_{\mu}(x)  U_{\nu}(x+\hat{\mu})
U^{\dagger}_{\mu}(x+\hat{\nu})  U^{\dagger}_{\nu}(x)
\right.
\nonumber \\
&+& U_{\nu}(x) U^{\dagger}_{\mu}(x+\hat{\nu}-\hat{\mu})
U^{\dagger}_{\nu}(x-\hat{\mu})  U_{\mu}(x-\hat{\mu})
\nonumber \\
&+& U^{\dagger}_{\mu}(x-\hat{\mu}) U^{\dagger}_{\nu}(x-\hat{\nu}-\hat{\mu})
U_{\mu}(x-\hat{\nu}-\hat{\mu})  U_{\nu}(x-\hat{\nu})
\nonumber \\
&+& U^{\dagger}_{\nu}(x-\hat{\nu}) U_{\mu}(x-\hat{\nu})
U_{\nu}(x-\hat{\nu}+\hat{\mu})  U^{\dagger}_{\mu}(x)
\nonumber \\
&-& \left. h.c. \right]\;\;.
\end{eqnarray} 

\end{appendix}


\clearpage           
{\bf \center \huge Acknowledgement } 

\vspace{1cm}
I wish to thank 

\parag
the referees
Dr. Karl Jansen,
Prof. Dr. Michael M\"uller-Preu{\ss}ker
and Prof. Dr. Gerrit Schierholz
for their taking on this extra work,

\parag
my colleagues 
Dr. Wolfgang Bock, 
Dr. Martin Hasenbusch,
Dr. Thomas Kalkreuter, 
Dr. Klaus Scharnhorst, 
and Andrea Voigt
in the Computational Physics group at the Humboldt Universit\"at,

Dr. Andreas Hoferichter, 
Dr. Roger Horsley and 
Dr. Ernst-Michael Ilgenfritz
in the Institut f\"ur Physik der Humboldt Universit\"at

for the general willingness to
collaborate, help and for many
fruitful discussions and explanations.

\parag
For support and help with all
computer problems
I am indebted to
Konrad-Zuse-Zentrum f\"ur 
Informationstechnik
and to
Deutsches Elektronensynchrotron -- 
Institut f\"ur Hochenergiephysik Zeuthen.
I would like to mention Dr. Hinnerk Stueben at ZIB
and Dr. Hannelies Novak, 
Dr. Wolfgang Friebel and Dr. Alf Wachsmann at the IfH. 

\parag
This work was supported by the
Deutsche Forschungsgemeinschaft under grants No. 
WO 389/3-1 and WO 389/3-2.

\end{document}

%% file: data_test_scale_all_br.tex
\setlength{\unitlength}{0.1bp}
\special{!
/gnudict 40 dict def
gnudict begin
/Color false def
/Solid false def
/gnulinewidth 5.000 def
/vshift -33 def
/dl {10 mul} def
/hpt 31.5 def
/vpt 31.5 def
/M {moveto} bind def
/L {lineto} bind def
/R {rmoveto} bind def
/V {rlineto} bind def
/vpt2 vpt 2 mul def
/hpt2 hpt 2 mul def
/Lshow { currentpoint stroke M
  0 vshift R show } def
/Rshow { currentpoint stroke M
  dup stringwidth pop neg vshift R show } def
/Cshow { currentpoint stroke M
  dup stringwidth pop -2 div vshift R show } def
/DL { Color {setrgbcolor Solid {pop []} if 0 setdash }
 {pop pop pop Solid {pop []} if 0 setdash} ifelse } def
/BL { stroke gnulinewidth 2 mul setlinewidth } def
/AL { stroke gnulinewidth 2 div setlinewidth } def
/PL { stroke gnulinewidth setlinewidth } def
/LTb { BL [] 0 0 0 DL } def
/LTa { AL [1 dl 2 dl] 0 setdash 0 0 0 setrgbcolor } def
/LT0 { PL [] 0 1 0 DL } def
/LT1 { PL [4 dl 2 dl] 0 0 1 DL } def
/LT2 { PL [2 dl 3 dl] 1 0 0 DL } def
/LT3 { PL [1 dl 1.5 dl] 1 0 1 DL } def
/LT4 { PL [5 dl 2 dl 1 dl 2 dl] 0 1 1 DL } def
/LT5 { PL [4 dl 3 dl 1 dl 3 dl] 1 1 0 DL } def
/LT6 { PL [2 dl 2 dl 2 dl 4 dl] 0 0 0 DL } def
/LT7 { PL [2 dl 2 dl 2 dl 2 dl 2 dl 4 dl] 1 0.3 0 DL } def
/LT8 { PL [2 dl 2 dl 2 dl 2 dl 2 dl 2 dl 2 dl 4 dl] 0.5 0.5 0.5 DL } def
/P { stroke [] 0 setdash
  currentlinewidth 2 div sub M
  0 currentlinewidth V stroke } def
/D { stroke [] 0 setdash 2 copy vpt add M
  hpt neg vpt neg V hpt vpt neg V
  hpt vpt V hpt neg vpt V closepath stroke
  P } def
/A { stroke [] 0 setdash vpt sub M 0 vpt2 V
  currentpoint stroke M
  hpt neg vpt neg R hpt2 0 V stroke
  } def
/B { stroke [] 0 setdash 2 copy exch hpt sub exch vpt add M
  0 vpt2 neg V hpt2 0 V 0 vpt2 V
  hpt2 neg 0 V closepath stroke
  P } def
/C { stroke [] 0 setdash exch hpt sub exch vpt add M
  hpt2 vpt2 neg V currentpoint stroke M
  hpt2 neg 0 R hpt2 vpt2 V stroke } def
/T { stroke [] 0 setdash 2 copy vpt 1.12 mul add M
  hpt neg vpt -1.62 mul V
  hpt 2 mul 0 V
  hpt neg vpt 1.62 mul V closepath stroke
  P  } def
/S { 2 copy A C} def
end
}
\begin{picture}(3600,2160)(0,0)
\special{"
gnudict begin
gsave
50 50 translate
0.100 0.100 scale
0 setgray
/Helvetica findfont 100 scalefont setfont
newpath
-500.000000 -500.000000 translate
LTa
600 251 M
0 1858 V
LTb
600 251 M
63 0 V
2754 0 R
-63 0 V
600 363 M
31 0 V
2786 0 R
-31 0 V
600 511 M
31 0 V
2786 0 R
-31 0 V
600 587 M
31 0 V
2786 0 R
-31 0 V
600 623 M
63 0 V
2754 0 R
-63 0 V
600 734 M
31 0 V
2786 0 R
-31 0 V
600 882 M
31 0 V
2786 0 R
-31 0 V
600 958 M
31 0 V
2786 0 R
-31 0 V
600 994 M
63 0 V
2754 0 R
-63 0 V
600 1106 M
31 0 V
2786 0 R
-31 0 V
600 1254 M
31 0 V
2786 0 R
-31 0 V
600 1330 M
31 0 V
2786 0 R
-31 0 V
600 1366 M
63 0 V
2754 0 R
-63 0 V
600 1478 M
31 0 V
2786 0 R
-31 0 V
600 1626 M
31 0 V
2786 0 R
-31 0 V
600 1701 M
31 0 V
2786 0 R
-31 0 V
600 1737 M
63 0 V
2754 0 R
-63 0 V
600 1849 M
31 0 V
2786 0 R
-31 0 V
600 1997 M
31 0 V
2786 0 R
-31 0 V
600 2073 M
31 0 V
2786 0 R
-31 0 V
600 2109 M
63 0 V
2754 0 R
-63 0 V
600 251 M
0 63 V
0 1795 R
0 -63 V
882 251 M
0 63 V
0 1795 R
0 -63 V
1163 251 M
0 63 V
0 1795 R
0 -63 V
1445 251 M
0 63 V
0 1795 R
0 -63 V
1727 251 M
0 63 V
0 1795 R
0 -63 V
2009 251 M
0 63 V
0 1795 R
0 -63 V
2290 251 M
0 63 V
0 1795 R
0 -63 V
2572 251 M
0 63 V
0 1795 R
0 -63 V
2854 251 M
0 63 V
0 1795 R
0 -63 V
3135 251 M
0 63 V
0 1795 R
0 -63 V
3417 251 M
0 63 V
0 1795 R
0 -63 V
600 251 M
2817 0 V
0 1858 V
-2817 0 V
600 251 L
LT0
2974 1626 D
3417 453 D
3276 455 D
3135 459 D
2994 462 D
2854 466 D
2713 471 D
2572 476 D
2431 482 D
2290 489 D
2149 497 D
2009 505 D
1868 515 D
1727 526 D
1586 539 D
1445 553 D
1304 568 D
1163 585 D
1023 603 D
882 622 D
741 642 D
600 663 D
LT1
2974 1526 A
3417 1079 A
3276 1087 A
3135 1095 A
2994 1103 A
2854 1112 A
2713 1121 A
2572 1130 A
2431 1140 A
2290 1150 A
2149 1161 A
2009 1172 A
1868 1184 A
1727 1196 A
1586 1209 A
1445 1222 A
1304 1236 A
1163 1251 A
1023 1266 A
882 1282 A
741 1299 A
600 1316 A
LT2
2974 1426 B
3417 1781 B
3276 1791 B
3135 1802 B
2994 1812 B
2854 1822 B
2713 1833 B
2572 1844 B
2431 1856 B
2290 1867 B
2149 1879 B
2009 1892 B
1868 1905 B
1727 1918 B
1586 1932 B
1445 1947 B
1304 1962 B
1163 1978 B
1023 1994 B
882 2011 B
741 2029 B
600 2047 B
stroke
grestore
end
showpage
}
\put(2854,1426){\makebox(0,0)[r]{146}}
\put(2854,1526){\makebox(0,0)[r]{86}}
\put(2854,1626){\makebox(0,0)[r]{30}}
\put(2008,51){\makebox(0,0){$s_{\rm min}$ in units of $\epsilon$}}
\put(100,1180){%
\special{ps: gsave currentpoint currentpoint translate
270 rotate neg exch neg exch translate}%
\makebox(0,0)[b]{\shortstack{$R_{\rm max}$}}%
\special{ps: currentpoint grestore moveto}%
}
\put(3417,151){\makebox(0,0){2}}
\put(3135,151){\makebox(0,0){1.8}}
\put(2854,151){\makebox(0,0){1.6}}
\put(2572,151){\makebox(0,0){1.4}}
\put(2290,151){\makebox(0,0){1.2}}
\put(2009,151){\makebox(0,0){1}}
\put(1727,151){\makebox(0,0){0.8}}
\put(1445,151){\makebox(0,0){0.6}}
\put(1163,151){\makebox(0,0){0.4}}
\put(882,151){\makebox(0,0){0.2}}
\put(600,151){\makebox(0,0){0}}
\put(540,2109){\makebox(0,0)[r]{1e+07}}
\put(540,1737){\makebox(0,0)[r]{1e+06}}
\put(540,1366){\makebox(0,0)[r]{100000}}
\put(540,994){\makebox(0,0)[r]{10000}}
\put(540,623){\makebox(0,0)[r]{1000}}
\put(540,251){\makebox(0,0)[r]{100}}
\end{picture}

%% file: data-3-d0.1-r.tex
\setlength{\unitlength}{0.1bp}
\special{!
/gnudict 40 dict def
gnudict begin
/Color false def
/Solid false def
/gnulinewidth 5.000 def
/vshift -33 def
/dl {10 mul} def
/hpt 31.5 def
/vpt 31.5 def
/M {moveto} bind def
/L {lineto} bind def
/R {rmoveto} bind def
/V {rlineto} bind def
/vpt2 vpt 2 mul def
/hpt2 hpt 2 mul def
/Lshow { currentpoint stroke M
  0 vshift R show } def
/Rshow { currentpoint stroke M
  dup stringwidth pop neg vshift R show } def
/Cshow { currentpoint stroke M
  dup stringwidth pop -2 div vshift R show } def
/DL { Color {setrgbcolor Solid {pop []} if 0 setdash }
 {pop pop pop Solid {pop []} if 0 setdash} ifelse } def
/BL { stroke gnulinewidth 2 mul setlinewidth } def
/AL { stroke gnulinewidth 2 div setlinewidth } def
/PL { stroke gnulinewidth setlinewidth } def
/LTb { BL [] 0 0 0 DL } def
/LTa { AL [1 dl 2 dl] 0 setdash 0 0 0 setrgbcolor } def
/LT0 { PL [] 0 1 0 DL } def
/LT1 { PL [4 dl 2 dl] 0 0 1 DL } def
/LT2 { PL [2 dl 3 dl] 1 0 0 DL } def
/LT3 { PL [1 dl 1.5 dl] 1 0 1 DL } def
/LT4 { PL [5 dl 2 dl 1 dl 2 dl] 0 1 1 DL } def
/LT5 { PL [4 dl 3 dl 1 dl 3 dl] 1 1 0 DL } def
/LT6 { PL [2 dl 2 dl 2 dl 4 dl] 0 0 0 DL } def
/LT7 { PL [2 dl 2 dl 2 dl 2 dl 2 dl 4 dl] 1 0.3 0 DL } def
/LT8 { PL [2 dl 2 dl 2 dl 2 dl 2 dl 2 dl 2 dl 4 dl] 0.5 0.5 0.5 DL } def
/P { stroke [] 0 setdash
  currentlinewidth 2 div sub M
  0 currentlinewidth V stroke } def
/D { stroke [] 0 setdash 2 copy vpt add M
  hpt neg vpt neg V hpt vpt neg V
  hpt vpt V hpt neg vpt V closepath stroke
  P } def
/A { stroke [] 0 setdash vpt sub M 0 vpt2 V
  currentpoint stroke M
  hpt neg vpt neg R hpt2 0 V stroke
  } def
/B { stroke [] 0 setdash 2 copy exch hpt sub exch vpt add M
  0 vpt2 neg V hpt2 0 V 0 vpt2 V
  hpt2 neg 0 V closepath stroke
  P } def
/C { stroke [] 0 setdash exch hpt sub exch vpt add M
  hpt2 vpt2 neg V currentpoint stroke M
  hpt2 neg 0 R hpt2 vpt2 V stroke } def
/T { stroke [] 0 setdash 2 copy vpt 1.12 mul add M
  hpt neg vpt -1.62 mul V
  hpt 2 mul 0 V
  hpt neg vpt 1.62 mul V closepath stroke
  P  } def
/S { 2 copy A C} def
end
}
\begin{picture}(3600,2160)(0,0)
\special{"
gnudict begin
gsave
50 50 translate
0.100 0.100 scale
0 setgray
/Helvetica findfont 100 scalefont setfont
newpath
-500.000000 -500.000000 translate
LTa
LTb
600 288 M
63 0 V
2754 0 R
-63 0 V
600 652 M
63 0 V
2754 0 R
-63 0 V
600 1016 M
63 0 V
2754 0 R
-63 0 V
600 1380 M
63 0 V
2754 0 R
-63 0 V
600 1745 M
63 0 V
2754 0 R
-63 0 V
600 2109 M
63 0 V
2754 0 R
-63 0 V
600 251 M
0 63 V
0 1795 R
0 -63 V
770 251 M
0 31 V
0 1827 R
0 -31 V
994 251 M
0 31 V
0 1827 R
0 -31 V
1109 251 M
0 31 V
0 1827 R
0 -31 V
1163 251 M
0 63 V
0 1795 R
0 -63 V
1333 251 M
0 31 V
0 1827 R
0 -31 V
1557 251 M
0 31 V
0 1827 R
0 -31 V
1672 251 M
0 31 V
0 1827 R
0 -31 V
1727 251 M
0 63 V
0 1795 R
0 -63 V
1896 251 M
0 31 V
0 1827 R
0 -31 V
2121 251 M
0 31 V
0 1827 R
0 -31 V
2236 251 M
0 31 V
0 1827 R
0 -31 V
2290 251 M
0 63 V
0 1795 R
0 -63 V
2460 251 M
0 31 V
0 1827 R
0 -31 V
2684 251 M
0 31 V
0 1827 R
0 -31 V
2799 251 M
0 31 V
0 1827 R
0 -31 V
2854 251 M
0 63 V
0 1795 R
0 -63 V
3023 251 M
0 31 V
0 1827 R
0 -31 V
3247 251 M
0 31 V
0 1827 R
0 -31 V
3362 251 M
0 31 V
0 1827 R
0 -31 V
3417 251 M
0 63 V
0 1795 R
0 -63 V
600 251 M
2817 0 V
0 1858 V
-2817 0 V
600 251 L
LT0
3174 1946 D
3417 449 D
3330 514 D
3242 530 D
3155 580 D
3068 623 D
2981 670 D
2893 687 D
2806 726 D
2719 775 D
2632 838 D
2544 892 D
2457 940 D
2370 978 D
2282 1025 D
2195 1116 D
2108 1129 D
2021 1216 D
1933 1260 D
1846 1305 D
1759 1382 D
1672 1436 D
1584 1513 D
1497 1615 D
1410 1675 D
1322 1761 D
1235 1868 D
1148 1991 D
LT1
3174 1846 A
3417 449 A
3330 462 A
3242 475 A
3155 580 A
3068 581 A
2981 594 A
2893 604 A
2806 726 A
2719 713 A
2632 743 A
2544 742 A
2457 930 A
2370 843 A
2282 886 A
2195 870 A
2108 1097 A
2021 1026 A
1933 1115 A
1846 983 A
1759 1185 A
1672 1096 A
1584 1168 A
1497 1052 A
1410 1257 A
1322 1162 A
1235 1238 A
1148 1183 A
stroke
grestore
end
showpage
}
\put(3054,1846){\makebox(0,0)[r]{Bitreversal}}
\put(3054,1946){\makebox(0,0)[r]{Subpolynomial}}
\put(2008,51){\makebox(0,0){$\epsilon$}}
\put(100,1180){%
\special{ps: gsave currentpoint currentpoint translate
270 rotate neg exch neg exch translate}%
\makebox(0,0)[b]{\shortstack{ratio $R$}}%
\special{ps: currentpoint grestore moveto}%
}
\put(3417,151){\makebox(0,0){0.1}}
\put(2854,151){\makebox(0,0){0.01}}
\put(2290,151){\makebox(0,0){0.001}}
\put(1727,151){\makebox(0,0){0.0001}}
\put(1163,151){\makebox(0,0){1e-05}}
\put(600,151){\makebox(0,0){1e-06}}
\put(540,2109){\makebox(0,0)[r]{1e+15}}
\put(540,1745){\makebox(0,0)[r]{1e+12}}
\put(540,1380){\makebox(0,0)[r]{1e+09}}
\put(540,1016){\makebox(0,0)[r]{1e+06}}
\put(540,652){\makebox(0,0)[r]{1000}}
\put(540,288){\makebox(0,0)[r]{1}}
\end{picture}

%% file: data-3-d0.1-m.tex
\setlength{\unitlength}{0.1bp}
\special{!
/gnudict 40 dict def
gnudict begin
/Color false def
/Solid false def
/gnulinewidth 5.000 def
/vshift -33 def
/dl {10 mul} def
/hpt 31.5 def
/vpt 31.5 def
/M {moveto} bind def
/L {lineto} bind def
/R {rmoveto} bind def
/V {rlineto} bind def
/vpt2 vpt 2 mul def
/hpt2 hpt 2 mul def
/Lshow { currentpoint stroke M
  0 vshift R show } def
/Rshow { currentpoint stroke M
  dup stringwidth pop neg vshift R show } def
/Cshow { currentpoint stroke M
  dup stringwidth pop -2 div vshift R show } def
/DL { Color {setrgbcolor Solid {pop []} if 0 setdash }
 {pop pop pop Solid {pop []} if 0 setdash} ifelse } def
/BL { stroke gnulinewidth 2 mul setlinewidth } def
/AL { stroke gnulinewidth 2 div setlinewidth } def
/PL { stroke gnulinewidth setlinewidth } def
/LTb { BL [] 0 0 0 DL } def
/LTa { AL [1 dl 2 dl] 0 setdash 0 0 0 setrgbcolor } def
/LT0 { PL [] 0 1 0 DL } def
/LT1 { PL [4 dl 2 dl] 0 0 1 DL } def
/LT2 { PL [2 dl 3 dl] 1 0 0 DL } def
/LT3 { PL [1 dl 1.5 dl] 1 0 1 DL } def
/LT4 { PL [5 dl 2 dl 1 dl 2 dl] 0 1 1 DL } def
/LT5 { PL [4 dl 3 dl 1 dl 3 dl] 1 1 0 DL } def
/LT6 { PL [2 dl 2 dl 2 dl 4 dl] 0 0 0 DL } def
/LT7 { PL [2 dl 2 dl 2 dl 2 dl 2 dl 4 dl] 1 0.3 0 DL } def
/LT8 { PL [2 dl 2 dl 2 dl 2 dl 2 dl 2 dl 2 dl 4 dl] 0.5 0.5 0.5 DL } def
/P { stroke [] 0 setdash
  currentlinewidth 2 div sub M
  0 currentlinewidth V stroke } def
/D { stroke [] 0 setdash 2 copy vpt add M
  hpt neg vpt neg V hpt vpt neg V
  hpt vpt V hpt neg vpt V closepath stroke
  P } def
/A { stroke [] 0 setdash vpt sub M 0 vpt2 V
  currentpoint stroke M
  hpt neg vpt neg R hpt2 0 V stroke
  } def
/B { stroke [] 0 setdash 2 copy exch hpt sub exch vpt add M
  0 vpt2 neg V hpt2 0 V 0 vpt2 V
  hpt2 neg 0 V closepath stroke
  P } def
/C { stroke [] 0 setdash exch hpt sub exch vpt add M
  hpt2 vpt2 neg V currentpoint stroke M
  hpt2 neg 0 R hpt2 vpt2 V stroke } def
/T { stroke [] 0 setdash 2 copy vpt 1.12 mul add M
  hpt neg vpt -1.62 mul V
  hpt 2 mul 0 V
  hpt neg vpt 1.62 mul V closepath stroke
  P  } def
/S { 2 copy A C} def
end
}
\begin{picture}(3600,2160)(0,0)
\special{"
gnudict begin
gsave
50 50 translate
0.100 0.100 scale
0 setgray
/Helvetica findfont 100 scalefont setfont
newpath
-500.000000 -500.000000 translate
LTa
LTb
600 288 M
63 0 V
2754 0 R
-63 0 V
600 652 M
63 0 V
2754 0 R
-63 0 V
600 1016 M
63 0 V
2754 0 R
-63 0 V
600 1380 M
63 0 V
2754 0 R
-63 0 V
600 1745 M
63 0 V
2754 0 R
-63 0 V
600 2109 M
63 0 V
2754 0 R
-63 0 V
600 251 M
0 63 V
0 1795 R
0 -63 V
770 251 M
0 31 V
0 1827 R
0 -31 V
994 251 M
0 31 V
0 1827 R
0 -31 V
1109 251 M
0 31 V
0 1827 R
0 -31 V
1163 251 M
0 63 V
0 1795 R
0 -63 V
1333 251 M
0 31 V
0 1827 R
0 -31 V
1557 251 M
0 31 V
0 1827 R
0 -31 V
1672 251 M
0 31 V
0 1827 R
0 -31 V
1727 251 M
0 63 V
0 1795 R
0 -63 V
1896 251 M
0 31 V
0 1827 R
0 -31 V
2121 251 M
0 31 V
0 1827 R
0 -31 V
2236 251 M
0 31 V
0 1827 R
0 -31 V
2290 251 M
0 63 V
0 1795 R
0 -63 V
2460 251 M
0 31 V
0 1827 R
0 -31 V
2684 251 M
0 31 V
0 1827 R
0 -31 V
2799 251 M
0 31 V
0 1827 R
0 -31 V
2854 251 M
0 63 V
0 1795 R
0 -63 V
3023 251 M
0 31 V
0 1827 R
0 -31 V
3247 251 M
0 31 V
0 1827 R
0 -31 V
3362 251 M
0 31 V
0 1827 R
0 -31 V
3417 251 M
0 63 V
0 1795 R
0 -63 V
600 251 M
2817 0 V
0 1858 V
-2817 0 V
600 251 L
LT0
3174 1946 D
3417 433 D
3330 469 D
3242 479 D
3155 524 D
3068 562 D
2981 596 D
2893 607 D
2806 656 D
2719 697 D
2632 733 D
2544 765 D
2457 806 D
2370 846 D
2282 889 D
2195 951 D
2108 978 D
2021 1034 D
1933 1064 D
1846 1100 D
1759 1152 D
1672 1176 D
1584 1213 D
1497 1268 D
1410 1302 D
1322 1345 D
1235 1403 D
1148 1457 D
LT1
3174 1846 A
3417 433 A
3330 460 A
3242 469 A
3155 524 A
3068 512 A
2981 539 A
2893 547 A
2806 656 A
2719 621 A
2632 673 A
2544 638 A
2457 839 A
2370 746 A
2282 810 A
2195 736 A
2108 973 A
2021 899 A
1933 1016 A
1846 793 A
1759 1045 A
1672 949 A
1584 1037 A
1497 898 A
1410 996 A
1322 970 A
1235 1076 A
1148 990 A
stroke
grestore
end
showpage
}
\put(3054,1846){\makebox(0,0)[r]{Bitreversal}}
\put(3054,1946){\makebox(0,0)[r]{Subpolynomial}}
\put(2008,51){\makebox(0,0){$\epsilon$}}
\put(100,1180){%
\special{ps: gsave currentpoint currentpoint translate
270 rotate neg exch neg exch translate}%
\makebox(0,0)[b]{\shortstack{maximum $M$}}%
\special{ps: currentpoint grestore moveto}%
}
\put(3417,151){\makebox(0,0){0.1}}
\put(2854,151){\makebox(0,0){0.01}}
\put(2290,151){\makebox(0,0){0.001}}
\put(1727,151){\makebox(0,0){0.0001}}
\put(1163,151){\makebox(0,0){1e-05}}
\put(600,151){\makebox(0,0){1e-06}}
\put(540,2109){\makebox(0,0)[r]{1e+15}}
\put(540,1745){\makebox(0,0)[r]{1e+12}}
\put(540,1380){\makebox(0,0)[r]{1e+09}}
\put(540,1016){\makebox(0,0)[r]{1e+06}}
\put(540,652){\makebox(0,0)[r]{1000}}
\put(540,288){\makebox(0,0)[r]{1}}
\end{picture}

%% file: data-3-d0.001-r.tex
\setlength{\unitlength}{0.1bp}
\special{!
/gnudict 40 dict def
gnudict begin
/Color false def
/Solid false def
/gnulinewidth 5.000 def
/vshift -33 def
/dl {10 mul} def
/hpt 31.5 def
/vpt 31.5 def
/M {moveto} bind def
/L {lineto} bind def
/R {rmoveto} bind def
/V {rlineto} bind def
/vpt2 vpt 2 mul def
/hpt2 hpt 2 mul def
/Lshow { currentpoint stroke M
  0 vshift R show } def
/Rshow { currentpoint stroke M
  dup stringwidth pop neg vshift R show } def
/Cshow { currentpoint stroke M
  dup stringwidth pop -2 div vshift R show } def
/DL { Color {setrgbcolor Solid {pop []} if 0 setdash }
 {pop pop pop Solid {pop []} if 0 setdash} ifelse } def
/BL { stroke gnulinewidth 2 mul setlinewidth } def
/AL { stroke gnulinewidth 2 div setlinewidth } def
/PL { stroke gnulinewidth setlinewidth } def
/LTb { BL [] 0 0 0 DL } def
/LTa { AL [1 dl 2 dl] 0 setdash 0 0 0 setrgbcolor } def
/LT0 { PL [] 0 1 0 DL } def
/LT1 { PL [4 dl 2 dl] 0 0 1 DL } def
/LT2 { PL [2 dl 3 dl] 1 0 0 DL } def
/LT3 { PL [1 dl 1.5 dl] 1 0 1 DL } def
/LT4 { PL [5 dl 2 dl 1 dl 2 dl] 0 1 1 DL } def
/LT5 { PL [4 dl 3 dl 1 dl 3 dl] 1 1 0 DL } def
/LT6 { PL [2 dl 2 dl 2 dl 4 dl] 0 0 0 DL } def
/LT7 { PL [2 dl 2 dl 2 dl 2 dl 2 dl 4 dl] 1 0.3 0 DL } def
/LT8 { PL [2 dl 2 dl 2 dl 2 dl 2 dl 2 dl 2 dl 4 dl] 0.5 0.5 0.5 DL } def
/P { stroke [] 0 setdash
  currentlinewidth 2 div sub M
  0 currentlinewidth V stroke } def
/D { stroke [] 0 setdash 2 copy vpt add M
  hpt neg vpt neg V hpt vpt neg V
  hpt vpt V hpt neg vpt V closepath stroke
  P } def
/A { stroke [] 0 setdash vpt sub M 0 vpt2 V
  currentpoint stroke M
  hpt neg vpt neg R hpt2 0 V stroke
  } def
/B { stroke [] 0 setdash 2 copy exch hpt sub exch vpt add M
  0 vpt2 neg V hpt2 0 V 0 vpt2 V
  hpt2 neg 0 V closepath stroke
  P } def
/C { stroke [] 0 setdash exch hpt sub exch vpt add M
  hpt2 vpt2 neg V currentpoint stroke M
  hpt2 neg 0 R hpt2 vpt2 V stroke } def
/T { stroke [] 0 setdash 2 copy vpt 1.12 mul add M
  hpt neg vpt -1.62 mul V
  hpt 2 mul 0 V
  hpt neg vpt 1.62 mul V closepath stroke
  P  } def
/S { 2 copy A C} def
end
}
\begin{picture}(3600,2160)(0,0)
\special{"
gnudict begin
gsave
50 50 translate
0.100 0.100 scale
0 setgray
/Helvetica findfont 100 scalefont setfont
newpath
-500.000000 -500.000000 translate
LTa
LTb
600 269 M
63 0 V
2754 0 R
-63 0 V
600 453 M
63 0 V
2754 0 R
-63 0 V
600 637 M
63 0 V
2754 0 R
-63 0 V
600 1005 M
63 0 V
2754 0 R
-63 0 V
600 1373 M
63 0 V
2754 0 R
-63 0 V
600 1741 M
63 0 V
2754 0 R
-63 0 V
600 2109 M
63 0 V
2754 0 R
-63 0 V
600 251 M
0 63 V
0 1795 R
0 -63 V
883 251 M
0 31 V
0 1827 R
0 -31 V
1048 251 M
0 31 V
0 1827 R
0 -31 V
1165 251 M
0 31 V
0 1827 R
0 -31 V
1256 251 M
0 31 V
0 1827 R
0 -31 V
1331 251 M
0 31 V
0 1827 R
0 -31 V
1394 251 M
0 31 V
0 1827 R
0 -31 V
1448 251 M
0 31 V
0 1827 R
0 -31 V
1496 251 M
0 31 V
0 1827 R
0 -31 V
1539 251 M
0 63 V
0 1795 R
0 -63 V
1822 251 M
0 31 V
0 1827 R
0 -31 V
1987 251 M
0 31 V
0 1827 R
0 -31 V
2104 251 M
0 31 V
0 1827 R
0 -31 V
2195 251 M
0 31 V
0 1827 R
0 -31 V
2270 251 M
0 31 V
0 1827 R
0 -31 V
2333 251 M
0 31 V
0 1827 R
0 -31 V
2387 251 M
0 31 V
0 1827 R
0 -31 V
2435 251 M
0 31 V
0 1827 R
0 -31 V
2478 251 M
0 63 V
0 1795 R
0 -63 V
2761 251 M
0 31 V
0 1827 R
0 -31 V
2926 251 M
0 31 V
0 1827 R
0 -31 V
3043 251 M
0 31 V
0 1827 R
0 -31 V
3134 251 M
0 31 V
0 1827 R
0 -31 V
3209 251 M
0 31 V
0 1827 R
0 -31 V
3272 251 M
0 31 V
0 1827 R
0 -31 V
3326 251 M
0 31 V
0 1827 R
0 -31 V
3374 251 M
0 31 V
0 1827 R
0 -31 V
3417 251 M
0 63 V
0 1795 R
0 -63 V
600 251 M
2817 0 V
0 1858 V
-2817 0 V
600 251 L
LT0
3174 1946 D
3417 439 D
3272 475 D
3126 510 D
2981 571 D
2835 636 D
2690 729 D
2544 799 D
2399 924 D
2253 1051 D
2108 1210 D
1962 1399 D
1817 1620 D
1672 1873 D
LT1
3174 1846 A
3417 390 A
3272 400 A
3126 409 A
2981 428 A
2835 437 A
2690 452 A
2544 460 A
2399 475 A
2253 491 A
2108 531 A
1962 576 A
1817 628 A
1672 689 A
1526 766 A
1381 857 A
1235 966 A
1090 1099 A
944 1260 A
799 1446 A
LT2
3174 1746 B
3417 390 B
3272 393 B
3126 449 B
2981 443 B
2835 465 B
2690 460 B
2544 544 B
2399 505 B
2253 546 B
2108 526 B
1962 595 B
1817 568 B
1672 615 B
1526 588 B
1381 688 B
1235 624 B
1090 655 B
944 636 B
799 718 B
LT3
3174 1646 C
3417 416 C
3272 433 C
3126 449 C
2981 470 C
2835 456 C
2690 490 C
2544 493 C
2399 508 C
2253 552 C
2108 551 C
1962 593 C
1817 582 C
1672 635 C
1526 652 C
1381 678 C
1235 730 C
1090 742 C
944 823 C
799 908 C
stroke
grestore
end
showpage
}
\put(3054,1646){\makebox(0,0)[r]{Montvay}}
\put(3054,1746){\makebox(0,0)[r]{Bit reversal}}
\put(3054,1846){\makebox(0,0)[r]{Pairing}}
\put(3054,1946){\makebox(0,0)[r]{Naive}}
\put(2008,51){\makebox(0,0){$\epsilon$}}
\put(100,1180){%
\special{ps: gsave currentpoint currentpoint translate
270 rotate neg exch neg exch translate}%
\makebox(0,0)[b]{\shortstack{ratio $R$}}%
\special{ps: currentpoint grestore moveto}%
}
\put(3417,151){\makebox(0,0){0.1}}
\put(2478,151){\makebox(0,0){0.01}}
\put(1539,151){\makebox(0,0){0.001}}
\put(600,151){\makebox(0,0){0.0001}}
\put(540,2109){\makebox(0,0)[r]{1e+30}}
\put(540,1741){\makebox(0,0)[r]{1e+24}}
\put(540,1373){\makebox(0,0)[r]{1e+18}}
\put(540,1005){\makebox(0,0)[r]{1e+12}}
\put(540,637){\makebox(0,0)[r]{1e+06}}
\put(540,453){\makebox(0,0)[r]{1000}}
\put(540,269){\makebox(0,0)[r]{1}}
\end{picture}

%% file: data-3-d0.001-m.tex
\setlength{\unitlength}{0.1bp}
\special{!
/gnudict 40 dict def
gnudict begin
/Color false def
/Solid false def
/gnulinewidth 5.000 def
/vshift -33 def
/dl {10 mul} def
/hpt 31.5 def
/vpt 31.5 def
/M {moveto} bind def
/L {lineto} bind def
/R {rmoveto} bind def
/V {rlineto} bind def
/vpt2 vpt 2 mul def
/hpt2 hpt 2 mul def
/Lshow { currentpoint stroke M
  0 vshift R show } def
/Rshow { currentpoint stroke M
  dup stringwidth pop neg vshift R show } def
/Cshow { currentpoint stroke M
  dup stringwidth pop -2 div vshift R show } def
/DL { Color {setrgbcolor Solid {pop []} if 0 setdash }
 {pop pop pop Solid {pop []} if 0 setdash} ifelse } def
/BL { stroke gnulinewidth 2 mul setlinewidth } def
/AL { stroke gnulinewidth 2 div setlinewidth } def
/PL { stroke gnulinewidth setlinewidth } def
/LTb { BL [] 0 0 0 DL } def
/LTa { AL [1 dl 2 dl] 0 setdash 0 0 0 setrgbcolor } def
/LT0 { PL [] 0 1 0 DL } def
/LT1 { PL [4 dl 2 dl] 0 0 1 DL } def
/LT2 { PL [2 dl 3 dl] 1 0 0 DL } def
/LT3 { PL [1 dl 1.5 dl] 1 0 1 DL } def
/LT4 { PL [5 dl 2 dl 1 dl 2 dl] 0 1 1 DL } def
/LT5 { PL [4 dl 3 dl 1 dl 3 dl] 1 1 0 DL } def
/LT6 { PL [2 dl 2 dl 2 dl 4 dl] 0 0 0 DL } def
/LT7 { PL [2 dl 2 dl 2 dl 2 dl 2 dl 4 dl] 1 0.3 0 DL } def
/LT8 { PL [2 dl 2 dl 2 dl 2 dl 2 dl 2 dl 2 dl 4 dl] 0.5 0.5 0.5 DL } def
/P { stroke [] 0 setdash
  currentlinewidth 2 div sub M
  0 currentlinewidth V stroke } def
/D { stroke [] 0 setdash 2 copy vpt add M
  hpt neg vpt neg V hpt vpt neg V
  hpt vpt V hpt neg vpt V closepath stroke
  P } def
/A { stroke [] 0 setdash vpt sub M 0 vpt2 V
  currentpoint stroke M
  hpt neg vpt neg R hpt2 0 V stroke
  } def
/B { stroke [] 0 setdash 2 copy exch hpt sub exch vpt add M
  0 vpt2 neg V hpt2 0 V 0 vpt2 V
  hpt2 neg 0 V closepath stroke
  P } def
/C { stroke [] 0 setdash exch hpt sub exch vpt add M
  hpt2 vpt2 neg V currentpoint stroke M
  hpt2 neg 0 R hpt2 vpt2 V stroke } def
/T { stroke [] 0 setdash 2 copy vpt 1.12 mul add M
  hpt neg vpt -1.62 mul V
  hpt 2 mul 0 V
  hpt neg vpt 1.62 mul V closepath stroke
  P  } def
/S { 2 copy A C} def
end
}
\begin{picture}(3600,2160)(0,0)
\special{"
gnudict begin
gsave
50 50 translate
0.100 0.100 scale
0 setgray
/Helvetica findfont 100 scalefont setfont
newpath
-500.000000 -500.000000 translate
LTa
LTb
600 269 M
63 0 V
2754 0 R
-63 0 V
600 453 M
63 0 V
2754 0 R
-63 0 V
600 637 M
63 0 V
2754 0 R
-63 0 V
600 1005 M
63 0 V
2754 0 R
-63 0 V
600 1373 M
63 0 V
2754 0 R
-63 0 V
600 1741 M
63 0 V
2754 0 R
-63 0 V
600 2109 M
63 0 V
2754 0 R
-63 0 V
600 251 M
0 63 V
0 1795 R
0 -63 V
883 251 M
0 31 V
0 1827 R
0 -31 V
1048 251 M
0 31 V
0 1827 R
0 -31 V
1165 251 M
0 31 V
0 1827 R
0 -31 V
1256 251 M
0 31 V
0 1827 R
0 -31 V
1331 251 M
0 31 V
0 1827 R
0 -31 V
1394 251 M
0 31 V
0 1827 R
0 -31 V
1448 251 M
0 31 V
0 1827 R
0 -31 V
1496 251 M
0 31 V
0 1827 R
0 -31 V
1539 251 M
0 63 V
0 1795 R
0 -63 V
1822 251 M
0 31 V
0 1827 R
0 -31 V
1987 251 M
0 31 V
0 1827 R
0 -31 V
2104 251 M
0 31 V
0 1827 R
0 -31 V
2195 251 M
0 31 V
0 1827 R
0 -31 V
2270 251 M
0 31 V
0 1827 R
0 -31 V
2333 251 M
0 31 V
0 1827 R
0 -31 V
2387 251 M
0 31 V
0 1827 R
0 -31 V
2435 251 M
0 31 V
0 1827 R
0 -31 V
2478 251 M
0 63 V
0 1795 R
0 -63 V
2761 251 M
0 31 V
0 1827 R
0 -31 V
2926 251 M
0 31 V
0 1827 R
0 -31 V
3043 251 M
0 31 V
0 1827 R
0 -31 V
3134 251 M
0 31 V
0 1827 R
0 -31 V
3209 251 M
0 31 V
0 1827 R
0 -31 V
3272 251 M
0 31 V
0 1827 R
0 -31 V
3326 251 M
0 31 V
0 1827 R
0 -31 V
3374 251 M
0 31 V
0 1827 R
0 -31 V
3417 251 M
0 63 V
0 1795 R
0 -63 V
600 251 M
2817 0 V
0 1858 V
-2817 0 V
600 251 L
LT0
3174 1946 D
3417 403 D
3272 427 D
3126 450 D
2981 493 D
2835 536 D
2690 599 D
2544 647 D
2399 733 D
2253 823 D
2108 932 D
1962 1064 D
1817 1217 D
1672 1393 D
1526 1610 D
1381 1870 D
LT1
3174 1846 A
3417 372 A
3272 377 A
3126 385 A
2981 404 A
2835 404 A
2690 415 A
2544 423 A
2399 432 A
2253 441 A
2108 464 A
1962 459 A
1817 467 A
1672 494 A
1526 536 A
1381 586 A
1235 646 A
1090 727 A
944 808 A
799 911 A
LT2
3174 1746 B
3417 372 B
3272 377 B
3126 422 B
2981 410 B
2835 436 B
2690 415 B
2544 508 B
2399 465 B
2253 512 B
2108 464 B
1962 550 B
1817 505 B
1672 555 B
1526 511 B
1381 638 B
1235 536 B
1090 609 B
944 530 B
799 655 B
LT3
3174 1646 C
3417 393 C
3272 408 C
3126 422 C
2981 441 C
2835 430 C
2690 460 C
2544 467 C
2399 481 C
2253 515 C
2108 514 C
1962 553 C
1817 551 C
1672 600 C
1526 616 C
1381 588 C
1235 646 C
1090 627 C
944 669 C
799 655 C
stroke
grestore
end
showpage
}
\put(3054,1646){\makebox(0,0)[r]{Montvay}}
\put(3054,1746){\makebox(0,0)[r]{Bit reversal}}
\put(3054,1846){\makebox(0,0)[r]{Pairing}}
\put(3054,1946){\makebox(0,0)[r]{Naive}}
\put(2008,51){\makebox(0,0){$\epsilon$}}
\put(100,1180){%
\special{ps: gsave currentpoint currentpoint translate
270 rotate neg exch neg exch translate}%
\makebox(0,0)[b]{\shortstack{maximum $M$}}%
\special{ps: currentpoint grestore moveto}%
}
\put(3417,151){\makebox(0,0){0.1}}
\put(2478,151){\makebox(0,0){0.01}}
\put(1539,151){\makebox(0,0){0.001}}
\put(600,151){\makebox(0,0){0.0001}}
\put(540,2109){\makebox(0,0)[r]{1e+30}}
\put(540,1741){\makebox(0,0)[r]{1e+24}}
\put(540,1373){\makebox(0,0)[r]{1e+18}}
\put(540,1005){\makebox(0,0)[r]{1e+12}}
\put(540,637){\makebox(0,0)[r]{1e+06}}
\put(540,453){\makebox(0,0)[r]{1000}}
\put(540,269){\makebox(0,0)[r]{1}}
\end{picture}